%% file: main.tex
\documentclass[fleqn,usenatbib]{mnras}

\usepackage{mathptmx}
\usepackage[T1]{fontenc}

\DeclareRobustCommand{\VAN}[3]{#2}
\let\VANthebibliography\thebibliography
\def\thebibliography{\DeclareRobustCommand{\VAN}[3]{##3}\VANthebibliography}

\usepackage{graphicx}	
\usepackage[countmax]{subfloat}
\usepackage{amsmath}
\usepackage{caption, subcaption}
\usepackage{graphicx}	
\usepackage{amsmath}

\usepackage{amssymb}	
\usepackage{mathtools}
\usepackage{siunitx}
\usepackage{textgreek}
\usepackage{multirow}
\usepackage{caption}
\usepackage[normalem]{ulem}
\usepackage[dvipsnames]{xcolor}
\DeclareSIUnit \h {\ensuremath{\mathit{h}}}
\DeclareSIUnit \parsec {pc}
\DeclareSIUnit \solarmass {\ensuremath{\mathrm{M}_\odot}}
\DeclareSIUnit \dex {dex}

\newcommand{\hMsun}{ ~h^{-1}\mathrm{M_\odot}}

\newcommand{\hkpc}{ ~h^{-1}\mathrm{kpc}}
\newcommand{\hMpc}{ ~h^{-1}\mathrm{Mpc}}
\newcommand{\hMpcInv}{ ~h\mathrm{Mpc}^{-1}}
\newcommand{\hGpc}{ ~h^{-1}\mathrm{Gpc}}

\newcommand{\kms}{ ~\mathrm{km~s^{-1}}}

\newcommand{\sqdeg}{ ~\mathrm{deg}^2}

\title[The \textsc{Uchuu}-\textsc{GLAM} BOSS and eBOSS LRG lightcones]{The \textsc{Uchuu}-\textsc{GLAM} BOSS and eBOSS LRG lightcones: Exploring clustering and covariance errors}

\author[Ereza et al.]{
Julia Ereza$^{1}$\thanks{E-mail: jferrer@iaa.es},
Francisco Prada$^{1}$, Anatoly Klypin$^{2}$, Tomoaki Ishiyama$^{3}$, Alex Smith$^{4}$, Carlton M. Baugh$^{4}$, 
\newauthor
Baojiu Li$^{4}$, C\'esar Hern\'andez-Aguayo$^{5,6}$, Jos\'e Ruedas$^{1}$
\vspace*{4pt} \\ 
\scriptsize $^{1}$Instituto de Astrof\'isica de Andaluc\'ia (CSIC), Glorieta de la Astronom\'ia, E-18080 Granada, Spain \vspace*{-2pt}\\
\scriptsize $^{2}$ Department of Astronomy, University of Virginia, Charlottesville, VA 22904, USA  \vspace*{-2pt}\\
\scriptsize $^{3}$ Digital Transformation Enhancement Council, Chiba University, 1-33, Yayoi-cho, Inage-ku, Chiba, 263-8522, Japan  \vspace*{-2pt}\\
\scriptsize ${4}$ Institute for Computational Cosmology, Department of Physics, Durham University, South Road, Durham DH1 3LE, U.K.
\vspace*{-2pt}\\
\scriptsize $^{5}$ Max-Planck-Institut fuer Astrophysik, Karl-Schwarzschild-Str. 1, D-85748 Garching, Germany
 \vspace*{-2pt}\\
\scriptsize $^{6}$Excellence Cluster ORIGINS, Boltzmannstrasse 2, D-85748 Garching, Germany
\vspace*{-2pt}\\%
}

\date{Accepted XXX. Received YYY; in original form ZZZ}
\pubyear{2024}

\begin{document}
\label{firstpage}
\pagerange{\pageref{firstpage}--\pageref{lastpage}}
\maketitle


\begin{abstract}
\input{abstract}
\end{abstract}




\section{Introduction}
\label{sec:intro}
\input{intro}

\begin{figure*}
    \centering
    \includegraphics[width=\linewidth]{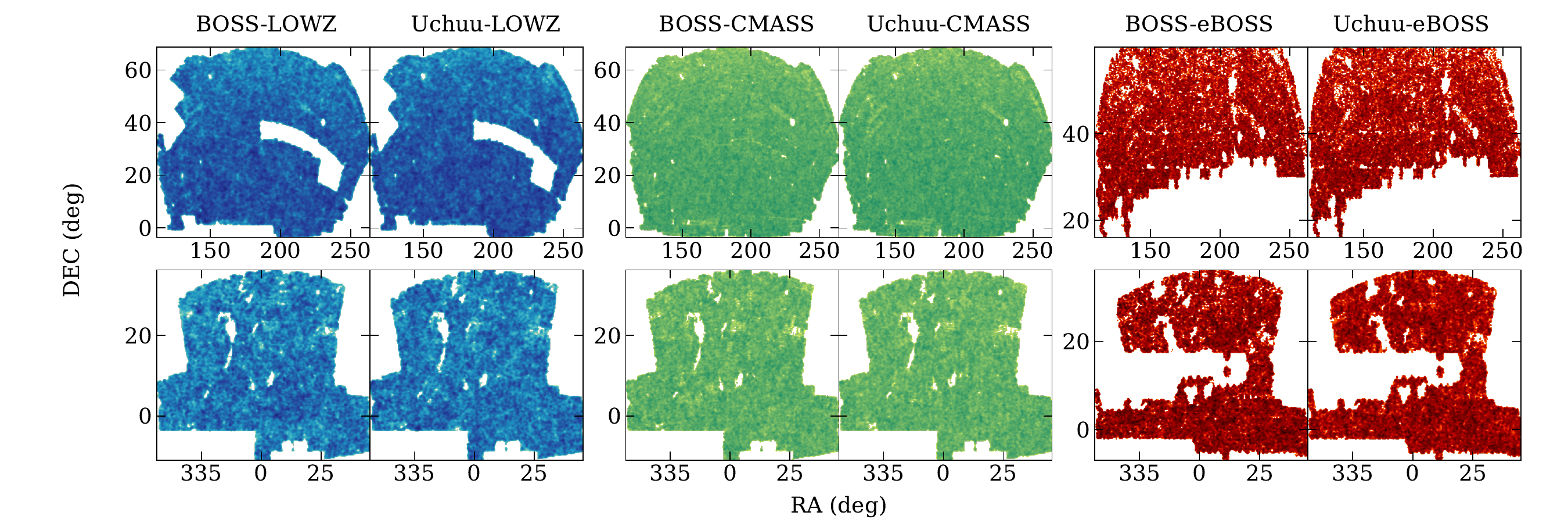}
    \caption{Sky area covered by the LOWZ, CMASS, and eBOSS data catalogues, and by the \textsc{Uchuu}-LOWZ, -CMASS, and -eBOSS lightcones. The upper and lower rows correspond to the northern and southern hemispheres, respectively. Colours indicate the angular number density normalized by the densest pixel. Darker colors represent regions of higher density.}
    \label{fig:radec_dens_boss}
\end{figure*}

\section{Galaxy data}
\label{sec:galaxy_data}

We make use of various publicly available observational datasets, including the BOSS-LOWZ, BOSS-CMASS and eBOSS-LRG samples, encompassing both the Northern (N) and Southern (S) hemispheres. 
Furthermore, we analyse the combined BOSS-LOWZCMASS samples for both hemispheres. For clarity, we will subsequently refer to these datasets as LOWZ, CMASS, COMB, and eBOSS. Below, we describe in detail each of these samples. A summary of their global properties is given in Table~\ref{tab:obs_info}.

\setlength{\arrayrulewidth}{0.3pt}
\begin{table}
\centering
    \begin{tabular}{lcccccc}
        \hline
        Name & $z$-range & $N_\mathrm{eff}$ & $A_\mathrm{eff}$ & $N_\mathrm{eff}/A_\mathrm{eff}$ & $V_\mathrm{eff}$ \\
        \hline
        LOWZ-N  & 0.2$-$0.4  & 196~186 & 5~790 & 33.9  & 0.3633   \\
        LOWZ-S  & 0.2$-$0.4  &  91~103 & 2~491 & 36.6  & 0.1621   \\
        CMASS-N & 0.43$-$0.7 & 605~884 & 6~821 & 88.8  & 1.2777   \\
        CMASS-S & 0.43$-$0.7 & 225~469 & 2~525 & 89.3  & 0.4745   \\
        COMB-N  & 0.2$-$0.75 & 807~900 & 5~790 & 139.5 & 1.7208   \\
        COMB-S  & 0.2$-$0.75 & 352~584 & 2~491 & 141.5 & 0.7468   \\
        eBOSS-N & 0.6$-$1.0  & 114~004 & 2~476 & 46.0  & 0.2584   \\
        eBOSS-S & 0.6$-$1.0  &  71~677 & 1~627 & 44.1  & 0.1602   \\
        \hline
    \end{tabular}
\caption{Properties of the BOSS and eBOSS samples studied in this work.
The first and second columns provide the sample name and the corresponding redshift range, respectively. The remaining columns list for each sample the (weighted) number of galaxies, the effective area (in $\sqdeg$), the galaxy density (in units of $\mathrm{deg}^{-2}$), and the effective volume (in $h^{-3}\mathrm{Gpc}^{3}$).}
\label{tab:obs_info}
\end{table}

\begin{figure}
    \centering
    \includegraphics[width=\linewidth]{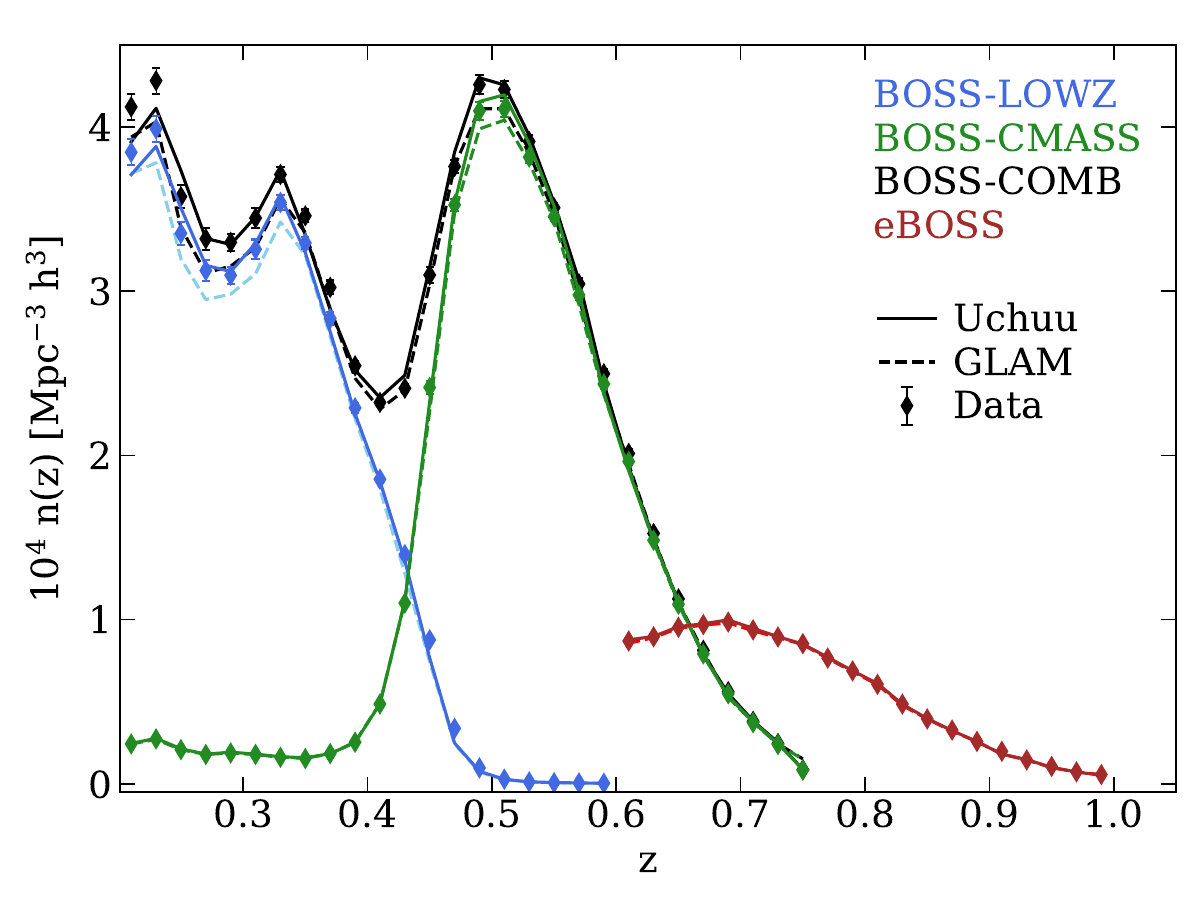}
    \caption{The comoving galaxy number density in the three samples studied: LOWZ in blue, CMASS in green, and eBOSS in red.
    The observational estimates are depicted using diamonds along with error bars. The mean $n(z)$ from both the \textsc{Uchuu} and \textsc{GLAM-Uchuu} lightcones are represented with solid and dashed lines, respectively (following the smoothing method described in the text).
    The error bars are obtained from the standard deviation of our set of 2000 \textsc{GLAM-Uchuu} lightcones for each of the samples.
    Additionally, for illustrative purposes, the $n(z)$ of the BOSS-COMB sample is represented by a solid black line.}
    \label{fig:nz}
\end{figure}

\subsection{BOSS samples}
\label{sec:cmass_lowz_info}

We use the final data release of SDSS-III BOSS LRG \citep[DR12;][]{Alam_clust}, which contains the redshifts of 1 million LRG. This dataset covers an area of approximately $10000$~deg$^2$, divided into two distinct subsets: LOWZ, designed for targeting LRG up to $z\sim0.4$, and CMASS, designed to target massive galaxies within the redshift range of $0.4<z<0.7$. Additionally, to maximize the effective volume covered by BOSS galaxies, a separate sample is constructed by combining the LOWZ and CMASS samples, denoted as the COMB sample. The spatial distribution and number density of these samples are shown in Figures~\ref{fig:radec_dens_boss}~and~\ref{fig:nz}.
The BOSS large-scale structure (LSS) catalogues are publicly available\footnote{\label{boss_public}Data catalogues and \textsc{mangle} masks are available at \href{https://data.sdss.org/sas/dr12/boss/lss/}{SDSS-BOSS Science Archive Server}}. For a more comprehensive understanding of the LOWZ and CMASS target selection criteria, and the creation of the COMB sample, we refer to \cite{Reid_lss}.

In these catalogues, as defined in \citet{Alam_clust}, a total weight is assigned to each galaxy $i$:
\begin{equation}
    w_{\textrm{tot},i}=w_{\textrm{systot},i}(w_{\textrm{cp},i}+w_{\textrm{noz},i}-1),
    \label{eq:boss_w}
\end{equation}
where $w_\textrm{systot}$ is the total angular systematic weight, $w_\textrm{cp}$ is the fibre collision correction weight, and $w_\textrm{noz}$ is the redshift failure weight. This total galaxy weight must be used to obtain an unbiased estimation of the galaxy density field.

The geometry of the sky area covered by each LRG sample is precisely outlined using \textsc{mangle} format masks\footref{boss_public}. Regarding the BOSS-COMB sample, the mask is constructed to include every sector (each area covered by a unique set of plates) contained within the CMASS mask into the LOWZ counterpart. 
Furthermore, information regarding the BOSS fibre completeness is recorded within the \textsc{mangle} mask files. The completeness of every observed sector (indexed by $j$) is then defined as follows:
\begin{equation}
    C_{\textrm{BOSS},j}=\frac{N_{\textrm{obs},j}+N_{\textrm{cp},j}}{N_{\textrm{star},j}+N_{\textrm{gal},j}+N_{\textrm{fail},j}+N_{\textrm{cp},j}+N_{\textrm{missed},j}},
    \label{eq:completeness}
\end{equation}
where:

\begin{description}
\item $N_{\textrm{gal},j}$ is the number of galaxies with redshifts from good BOSS spectra.   
\item $N_{\textrm{star},j}$ is the number of spectroscopically confirmed stars.
\item $N_{\textrm{fail},j}$ is the number of objects with BOSS spectra for which either stellar classification or redshift determination failed.
\item $N_{\textrm{obs},j}$ is defined as $N_{\textrm{obs},j} = N_{\textrm{gal},j}+N_{\textrm{star},j}+N_{\textrm{fail},j}$.
\item $N_{\textrm{cp},j}$ is the number of objects with no spectra, in a fiber collision group where at least one other object of the same target class has a spectrum.
\item $N_{\textrm{missed},j}$ is the number of objects with no spectra, and if in a fibre collision group, there are no other objects from the same target class.
\end{description}

\noindent On the edges of the survey there are regions of high incompleteness. To avoid these sectors, where a substantial fraction of redshifts is absent, sectors with $C_{\textrm{BOSS},j}<0.7$ are excluded from analysis and are not considered within the scope of this work.

The LSS catalogues for LOWZ and CMASS do not include galaxy stellar masses. We obtain this information from the lightcones presented in \citet{Sergio} and \citet{Kitaura_2016}. 
The observed LRG stellar mass functions (SMFs) are shown in the left and center panels of Fig.~\ref{fig:smf} for LOWZ and CMASS, respectively.

\subsection{eBOSS LRG sample}
\label{sec:eboss_info}

We use the final data release of SDSS-IV eBOSS LRG \citep[DR16;][]{eboss_lrgs}, which contains $\sim200000$ LRG over $4000$~deg$^2$. This sample spans a higher redshift range than the CMASS sample, covering $0.6<z<1$ \citep{Ashely_eboss_lss}. The spatial distribution and number density of this eBOSS sample can be seen in Figures~\ref{fig:radec_dens_boss}~and~\ref{fig:nz}. 
The final LSS catalogues for eBOSS are publicly available\footnote{\label{eboss_public}Data catalogues and \textsc{mangle} masks are available at \href{https://dr15.sdss.org/sas/dr16/eboss/lss/catalogs/DR16/}{SDSS-eBOSS Science Archive Server}}.

In these catalogues, the total weight for every galaxy, $i$, as described in \citet{Ashely_eboss_lss} is:
\begin{equation}
w_{\textrm{tot},i}=w_{\textrm{systot},i}~w_{\textrm{cp},i}~w_{\textrm{noz},i}.
    \label{eq:eboss_w}
\end{equation}
where $w_{\textrm{systot},i}$, $w_{\textrm{cp},i}$, and 
$w_{\textrm{noz},i}$ are the same weights as described for BOSS.

The geometry of the sky area covered by the eBOSS sample is detailed within a \textsc{mangle} format mask\footref{eboss_public}. The completeness for each sector is determined using the same methodology as employed for the BOSS samples.  
Following the observational systematics processing described in \citet{Ashely_eboss_lss}, we apply a spectroscopic completeness threshold of $C_{\textrm{eBOSS},j}<0.5$ to exclude sectors with low completeness.  

The SMF of the eBOSS galaxies is shown in the right panel of Fig.~\ref{fig:smf}. Stellar masses are obtained from the data outputs provided by \citet{comparat_maraston} through the use of the \texttt{FIREFLY\footnote{\url{http://www.icg.port.ac.uk/firefly/}}} code applied to the DR16 eBOSS spectra. To avoid any overlap with the BOSS-CMASS LRG, eBOSS adopted a specific magnitude cut of $i \geq 19.9$. This limit results in the most massive LRG being excluded from the eBOSS sample. For details regarding the LRG selection, refer to \citet{eboss_lrgs}.

\section{Cosmological simulations} 
\label{sec:cosmo_simu}

\begin{table*}
\centering
    \begin{tabular}{lccccccccccccc}
        \hline
        Simulation & $L_\mathrm{box}$ & $N_\mathrm{part}$ & $m_\mathrm{part}$ & $h$ & $\Omega_\mathrm{m_0}$ & $\Omega_\mathrm{\Lambda_0}$ & $\Omega_\mathrm{b_0}$ & $\sigma_8$ & $N_\mathrm{g}$ & $\epsilon$ & $z_\mathrm{init}$ & $N_\mathrm{s}$ & $N_\mathrm{r}$ \\
        \hline
        \textsc{Uchuu}           & 2~000 & $12~800^3$ & $3.27\times10^{8}$  & 0.677 & 0.309 & 0.691 & 0.0486 & 0.816 & 468~384 & 0.00427 & 127 & 5~570 & 1 \\
        \textsc{GLAM-Uchuu}      & 1~000 &  $2~000^3$ & $1.06\times10^{10}$ & 0.677 & 0.309 & 0.691 & 0.0486 & 0.816 &   5~800 &   0.172 & 104   & 136 & 2~000    \\
        \textsc{GLAM-PMILL}      & 1~000 &  $2~000^3$ & $1.06\times10^{10}$ & 0.678 & 0.307 & 0.693 & 0.0481 & 0.828 &   4~000 &   0.250 & 104   & 136 & 100    \\
        \textsc{GLAM-PMILLnoBAO} & 1~000 &  $2~000^3$ & $1.06\times10^{10}$ & 0.674 & 0.307 & 0.693 & 0.0487 & 0.828 &   4~000 &   0.250 & 104   & 136 & 100    \\
        \textsc{GLAM-Abacus}     & 1~000 &  $2~000^3$ & $1.09\times10^{10}$ & 0.674 & 0.315 & 0.685 & 0.0493 & 0.812 &   7~500 &   0.133 & 104   & 136 & 100    \\
        \hline
    \end{tabular}
\caption{Numerical and cosmological parameters of the \textsc{Uchuu} and \textsc{GLAM-Uchuu} simulations used in this work. 
The columns, from left to right, include: the simulation name, size of the simulated box $L_\mathrm{box}$ in $\hMpc$, the number of particles $N_\mathrm{part}$, the mass of the simulation particle $m_\mathrm{part}$ in $\hMsun$ units, the cosmological parameters adopted in the simulation ($h$, $\Omega_\mathrm{m_0}$, $\Omega_\mathrm{\Lambda_0}$, $\Omega_\mathrm{b_0}$, $\sigma_8$), the mesh size $N_\mathrm{g}$, the gravitational softening length $\epsilon$ in $\hMpc$ units, the redshift of the initial linear power spectrum $z_\mathrm{init}$, the number of time-steps $N_\mathrm{s}$, and number of realizations $N_\mathrm{r}$.}
\label{tab:simus_info}
\end{table*}

Using the \textsc{Uchuu} and \textsc{GLAM} simulations listed in Table~\ref{tab:simus_info}, we generate galaxy lightcones as well as lightcones for covariance errors for each of the analysed samples: LOWZ, CMASS, COMB, and eBOSS. The goal of these lightcones is to faithfully reproduce the clustering statistics observed in the survey data. 

\begin{table*}
    \centering
    \begin{tabular}{cccccccc}
        \hline
        Simulation & Sample & $N_\mathrm{light}$ & $z$-range & $N_\mathrm{eff}$ & $A_\mathrm{eff}$ & $N_\mathrm{eff}/A_\mathrm{eff}$ & $V_\mathrm{eff}$ \\
        \hline 
        \multirow{8}{*}{\textsc{Uchuu}} & LOWZ-N  & 2 & 0.2$-$0.4  & 193~973  & 5~751 & 33.7  & 0.36   \\
        & LOWZ-S  & 2 & 0.2$-$0.4  &  89~944  & 2~470 & 36.4  & 0.16   \\
        & CMASS-N & 2 & 0.43$-$0.7 & 602~206  & 6~798 & 88.8  & 1.27   \\
        & CMASS-S & 2 & 0.43$-$0.7 & 221~921  & 2~482 & 89.3  & 0.47   \\
        & COMB-N  & 2 & 0.2$-$0.75 & 806~468  & 5~751 & 140.2 & 1.71  \\
        & COMB-S  & 2 & 0.2$-$0.75 & 351~348  & 2~470 & 142.3 & 0.74  \\
        & eBOSS-N & 2 & 0.6$-$1.0  & 113~400  & 2~462 & 46.1  & 0.26   \\
        & eBOSS-S & 2 & 0.6$-$1.0  &  71~424  & 1~621 & 44.1  & 0.16   \\
        \hline
        \multirow{8}{*}{\textsc{GLAM-Uchuu}} & LOWZ-N  & 2~000 & 0.2$-$0.4  & 188~160 & 5~751 & 32.7  & 0.36  \\
        & LOWZ-S  & 2~000 & 0.2$-$0.4  &  86~194 & 2~470 & 34.9  & 0.16  \\
        & CMASS-N & 2~000 & 0.43$-$0.7 & 593~339 & 6~798 & 87.3  & 1.26  \\
        & CMASS-S & 2~000 & 0.43$-$0.7 & 217~727 & 6~798 & 87.7  & 0.46  \\
        & COMB-N  & 2~000 & 0.2$-$0.75 & 787~474 & 5~751 & 136.9 & 1.69  \\
        & COMB-S  & 2~000 & 0.2$-$0.75 & 341~650 & 2~470 & 138.3 & 0.73  \\
        & eBOSS-N & 2~000 & 0.6$-$1.0  & 112~259 & 2~462 & 45.6  & 0.25  \\
        & eBOSS-S & 2~000 & 0.6$-$1.0  &  70~510 & 1~621 & 43.5  & 0.17  \\
        \hline
    \end{tabular}
    \caption{Properties of the \textsc{Uchuu} and \textsc{GLAM-Uchuu} lightcones generated for each observational sample.
    The first column indicates the simulation used.
    The second and third columns provide the name and number of lightcones ($N_\mathrm{light}$) defining each sample dataset.
    The fourth column gives the redshift range considered for clustering studies. Subsequent columns provide the (weighted) mean number of galaxies, the mean effective area (in $\sqdeg$), the mean density of galaxies per unit area (in $\mathrm{deg}^{-2}$), and the mean effective volume (in $h^{-3}\mathrm{Gpc}^{3}$).}
    \label{tab:Uchuu_GUchuu_info}
\end{table*}

\subsection{The \textsc{Uchuu} Simulation} 
\label{sec:uchuu_simu}

The \textsc{Uchuu} simulation\footnote{All \textsc{Uchuu} data products are publicly available at \href{http://www.skiesanduniverses.org/Simulations/Uchuu/}{Skies \& Universes, \textsc{Uchuu} Simulation}, including simulated catalogues constructed using various methods~\citep{chian,Aung2023,Oogi2023,Prada2023}.} is the largest run within the \textsc{Uchuu} suite \citep{Ishiyama21}, tracing the evolution of $12,800^3$ dark matter particles, each with a mass of \SI{3.27e8}{\per\h\solarmass}, within a periodic box of \SI{2.0}{\per\h\giga\parsec}. The cosmological parameters adopted are $\Omega_\mathrm{m_0}=0.309$, $\Omega_\mathrm{b_0}=0.0486$, $\Omega_\mathrm{\Lambda_0}=0.691$, $h=0.677$, $n_\mathrm{s}=0.9667$, and $\sigma_8=0.816$, representing the best fitting \textLambda CDM parameters corresponding to the Planck 2015 cosmology \citep[][]{planck15}. Employing the TreePM code \textsc{GreeM} \citep{Ishiyama09, Ishiyama12} with a gravitational softening length of \SI{4.27}{\per\h\kilo\parsec}, \textsc{Uchuu} tracks the gravitational evolution of particles from redshift $z=127$ to $z=0$. A total of 50 particle snapshots were stored over the redshift interval from $z=0$ to $z=14$. The identification of bound structures was performed using the \textsc{Rockstar} phase-space halo/subhalo finder \citep{Behroozi13}; 
subhalos in \textsc{Uchuu} are $90\%$ complete down to $V_\mathrm{peak} \gtrsim \SI{70}{\kilo\meter\per\second}$, while distinct halos are $90\%$ complete down to $V_\mathrm{peak} \gtrsim \SI{50}{\kilo\meter\per\second}$.
Merger trees were then constructed using the \textsc{Consistent Trees} code \citep{Behroozi2013b}. The properties of the \textsc{Uchuu} simulation are summarized in Table~\ref{tab:simus_info}.

The high resolution of the \textsc{Uchuu} simulation makes it optimal for generating simulated galaxy lightcones. \textsc{Uchuu} is able to resolve dark matter halos and subhalos down to small masses, a key requirement for employing the \textsc{SHAM}. This capability results in high completeness, particularly in terms of the fraction of galaxies resolved at lower redshift. Furthermore, the large volume of \textsc{Uchuu} allows us to study large-scale clustering features, such as the BAO.

\subsection{The \textsc{GLAM} simulations}
\label{sec:glam_simu}

The \textsc{GLAM} simulations \citep{GLAMsimu} are $N$-body cosmological simulations that follow the evolution of $2000^3$ dark matter particles, each having a particle mass  of \SI{1.06e10}{\per\h\solarmass}. All \textsc{GLAM} boxes are \SI{1.0}{\per\h\giga\parsec} comoving periodic boxes, with $N_{\rm s}=136$ time-steps and a mesh of $N_{\rm g}=5800$ cells per side, resulting in a spatial resolution of $0.172\hMpc$. The initial conditions are generated using the Zeldovich approximation starting at $z_\mathrm{ini}=104$. Because of the lower resolution of these simulations compared to \textsc{Uchuu}, the \textsc{GLAM} simulations are only capable of resolving distinct halos (not subhalos) with virial masses greater than $10^{12}\hMsun$ \citep{CesarGLAM}.

The \textsc{GLAM} simulations have been run assuming different cosmologies. The numerical and cosmological parameters adopted in these simulations are listed in Table~\ref{tab:simus_info} and summarized below:
\begin{itemize}
    \item \textsc{GLAM-Uchuu}: Adopts the same Planck cosmology (PL15) and linear power spectrum as the \textsc{Uchuu} simulation.
    \item \textsc{GLAM-PMILL} and -\textsc{PMILLnoBAO}: Adopt the cosmological parameters used in the Planck Millennium simulation \citep[PMILL;][]{Pmill_cosmo}, which uses the best-fitting \textLambda CDM parameters from the first Planck 2013 data release \citep[PL13;][]{planck13}. In \textsc{PMILLnoBAO}, an initial power spectrum that models a matter distribution without baryonic acoustic oscillations is used, leading to the absence of the BAO feature in these realisations.
    \item \textsc{GLAM-Abacus}: Adopts the \texttt{abacus$\_$cosm000} cosmology from the \textsc{AbacusSummit} $N$-body simulation suite \citep{Abacus}, which includes the effect of massive neutrinos.
\end{itemize}

\section{Constructing the \textsc{Uchuu-LRG} lightcones}
\label{sec:uchuu_light}

\begin{figure*}
    \centering
    \includegraphics[width=\linewidth]{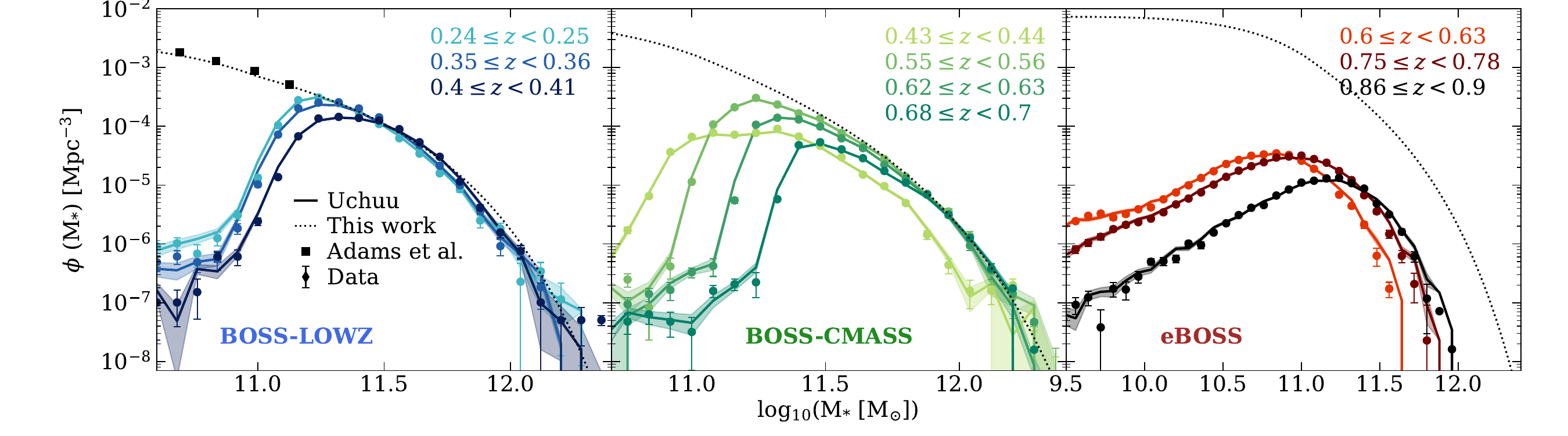}
    \caption{The observed LRG SMF (points) in the LOWZ (left panel), CMASS (middle panel), and eBOSS (right panel) samples.
    The mean SMF of the \textsc{Uchuu-LRG} lightcones is shown using solid lines.
    We show the SMF for several redshift bins within the relevant redshift ranges for each sample, using a colour scheme indicated in each plot.
    The dotted curves show the complete SMFs adopted for the implementation of the \textsc{SHAM} method (see Table~\ref{tab:sch_params}), while the square points denote the PRIMUS data taken from \citet{Moustakas_Primus} (see Section~\ref{sec:sham}).
    Data error bars represent the standard deviation, 1-$\sigma$, of our set of 2000 \textsc{GLAM-Uchuu} lightcones, while the shaded area for \textsc{Uchuu} represents the error on the mean, $\sigma/\sqrt{2}$.
    }
    \label{fig:smf}
\end{figure*}

\begin{figure*}
    \centering
    \includegraphics[width=\columnwidth]{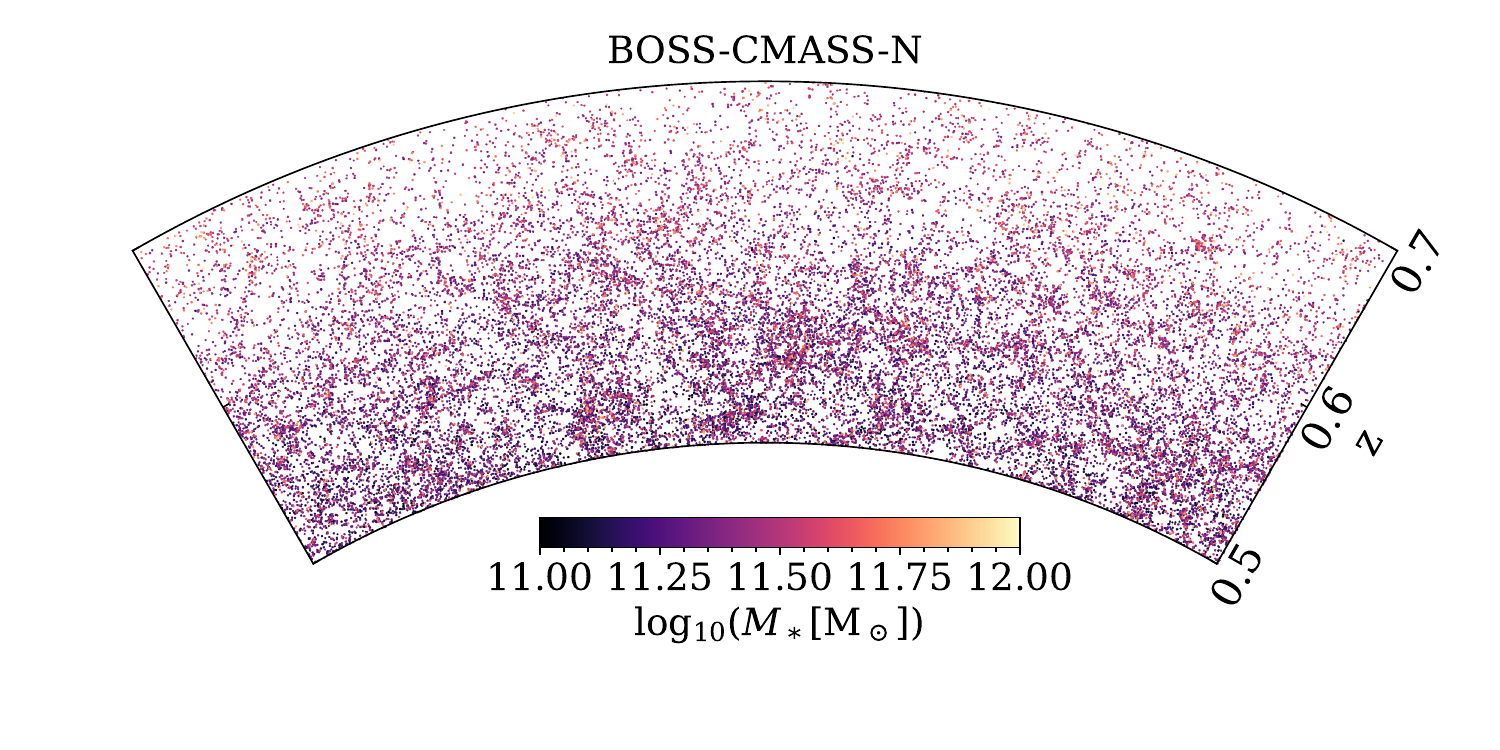}
    \includegraphics[width=\columnwidth]{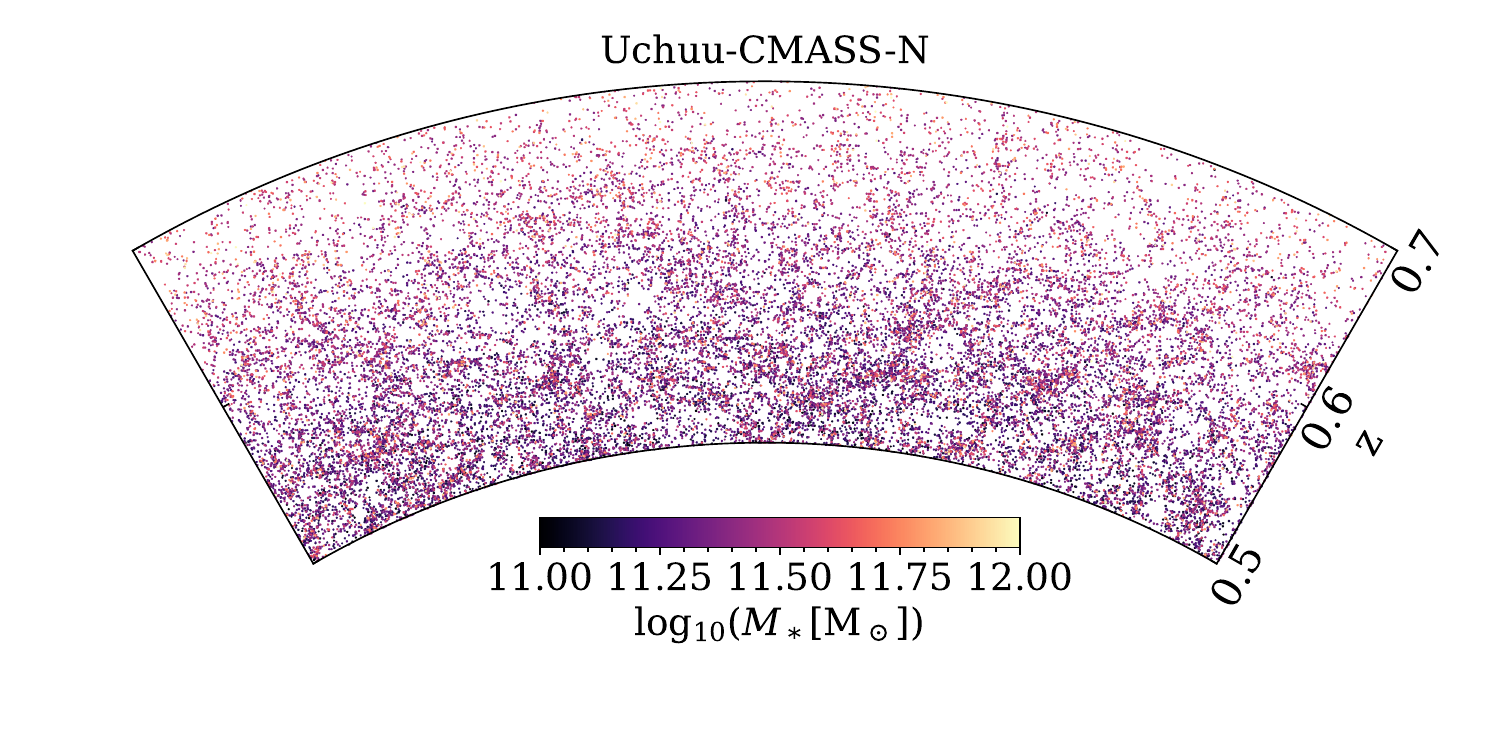}
    \caption{Slices through the CMASS-N sample (left), and the \textsc{Uchuu}-CMASS-N lightcone (right), with galaxies colour-coded according to their stellar mass as indicated in the legend. The displayed galaxies fall within the range of $10 < \mathrm{dec} < 30~\deg$, and $150 < \mathrm{RA} < 210~\deg$.}
    \label{fig:mest_cheese}
\end{figure*}

We will now outline the procedure employed to create LRG lightcones from the \textsc{Uchuu} simulation. This includes details about the \textsc{SHAM} method, modeling of the LRG stellar mass incompleteness, handling angular masks, and addressing fiber collisions. For each of the 8 analyzed samples, we generate 2 independent lightcones by duplicating the survey footprint across the sky, resulting in a total of 16 lightcones. The properties of the \textsc{Uchuu} lightcones are shown in Table~\ref{tab:Uchuu_GUchuu_info}.

To capture the redshift evolution of galaxy clustering, we use a series of \textsc{Uchuu} snapshots. For each sample, the redshifts of the boxes used are:
\begin{description}
    \item LOWZ:~~~$z=0.19, 0.25, 0.30, 0.36, 0.43$
    \item CMASS: $z=0.43, 0.49, 0.56, 0.63, 0.7$
    \item eBOSS:~~~$z=0.63, 0.7, 0.78, 0.86$
\end{description}

\noindent Fig.~\ref{fig:mest_cheese} shows a comparison between a slice of the CMASS-N galaxy distribution in the observations (left panel) and the corresponding \textsc{Uchuu-LRG} lightcone (right panel). Individual galaxies are colour-coded based on their stellar mass, highlighting the consistency between our simulated lightcones and the observed data. The decrease in galaxy density with increasing redshift matches the observed redshift distribution $n(z)$ in this particular sample (see Fig.~\ref{fig:nz}).

\subsection{Stellar mass assignment}
\label{sec:sham}

We adopt the \textsc{SHAM} method to populate galaxies within the dark matter halos and subhalos in each \textsc{Uchuu} simulation box. This technique assumes a monotonic relation between the stellar mass of galaxies and a specific property of the associated halo, which is a proxy of its mass, thereby assigning more massive galaxies to more massive (sub)halos \citep{Guo-HAM_vs_HOD}. In the standard \textsc{SHAM} approach, there is no distinction between halos and subhalos, since the assignment is only based on a subhalo property and not the environment of the subhalo (although the method has been extended to include such information, see \citealt{Contreras:2021}).

For the SHAM technique, we use the peak value of the maximum circular velocity over the history of the (sub)halo, $V_\mathrm{peak}$, as a proxy for the mass. This choice of property has been shown to yield galaxy catalogues that exhibit better agreement with observational data \citep[as discussed in][]{Trujillo,Reddick_2013, Guo-HAM_vs_HOD}. Several reasons underpin this improvement: 
i) circular velocity parameters are typically reached at one-tenth of the (sub)halo's radius, which provides a more accurate characterization of the scales directly influencing galaxies \citep{Jonas}; 
ii) $V_\mathrm{peak}$ is better defined for both halos and subhalos when compared to halo mass, as the latter becomes ambiguous for subhalos. Furthermore, subhalo mass is also influenced by the choice of halo finder, as it depends on a code-defined truncation radius at which the particles are defined as unbound and are removed \citep{Trujillo}.
iii) the circular velocity is less affected than a mass measurement by the mass loss experienced by a halo or subhalo after it becomes part of a large structure \citep{Kravtsov, Hayashi_2003}.

It is important to note that while the original \textsc{SHAM} approach assumes a monotonic relation between the galaxy stellar mass and $V_\mathrm{peak}$, observations suggest that this assignment cannot be strictly one-to-one. To achieve more realistic results, it becomes necessary to introduce scatter into the abundance matching process \citep[e.g.][]{Skibba_2010, Trujillo}. We adopt the  scatter scheme proposed in \citet{Sergio}, which defines the scattered $V_\mathrm{peak}$ as the abundance matching parameter:
\begin{equation}
V_\mathrm{peak}^\mathrm{scat}=(1+ \mathcal{N}(0,\sigma_\mathrm{SHAM}))V_\mathrm{peak},
\label{eq:sigma_sham}
\end{equation}
where $\mathcal{N}$ represents a Gaussian-distributed random number with a mean of zero and standard deviation $\sigma_\mathrm{SHAM}$ (the SHAM scatter parameter). 
The optimal values for the scatter parameters are determined by optimizing the goodness-of-fit between the two-point correlation functions measured from the observational samples and those obtained from the \textsc{Uchuu} lightcones. 
Note that the footprint and number density of these lightcones are set beforehand to match the data, as described in Section~\ref{sec:nz_and_mask}.
The mean \textsc{SHAM} scatter parameters over the boxes of each sample are as follows: $\overline{\sigma}_\mathrm{LOWZ}=0.26$, $\overline{\sigma}_\mathrm{CMASS}=0.30$, $\overline{\sigma}_\mathrm{eBOSS}=0.65$, all in agreement with previous work \citep[see][]{Sergio, nuza_2013, Yu_2022}. 
We note that these values increase with increasing redshift. The scatter is not necessarily directly linked to redshift. 
However, redshift could influence the halo-galaxy connection due to the dynamic processes of galaxy formation and evolution. In addition, as redshift values increase, observations might encounter higher levels of noise and limitations from instruments. 
The inferred value for the eBOSS scatter parameter seems substantially higher than the values recovered for the other samples which may indicate that it is influenced by the high incompleteness in the SMF for this sample.

In order to implement the \textsc{SHAM} technique in the \textsc{Uchuu} boxes, we require a complete SMF, which accurately represents a complete population of galaxies prior to applying any of the magnitude and colour selections specific to the LRG samples considered here.  
Due to the selection function in the BOSS and eBOSS surveys, the SMF becomes incomplete in these samples at low masses  (see Fig.~\ref{fig:smf}).
For CMASS and eBOSS we adopt the Schechter function taken from \citet{Sergio} as the complete SMF, which is shown by the dashed line in Fig.~\ref{fig:smf}.
These authors combine the CMASS SMF, which is complete at higher stellar masses, with the SMF from the PRIMUS survey at lower masses \citep{Moustakas_Primus}.
In the case of LOWZ, we also incorporate the averaged SMF measurements from PRIMUS for the redshift ranges $0.2<z<0.3$ and $0.3<z<0.5$ (square symbols). Subsequently, we perform a fit to both data sets using a Schechter function. 
All the fits are shown as dotted lines in Fig.~\ref{fig:smf}. The fitted parameters are given in Table~\ref{tab:sch_params}.
\begin{table}
\centering
    \begin{tabular}{ccccc}
        \hline
        Sample & Mass range & $10^3~\phi_\ast$ & $\alpha$ & log$_{10}$~$M_\ast$\\
        \hline
        \multirow{2}{*}{LOWZ} & $M_\ast\leq~10^{11.3}$& $2.894$& $-0.996$& $10.82$\\
                              & $M_\ast>~10^{11.3}$& $0.476$& $-1.557$& $11.23$\\
        \multirow{2}{*}{CMASS $\&$ eBOSS} & $M_\ast\leq~10^{11.0}$ & $4.002$ & $-0.938$ & $10.76$ \\
                                         & $M_\ast>~10^{11.0}$    & $0.266$ & $-2.447$ & $11.42$ \\
        \hline
    \end{tabular}
\caption{Fitted parameters of the Schechter functions adopted in this work as complete SMFs for the LOWZ, CMASS, and eBOSS samples. The characteristic stellar mass, $M_\ast$, is given in units of $M_\odot$, and $\phi_\ast$ is given in units of Mpc$^{-3}~(\log_{10}M_\ast)^{-1}$.}
\label{tab:sch_params}
\end{table}

Once we have the complete SMF fits and the values of  $V_\mathrm{peak}^\mathrm{scat}$ for each halo and subhalo in \textsc{Uchuu} (i.e. after including the scatter), we are able to apply the \textsc{SHAM} technique. The procedure is as follows:
\begin{enumerate}
    \item Compute the cumulative number density of (sub)halos as a function of decreasing $V_\mathrm{peak}^\mathrm{scat}$, i.e., the scattered value of the circular velocity. 
    \item Compute the cumulative number density of galaxies as a function of decreasing stellar mass, $M_{\ast}$, using the adopted complete SMF.
    \item Construct a monotonic relation between the cumulative number density functions from steps (i) and (ii):
    \begin{equation}
        n_\mathrm{gal}(>M_{\ast})=n_\mathrm{halo}(>V^\mathrm{scat}_\mathrm{peak}),
    \end{equation}
    which implies that a (sub)halo with $V^\mathrm{scat}_{\mathrm{peak},i}$ will contain a galaxy with stellar mass $M_{\ast,i}$, assigning the most massive galaxy to the (sub)halo with the highest $V^{\rm scat}_{\rm peak}$. We note that this assignment is monotonic between $V^\mathrm{scat}_\mathrm{peak}$ and $M_{\ast}$, but not between $V_\mathrm{peak}$ and $M_{\ast}$. 
    \item Repeat this sequence for every \textsc{Uchuu} box.
\end{enumerate}

The next step is to model the incomplete stellar mass distribution observed in all samples.
For \textsc{Uchuu}-LOWZ and -CMASS galaxy boxes, we do this by randomly down-sampling galaxies from the complete SMFs.
The approach is different for the case of eBOSS, as the SMF is highly incomplete in this instance for the whole stellar mass range (see the right panel of Fig.~\ref{fig:smf}). To account for this effect, we rely on the results presented in \citet{Alam_2020}, where they report that the probability of a distinct halo containing a central LRG is at most 30$\%$.
Therefore, after applying the \textsc{SHAM} on the \textsc{Uchuu}-eBOSS boxes, we randomly select 30$\%$ of the galaxies. 
The remainder of the observed incompleteness, primarily located at the lower end of stellar masses, is acquired through the same methodology as applied in the LOWZ and CMASS samples.
The final SMFs of our \textsc{Uchuu-LRG} lightcones are shown by the solid curves in Fig.~\ref{fig:smf}.
We validate our scheme for assigning stellar masses to galaxies by measuring the galaxy clustering in the lightcones as a function of stellar mass, as described below.

\subsection{Creating the lightcones by joining the cubic boxes}
\label{sec:clus_evol_z}

The next step in the production of the \textsc{Uchuu} lightcones involves joining the cubic boxes that have been populated with galaxies, into spherical shells  \citep[see][for a detailed description of this method]{Smith22b}. 

First, we position the observer at $(0, 0, 0)\hGpc$ in comoving coordinates and transform the Cartesian coordinates (x,y,z) of each galaxy within the cubic box to redshift-space coordinates (RA,dec,z), considering the effect of galaxy peculiar velocities.
Each box is then divided into spherical shells, where the comoving distance between the observer and the inner/outer edges of the shell corresponds to the redshift that lies midway between this snapshot and the next/previous snapshot. These spherical shells are then joined together to build three full-sky lightcones: \textsc{Uchuu}-LOWZ, -CMASS and -eBOSS.
To cover their sky areas, it becomes necessary to periodically replicate the corresponding galaxy cubic boxes at specific redshift ranges.
If the centre of a halo lies within a given spherical shell, the entire halo is included in the lightcone. However, if a halo partially extends into a shell but its center falls outside the shell, it is excluded from the lightcone.

\subsection{Radial selection function and angular mask}
\label{sec:nz_and_mask}

We randomly downsample the galaxy population in the redshift ranges where the lightcone's $n(z)$ surpasses the observed one. By doing this, we also implement a smooth transition between the spherical shells, avoiding a sudden step-like behaviour. 
The redshift distribution of galaxies in the \textsc{Uchuu} lightcones resulting from this process is shown by the solid lines in Fig.~\ref{fig:nz}.

We proceed to match the sky area of the full-sky \textsc{Uchuu} lightcones with that of the observed samples, using the available survey masks (see Sections~\ref{sec:cmass_lowz_info}~and~\ref{sec:eboss_info}). 
Additionally, we apply various veto masks. These include masks for bright stars, bright objects, and non-photometric conditions (in the real data). Moreover, we incorporate veto masks based on the seeing during the imaging data acquisition and Galactic extinction (avoiding high-extinction areas), see \citealp{Ashely_boss_lss} and \citealp{Ashely_eboss_lss} for details.
The result of applying these masks and the comparison with observed data is shown in Fig.~\ref{fig:radec_dens_boss}. 
To improve our statistics and benefiting from our lightcones (without masks) covering the whole sky area, we repeat this procedure once more by shifting the sky position of the masks by $180$~deg. By doing this, we are able to obtain two independent \textsc{Uchuu} lightcones for each sample.

As shown in Table~\ref{tab:Uchuu_GUchuu_info}, the effective area of the \textsc{Uchuu} lightcones is slightly smaller than that of the actual data ($\sim1\%$ lower). The angular geometry of the samples is represented by \textsc{mangle} polygons, which lack the finest angular footprint details, contributing to this minor variation in the effective area.

\subsection{Completeness and fibre collisions}
\label{sec:compl_fib}

A final step is required to fully replicate the observed samples, which involves the incorporation of survey systematics. This includes the angular completeness of the masks, $C$, defined in Equation~\ref{eq:completeness}. To account for this property, we downsample the regions where the completeness falls below one, retaining only those regions where $C\geq~0.7$ for \textsc{Uchuu}-LOWZ and -CMASS, and $C\geq~0.5$ for \textsc{Uchuu}-eBOSS. 

Another observational artefact is the so-called fiber collisions. This is due to the finite size of the fiber-head on the plate. If two galaxies are separated by less than 62 arcsec apart (which represents the fiber collision radius), it is possible that only one of these galaxies will be assigned a spectroscopic fiber, resulting in the inability to determine the redshift of the other galaxy \citep{Reid_lss}. The distribution of such closely spaced galaxy pairs is expected to be correlated with galaxy density. To model this effect, we adopt the method described in \citet{fibre_coll}, where the main goal is to divide the complete galaxy sample into two distinct populations: 

\textit{Population 1 (P1)}: A subsample where no galaxy is within the fiber collision radius of any other galaxy within this same subsample. 

\textit{Population 2 (P2)}: A subsample including all galaxies that are not of P1. Every galaxy in this population lies within the fiber collision radius of a galaxy in P1. All the galaxies affected by fiber collisions belong to this subsample. 

\noindent The procedure for creating these two populations is as follows:
\begin{enumerate}
    \item Initially, all galaxies in the lightcone are designated as members of P1. 
    \item For each pair of collided galaxies in P1, we randomly select one galaxy to be reassigned to P2.
    \item For each group of three or more collided galaxies in P1, we chose the galaxy that collides with the most P1 galaxies and assign it to P2.
    \item We repeat steps (ii) and (iii) until no galaxies in P1 collides with each other.
\end{enumerate}
Once the two populations have been created, we proceed to randomly select from P2 the galaxies that will not be observed. We do this for all the \textsc{Uchuu} lightcones, resulting in the final percentages of fiber-collided galaxies as follows: $1.3\%$ ($2.9\%$) for LOWZ-N (S) lightcones, $5.0\%$ ($4.3\%$) for CMASS, and $3.4\%$ ($3.5\%$) for eBOSS, which match the values in the observed samples  \citep{Reid_lss}. Once the unobserved galaxies are determined, we simply assign higher weights to the nearest galaxies that were assigned fibers, accounting for collided galaxies that were not assigned fibers (Nearest Neighbour Weights, NNW). 

The NNW method, however, has a drawback: it ignores the correlation between observed and unobserved targets.
To address this correlation properly, an approach known as the `pairwise inverse probability' (PIP) weighting scheme was proposed by \citet*{Bianchi_2017}.
This method has been applied to the eBOSS data as demonstrated in \citet{Mohammad_2020}. While the authors acknowledged the challenges involved in the PIP process, they also noted that approximate methods like NNW yield sufficiently accurate results considering the statistical uncertainties present in the data.
Moreover, applying PIP weights for the BOSS samples would require further information that is not publicly available. Given these considerations, we have decided not to apply PIP weights to our lightcones. 
It is on scales smaller than $1\hMpc$ where these weights could have the most significant impact. Consequently, we may expect discrepancies in the clustering of galaxies at these pair separations between the results from observations and simulations.

\subsection{The \textsc{Uchuu}-COMB lightcone}
\label{sec:uchuu_comb}

We now describe the process of generating the  \textsc{Uchuu}-COMB lightcones. To achieve this, we combine our \textsc{Uchuu}-LOWZ and -CMASS lightcones and apply the corresponding COMB \textsc{mangle} mask. When combining galaxy populations with different clustering amplitudes, it is optimal to assign weights to each sample to account for these differences \citep{Percival_weights}. In this work, we count the number of galaxy pairs for various redshift bins within each sample. We then normalize these values and use them as the weight for galaxies within that redshift range for each sample.

\section{Constructing the \textsc{GLAM-Uchuu} covariance lightcones}
\label{sec:glam_light}

To estimate covariance errors, we need to significantly increase the number of simulated lightcones. In this section, we detail the construction of the \textsc{GLAM-Uchuu} lightcones: a set of 16000 lightcones (2000 lightcones for each of the 8 studied samples) derived from the \textsc{GLAM-Uchuu} $N$-body simulations. The methodology employed for generating these lightcones is described below, while the properties of the produced covariance lightcones are presented in Table~\ref{tab:Uchuu_GUchuu_info}.

Similarly to the \textsc{Uchuu} lightcones, we use several snapshots from the \textsc{GLAM} simulations to reproduce the redshift-dependent evolution of galaxy clustering. We select the \textsc{GLAM} boxes that closely match the redshifts of the corresponding \textsc{Uchuu} snapshots:
\begin{description}
    \item LOWZ:~~~$z=0.20, 0.27, 0.31, 0.37, 0.46$ 
    \item CMASS: $z=0.46, 0.49, 0.564, 0.62, 0.72$  
    \item eBOSS:~~~$z=0.62, 0.72, 0.79, 0.86$
\end{description}

\subsection{Halo occupation distribution}
\label{sec:glam_hod}

As described in Section~\ref{sec:glam_simu}, the \textsc{GLAM} simulations do not resolve substructure inside distinct halos. Therefore, the \textsc{SHAM} method discussed in Section~\ref{sec:sham} cannot be directly applied to populate galaxies within halos in the \textsc{GLAM} simulations. Instead, a statistical method using the halo occupancy of galaxies from the \textsc{Uchuu} LRG catalogues must be employed. The HOD describes the mean number of galaxies within a distinct halo of virial mass $M_\mathrm{halo}$, which can be written as the sum of the mean number of central and satellite galaxies:
\begin{equation}
    \langle N_\mathrm{gal}(M_\mathrm{halo})\rangle=\langle N_\mathrm{cen}(M_\mathrm{halo})\rangle + \langle N_\mathrm{sat}(M_\mathrm{halo})\rangle.
    \label{eq:hod_sum}
\end{equation}

In this context, we obtain the HOD from the \textsc{Uchuu} galaxy cubic boxes we construct in Section~\ref{sec:sham}. This is possible because each galaxy provides information whether it is a central or satellite, as well as the mass of the distinct halo it inhabits. We then use the set of 2000 \textsc{GLAM-Uchuu} halo cubic boxes available at each redshift to generate galaxy catalogues. In this process, we randomly select galaxies for each distinct halo, following the corresponding HOD as determined from the \textsc{Uchuu-LRG} samples. The \textsc{Uchuu-LRG} HODs for LOWZ, CMASS and eBOSS are shown in Section~\ref{sec:glam_vs_other}. 
\textsc{GLAM-Uchuu} simulations resolve halos with masses within $10^{12}$ and $10^{15}$. While the HODs obtained from the \textsc{Uchuu-LRG} samples do not extend significantly outside this mass range, our \textsc{GLAM-Uchuu} boxes end up having a marginal deficit of galaxies compared to \textsc{Uchuu} due to their incompleteness (and not-well resolved halos) below $\sim10^{12.5}$ and above $\sim10^{14.5}$. This effect is visible in Table~\ref{tab:Uchuu_GUchuu_info} (and Fig.~\ref{fig:nz}), where the \textsc{GLAM} $N_\mathrm{eff}$ ($n(z)$) is slightly below that of the data and \textsc{Uchuu}.

\subsection{Stellar mass assignment and generation of lightcones}
\label{sec:GLAM_wcp}

The galaxies added into the \textsc{GLAM-Uchuu} boxes do not have assigned stellar masses. Unlike the \textsc{SHAM} approach, which inherently accounts for this property, we must incorporate stellar masses independently. To achieve this, we derive the  probability distribution of central and satellite stellar masses from \textsc{Uchuu} galaxy boxes across various ranges of distinct halo masses. These distributions are then employed to determine the stellar masses of the \textsc{GLAM} galaxies. 
The resulting SMFs are, as designed, consistent with those of \textsc{Uchuu}. 

The subsequent step in generating the \textsc{GLAM-Uchuu} LRG lightcones involves joining the galaxy boxes into spherical shells and smoothing their number density, $n(z)$. This procedure is explained in Sections~\ref{sec:clus_evol_z}~and~\ref{sec:nz_and_mask}. By applying the available survey and veto masks, we match the area of the lightcones to that of the observed samples (without shifting the sky position of the masks). As a last step, following the process described in Section~\ref{sec:compl_fib}, we account for survey systematics. The creation of the \textsc{GLAM-Uchuu}-COMB lightcones follows the steps described in Section~\ref{sec:uchuu_comb}.

\section{Results}
\label{sec:results}

Here we present the results of the galaxy clustering analysis conducted both in configuration and Fourier space. These analyses require the application of appropriate weights to both the data and the simulated lightcones.

For the LOWZ, CMASS and COMB data, we apply the $w_{\textrm{tot},i}$ weights as outlined in Equation~\ref{eq:boss_w}. The weights used for the eBOSS data are specified in Equation~\ref{eq:eboss_w}.
Incorporating the fibre collision weights, described in Sections~\ref{sec:compl_fib}~\&~\ref{sec:GLAM_wcp}, is essential for the \textsc{Uchuu} and \textsc{GLAM-Uchuu} lightcones. To enhance the signal-to-noise ratio of the clustering measurements, galaxies from both the data and lightcones are additionally weighted based on the galaxy number density, $n(z)$, using the weights introduced by \cite{Feldman94} (hereafter FKP weights):
\begin{equation}  
    w_{\textrm{FKP}}=\frac{1}{1+n(z)P_{\textrm{FKP}}}.
    \label{eq:fkp_weigths}
\end{equation}
We set $P_{\textrm{FKP}} = 10^4~h^{-3}\mathrm{Mpc}^{3}$, which is close to the value of the BOSS and eBOSS LRG power spectrum at $\mathrm{k} = 0.1\hMpc$, the scale where we aim to minimize the variance \citep[see][]{Ashely_boss_lss, Ashely_eboss_lss}. The $n(z)$ is estimated for each case from either the data or the lightcones.
The random catalogs required for these analyses are obtained from the \href{https://data.sdss.org/sas/dr12/boss/lss/}{SDSS-BOSS Science Archive Server} and \href{https://dr15.sdss.org/sas/dr16/eboss/lss/catalogs/DR16/}{SDSS-eBOSS Science Archive Server}. Randoms are weighted only by FKP weights.

The redshift ranges considered for each sample, as well as those shown in the following figures, are listed in Table~\ref{tab:obs_info}: $0.2<z<0.4$ for LOWZ, $0.43<z<0.7$ for CMASS, and $0.6<1.0$ for eBOSS. 
In the case of COMB galaxies, we adopt the same analysis scheme as presented by \citet{Alam_clust}, and consider three redshift ranges: $0.2<z<0.5$, $0.4<z<0.6$ and $0.5<z<0.75$.

\subsection{Testing the Planck cosmology against the data with \textsc{Uchuu}}
\label{sec:provingPL15}

In this section, we assess the performance of our theoretical predictions derived from the Planck cosmology, as determined from our \textsc{Uchuu} lightcones, to reproduce the observed data. We start by studying the LOWZ-, CMASS-, and eBOSS-N samples and conclude with an evaluation of the COMB-N$+$S sample.

\subsubsection{Clustering in configuration space}
\label{sec:red}

We first study the anisotropic two-point correlation function (2PCF) in redshift space, $\xi(s,\mu)$, which is computed using the \citet*{Landy93} estimator. We refer to the observed or simulated data galaxies as $D$ and the random galaxies as $R$:
\begin{equation}
    \xi(s,\mu) = \frac{\mathrm{DD} - 2\mathrm{DR} + \mathrm{RR}}{\mathrm{RR}},
    \label{eq:tpcf_estimator}
\end{equation}
where $s$ represents the redshift-space separation between a pair of objects in units of $\hMpc$ and $\mu$ is the cosine of the angle between the pair separation vector and the line-of-sight. Quantities $\mathrm{DD}$, $\mathrm{DR}$ and $\mathrm{RR}$ are the normalized pair counts of the galaxy-galaxy, galaxy-random and random-random catalogues, respectively.
For a more exhaustive analysis, we decompose $\xi(\mathrm{s},\mu)$ into its Legendre multipoles:
\begin{equation}
    \xi_\ell(\mathrm{s}) = \frac{2\ell+1}{2} \int^1_{-1} \xi(\mathrm{s},\mu)P_\ell(\mu)\mathrm{d}\mu,
    \label{eq:multipoles}
\end{equation}
where $P_\ell(\mu)$ is the $\ell th$ Legendre Polynomial.

\begin{figure*}
    \centering
    \begin{subfigure}[c]{\linewidth}
        \includegraphics[width=\linewidth]{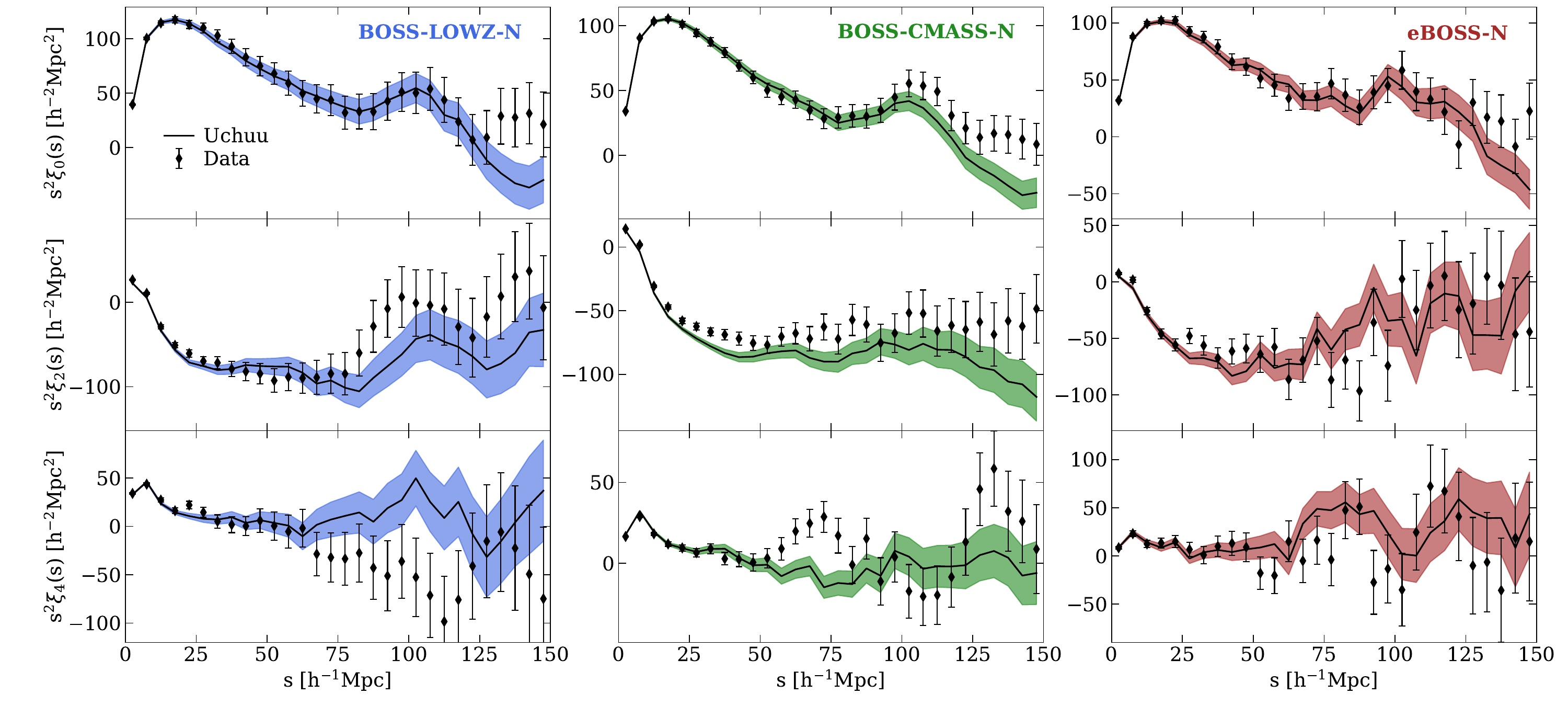}
        \caption{Large scale.}
    \end{subfigure}
    \begin{subfigure}[c]{\linewidth}
        \includegraphics[width=\linewidth]{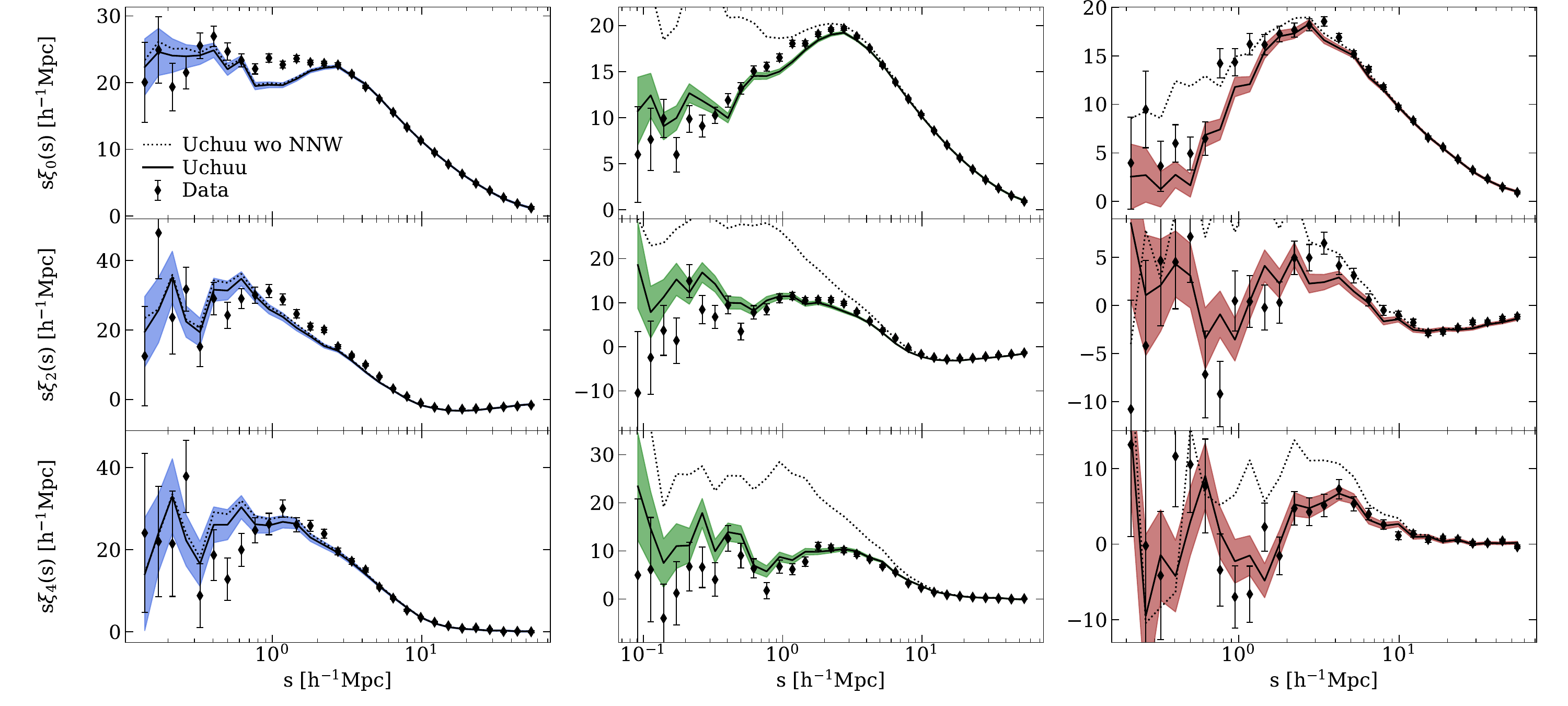}
        \caption{Small scale.}
    \end{subfigure}
    \caption{Measurements of the monopole, quadrupole and hexadecapole moments of the 2PCF are presented in the first, middle and bottom rows, respectively, for CMASS-N, LOWZ-N and eBOSS-N, arranged from left to right in the columns. 
    Observational measurements are shown by data points, while the solid lines show the mean 2PCF derived from the \textsc{Uchuu} lightcones.
    Data error bars represent the standard deviation of our set of 2000 \textsc{GLAM-Uchuu} lightcones at the 1-$\sigma$ level, whereas the shaded area for \textsc{Uchuu} indicates the error on the mean $\sigma/\sqrt{2}$.
    Panel (a) shows the 2PCF measurements spanning from $0$ to $150\hMpc$ in bins of $5\hMpc$. Panel (b) presents the results up to $60\hMpc$ in bins equally spaced in the logarithm of pair separation. For reference, the mean \textsc{Uchuu} 2PCF without fibre collision correction (NNW) is depicted with dotted lines. }
    \label{fig:Uchuu_TPCF}
\end{figure*}
In Fig.~\ref{fig:Uchuu_TPCF} we present the monopole ($\ell=0$), quadrupole ($\ell=2$) and hexadecapole ($\ell=4$) moments of the 2PCF for the three samples: LOWZ, CMASS and eBOSS-N. The results include both observed data and theoretical predictions determined from the mean of the \textsc{Uchuu} lightcones.
Overall, our \textsc{Uchuu} lightcones are able to reproduce the clustering measurements obtained from the observed data, showcasing remarkable accuracy in certain samples. 
Residuals from the 2PCF monopole, denoted as $r=\xi_0^\mathrm{Uchuu}/\xi_0^\mathrm{data} -1$, maintain values under $2.5\%$ across all samples ranging from the fibre collision scale\footnote{We define the fibre collision scale as the $s$-value at which the ratio between the multipoles of \textsc{Uchuu} with and without NNW is greater than $2\%$. We set these scales as the minimum, since below this value the accuracy of our results decreases. For 2PCF monopole, it is $2.3\hMpc$ in LOWZ, $4.2\hMpc$ in CMASS, and $6.5\hMpc$ in eBOSS.} up to $25\hMpc$. Within the range of $100\hMpc$, residuals remain below $10\%$ for LOWZ and CMASS, and under $15\%$ for eBOSS.

Beyond the BAO scale ($\mathrm{s}\geq120\hMpc$), significant fluctuations become apparent due to cosmic variance. 
At the smallest scales, below $\sim1\hMpc$, deviations between the measurements from the lightcones and the observed data can be attributed to the non-application of the PIP weights in the lightcones, as mentioned in Section~\ref{sec:compl_fib}.
Notably, there are differences between the results by more than 1-$\sigma$ for some scales above $60\hMpc$. Moreover, while our \textsc{Uchuu} lightcones monopole cross zero after BAO peak, the observations do not. It is widely recognized that, at large scales, the measurements are significantly influenced by observational systematics. \citet{Huterer13} conducted an exhaustive investigation into photometric calibration errors and their repercussions on clustering measurements, illustrating that calibration uncertainties consistently give rise to large-scale power variations. The significance and potential causes of the large-scale excess are studied also in \citet{Ashely_boss_lss}, where they demonstrate that it has no significant impact on BAO measurements. Their findings indicate that these measurements remain resilient, both in lightcones and the data, regardless of whether or not any weights are included.

One may also discern some differences between the results for the multipole moments that cannot be accounted for by the aforementioned reasons, such as the CMASS quadrupole. Exploring variations of the only free parameter in the SHAM, $\sigma_\mathrm{SHAM}$, does not seem to meaningfully impact the shape of multipoles, suggesting that the observed discrepancies are most likely a result of statistical fluctuation. This is supported by the improved performance of our 2000 \textsc{GLAM-Uchuu} lightcone sets (see Section~\ref{sec:glam_vs_other:tpcf}).
All the highlighted trends (including the CMASS quadrupole behavior) are in agreement with results presented in e.g. \citet{Sergio}, \citet{Kitaura_2016} or \citet{Ashely_boss_lss}.

\begin{figure*}
    \hspace*{-1cm}
    \includegraphics[width=\linewidth]{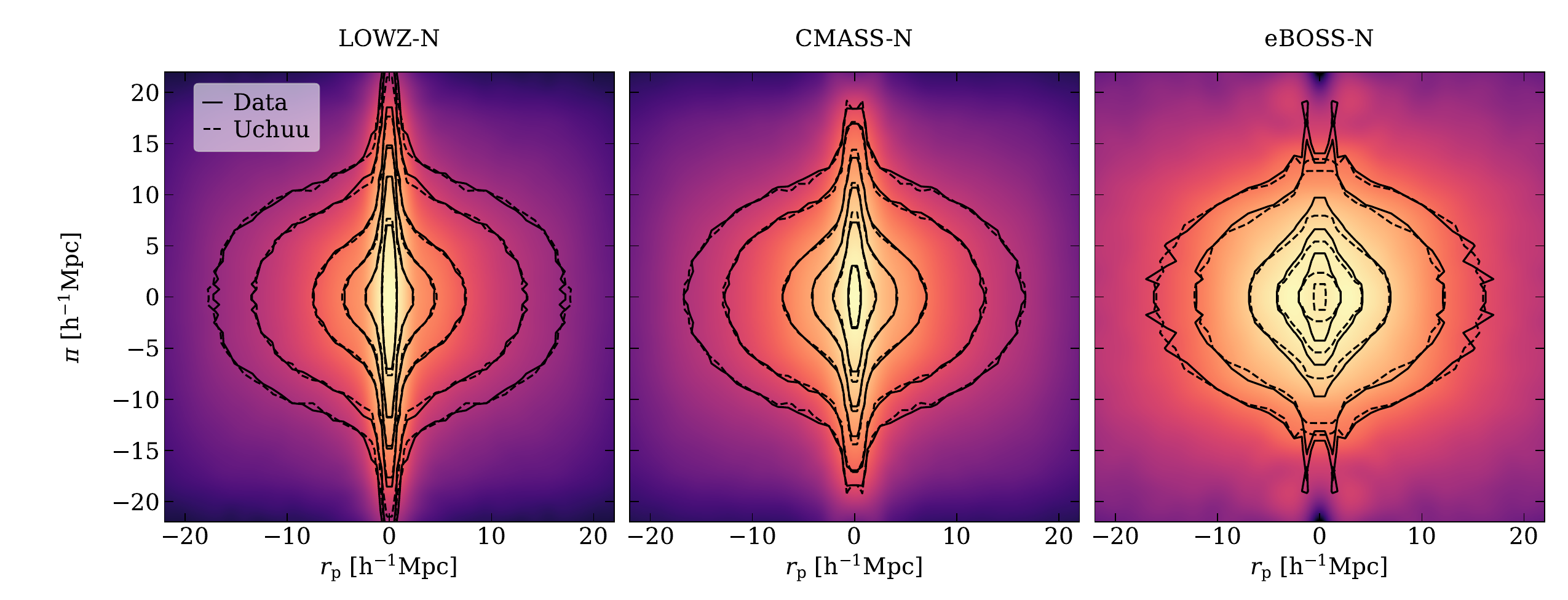}
    \caption{The mean 2PCF$_\mathrm{2d}$ from \textsc{Uchuu}-LOWZ-N (left column), -CMASS-N (middle column), and -eBOSS-N lightcones (right column) is plotted as a function of pair separation perpendicular ($r_\mathrm{p}$) and parallel ($\pi$) to the line-of-sight. To illustrate deviations from circular symmetry, the first quadrant is replicated with reflections in both axes.
    Contours of the 2PCF$_\mathrm{2d}$ estimated from the observations and from \textsc{Uchuu} are displayed as black solid and dashed curves, respectively. The contour levels shown are $[0.5,0.8,2,4,8,20]$.}
    \label{fig:Uchuu_RSD}
\end{figure*}
Using the Landy-Szalay estimator, we now compute the two-dimensional correlation function 2PCF$_\mathrm{2d}= \xi(r_\mathrm{p},\pi)$, where $r_\mathrm{p}$ and $\pi$ represent the pair separations perpendicular and parallel to the line-of-sight, respectively.
The results are presented in Fig.~\ref{fig:Uchuu_RSD}, where the mean 2PCF$_\mathrm{2d}$ derived from \textsc{Uchuu}-LOWZ, -CMASS, and -eBOSS-N lightcones is shown in the left, middle, and right columns, respectively. Solid curves represent the contours estimated from the observed data, while dashed curves depict the contours from \textsc{Uchuu}. 

The 2PCF$_\mathrm{2d}$ is the simplest statistical measure of peculiar velocities the in cosmological structure. This function remains directionally independent in an isotropic universe. However, this property does not hold in redshift space, where only radial separations, $\pi$, are distorted by the peculiar velocities of the galaxies. This distortion is commonly referred to as redshift space distortions (RSD). The impact of RSD is particularly prominent on small scales, such as within galaxy groups and clusters, leading to the well-known ``fingers-of-God'' effect visible in cone plots showing galaxy positions  \citep{Peacock_2001}.
The results shown in Fig.~\ref{fig:Uchuu_RSD} clearly exhibit the presence of RSD, evident by the elongations along the $\pi$-axis at scales within $5\hMpc$. The remarkable agreement between our \textsc{Uchuu} lightcones and the BOSS/eBOSS data is noteworthy. Minor deviations on small scales, related to the non-use of PIP weights, are consistent with the previous results shown in Fig.~\ref{fig:Uchuu_TPCF}.

\begin{figure*} 
    \centering
    \includegraphics[width=\linewidth]{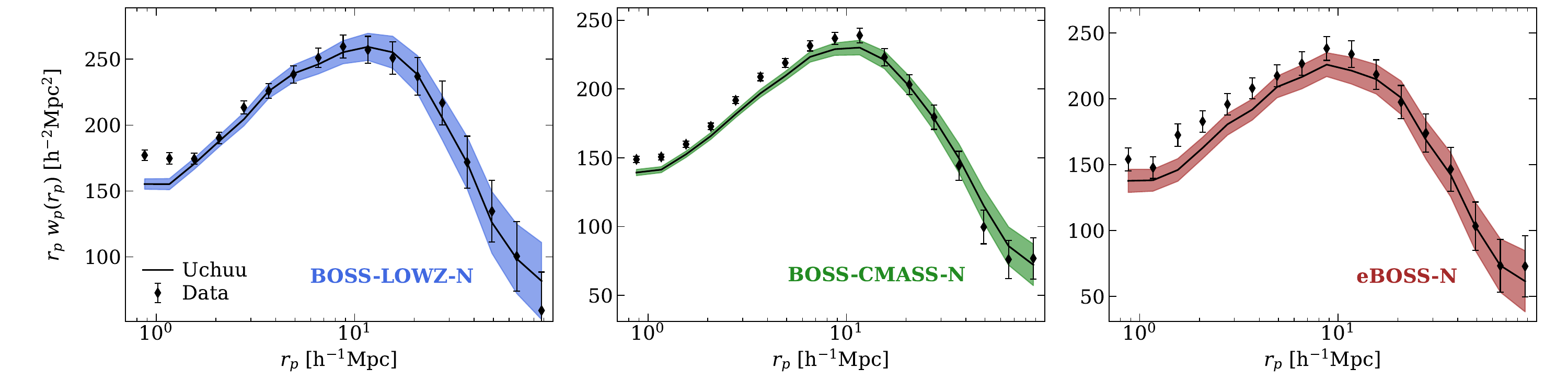}
    \caption{Measurements of the 2PCF for CMASS-N (left column), LOWZ-N (middle column) and eBOSS-N (right column). 
    Measurements from observations are shown by data points, while the solid lines represent the mean 2PCF derived from the \textsc{Uchuu} lightcones.
    Data error bars represent the standard deviation of our set of 2000 \textsc{GLAM-Uchuu} lightcones, 1-$\sigma$, while the \textsc{Uchuu} shaded area is the error on the mean, calculated as $\sigma/\sqrt{2}$.
    The 2PCF is measured up to $100\hMpc$ in equally-spaced logarithmic bins.}
    \label{fig:Uchuu_pCF}
\end{figure*}
As mentioned above, the impact of RSD only affects the pair separation along the line-of-sight ($\pi$). Therefore, these distortions can be removed by projecting $\xi(r_\mathrm{p},\pi)$ onto the $\pi$-axis \citep*{Peebles,norberg2009} as follows:
\begin{equation}
    w_\mathrm{p}(r_\mathrm{p})=\int_{-\infty}^{+\infty} \xi(r_\mathrm{p},\pi)\mathrm{d}\pi.
\end{equation}
In practice, this integral is truncated at a certain $\pi_\mathrm{max}$. In this work, we adopt $\pi_\mathrm{max}=80\hMpc$.
The projected correlation functions (pCF) are shown in Fig.~\ref{fig:Uchuu_pCF}. 
The agreement between the theoretical predictions determined from the mean of the \textsc{Uchuu} lightcones and the observational estimates aligns with the expectations from Fig.~\ref{fig:Uchuu_RSD}. 

Finally, we analyze the three-point correlation function (3PCF) in redshift space, $\zeta(r_{12}, r_{23}, r_{13})$, which provides a description of the probability of finding triplets of galaxies as a function of their triangular geometry. We compute the 3PCF using \textsc{encore} code\footnote{\url{https://github.com/oliverphilcox/encore}}, which implements \citet*{3pcf_estim} estimator:
\begin{equation}
    \zeta(r_{12}, r_{23}, r_{13}) = \frac{\mathrm{DDD} - 3\mathrm{DDR} + 3\mathrm{DRR} - \mathrm{RRR}}{\mathrm{RRR}},
    \label{eq:3pcf_estimator}
\end{equation}
where $r_{ij}=r_i - r_j$ represents the redshift-space separation between two objects, $i$ and $j$, in units of $\hMpc$.

\begin{figure*}
    \centering
    \includegraphics[width=\linewidth]{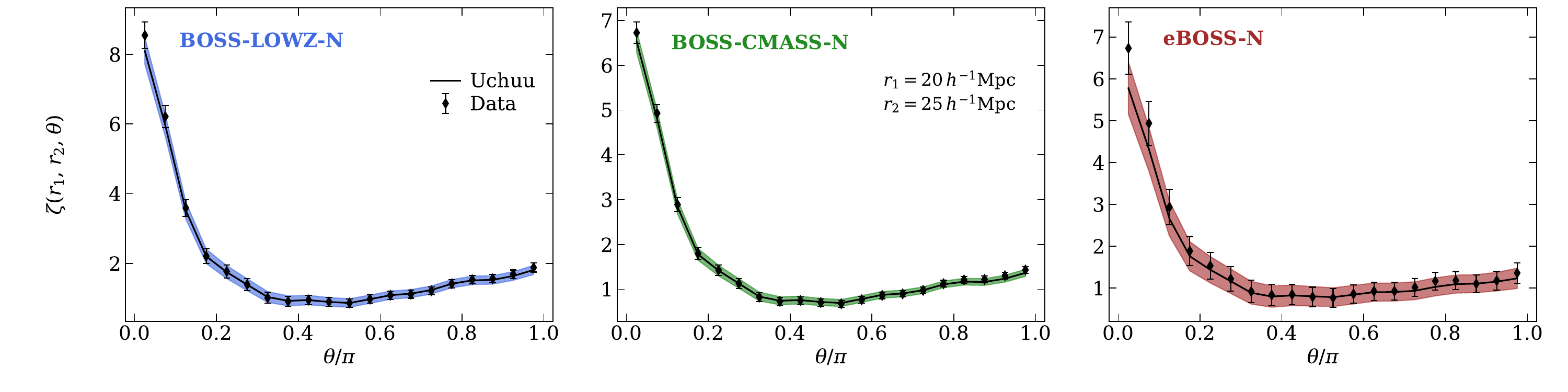}
    \caption{Measurements of the 3PCF for LOWZ-N (left column), CMASS-N (middle column) and eBOSS-N (right column). 
    Measurements from observations are shown by data points, while the solid lines represent the mean 2PCF derived from the \textsc{Uchuu} lightcones.
    Data error bars represent the standard deviation of our set of 2000 \textsc{GLAM-Uchuu} lightcones, 1-$\sigma$, while the \textsc{Uchuu} shaded area is the error on the mean, calculated as $\sigma/\sqrt{2}$.
    The 3PCF is presented for $r_{1}=20\hMpc$ and $r_{2}=25\hMpc$, and $\theta$ is the angle between $r_{1}$ and $r_{2}$, in radians.}
    \label{fig:Uchuu_3PCF}
\end{figure*}
Fig.~\ref{fig:Uchuu_3PCF} displays the 3PCF for both observed data and theoretical predictions, for LOWZ, CMASS and eBOSS-N samples. We find a good agreement between our theoretical predictions determined from the mean of the \textsc{Uchuu} lightcones and the observational estimates, with all the data points in agreement within 1-$\sigma$ errors.

\subsubsection{Clustering in Fourier space}
\label{sec:four}

\begin{figure*}
    \centering
    \includegraphics[width=\linewidth]{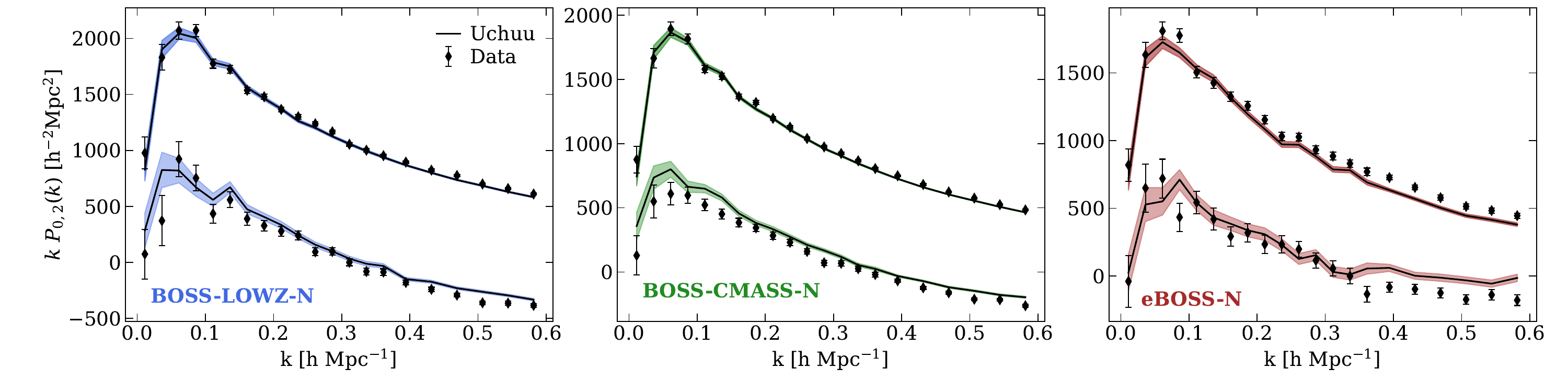}
    \caption{Measurements of the power spectrum monopole (upper curve) and quadrupole (lower curve, lighter shading) for LOWZ-N (left column), CMASS-N (middle column) and eBOSS-N (right column). 
    Observed data measurements are represented by data points. The solid lines represent the mean power spectrum derived from the \textsc{Uchuu} lightcones.
    Data error bars represent the standard deviation ($1\sigma$) using the set of 2000 \textsc{GLAM-Uchuu} lightcones, while the \textsc{Uchuu} shaded area indicates the error on the mean, $\sigma/\sqrt{2}$.
    The power spectrum are measured from $\mathrm{k}=0.005$ to $0.6\hMpcInv$ in $0.03\hMpcInv$ bins.}
    \label{fig:Uchuu_Pk}
\end{figure*}
We measure the power spectrum monopole, $P_{0}(k)$ and quadrupole, $P_{2}(k)$, which can be calculated following Equations~2 to 7 from \citet*{Feldman94}.
These calculations have been carried out using the Python package \textsc{pypower}\footnote{\url{https://pypower.readthedocs.io/en/latest/api/api.html}}, which implements the \citet{Hand2017} estimator. This estimator takes coordinate-space positions at mesh nodes and directly computes the normalisation factor from the mesh.
To mitigate the impact of aliasing, we employ a piecewise cubic spline (PCS) mesh assignment scheme with interlacing, as described by \citep{Sefusatti2016}, and use a grid resolution of $N_\mathrm{grid} = 1024$ in each dimension. 

Fig.~\ref{fig:Uchuu_Pk} shows the power spectrum monopole and quadrupole for both observed data and the mean of the \textsc{Uchuu} lightcones, with shot noise subtracted from the $\mathrm{P}_{0}(\mathrm{k})$ measurements.
Our theoretical predictions based on the Planck cosmology successfully reproduce the observed power spectrum monopole, exhibiting residuals below $5.1\%$ for $\mathrm{k}>0.03$ and $\mathrm{k}<0.6\hMpcInv$ in the LOWZ sample, under $4.0\%$ for CMASS, and within $12.5\%$ in eBOSS.

Many of the observed deviations can be attributed to statistical fluctuations. With our extensive set of 2000 \textsc{GLAM-Uchuu} lightcones, we are better equipped to accurately measure the noise (see Section~\ref{sec:glam_vs_other:pk}). This enables us to discern where \textsc{Uchuu} predictions align with observations.
We find some predictions that slightly differ from the observational data. One instance concerns the \textsc{Uchuu}-eBOSS monopole for $0.2<\mathrm{k}<0.4\hMpcInv$. Notably, a similar trend has been previously reported for the eBOSS data \citep[see][]{Zhao_2021}. 
Another case involves the quadrupole for $\mathrm{k}>0.35\hMpcInv$. However, as illustrated in Fig.~\ref{fig:Uchuu_TPCF} (b), effects of fiber collisions on quadrupole extends to scales up to $15\hMpc$. This would explain the observed discrepancies in the \textsc{Uchuu} power spectrum quadrupole at $\mathrm{k}\geq0.4\hMpcInv$, since the accuracy of our results diminishes above these scales (below s$=15\hMpc$).

\subsubsection{Dependence of clustering on stellar mass}

\begin{figure*}
    \centering
    \begin{subfigure}[c]{\linewidth}
        \includegraphics[width=\linewidth]{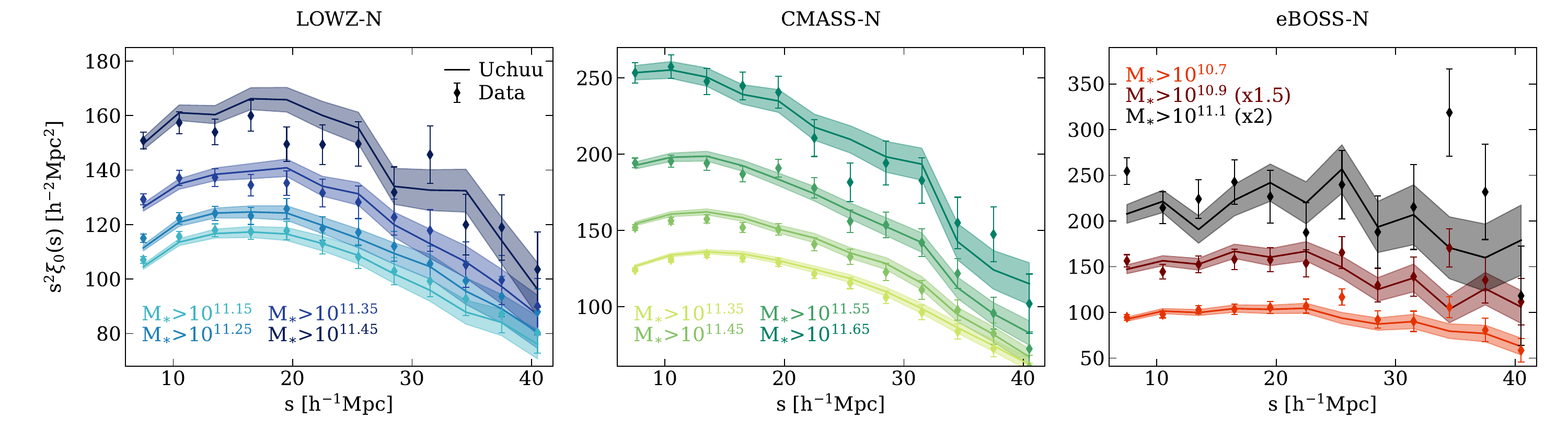}
        \caption{Monopole moment of the 2PCF.}
    \end{subfigure}
    \begin{subfigure}[c]{\linewidth}
        \includegraphics[width=\linewidth]{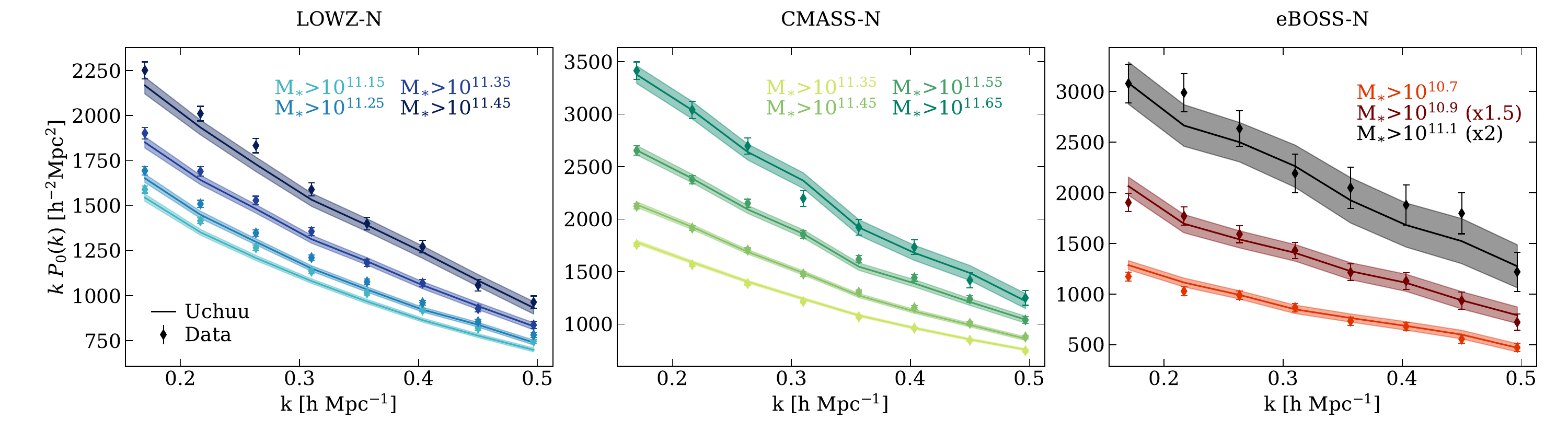}
        \caption{Monopole moment of the power spectrum.}
    \end{subfigure}
    \caption{Measurements of the 2PCF (a) and the power spectrum (b) monopoles for different stellar mass thresholds in LOWZ-N (left column), CMASS-N (middle column), and eBOSS-N (right column). 
    The various stellar mass thresholds are indicated by the colour scheme in the legend.
    For a better visualization of the eBOSS sample, we have multiplied the $10.9$ and $11.1$ stellar mass thresholds measurements by $1.5$ and $2$ respectively.
    Observational measurements are shown by data points. The solid lines show the mean derived from the \textsc{Uchuu} lightcones. 
    Data error bars represent the standard deviation of our set of 2000 \textsc{GLAM-Uchuu} lightcones at 1-$\sigma$ level, while the shaded area for \textsc{Uchuu} represents the error on the mean, $\sigma/\sqrt{2}$.
    We measure the 2PCF monopole from s$=5$ to $45\hMpc$ in $5\hMpc$ bins, and the power spectrum monopole from $\mathrm{k}=0.15$ to $0.5\hMpcInv$} in $0.05\hMpcInv$ bins. 
    \label{fig:Uchuu_mass}
\end{figure*}

We now investigate the completeness of the observed galaxy populations by analysing the dependence of the 2PCF and power spectrum monopoles on stellar mass. 
The results are shown in Fig.~\ref{fig:Uchuu_mass}, where we observe that, for both the LOWZ and CMASS samples, a significant overlap exists between the observed data and the mean values obtained from the \textsc{Uchuu} lightcones across all the stellar mass thresholds. 
In agreement with previous studies \citep[see][]{Maraston_2013}, these results indicate that the LRG population in both BOSS samples is complete at the massive end.
We conclude that the complete SMF assumed in the \textsc{SHAM} method for BOSS accurately describes the real galaxy populations.

For the eBOSS sample, consistent with results from \citet{comparat_maraston}, we consider an incomplete population of galaxies when generating the \textsc{Uchuu}-eBOSS lightcones. This is achieved by using a complete SMF that goes exceeds the SMF estimated from the data (see Section~\ref{sec:sham}). 
Similar to the other samples, our goal is to assess whether the clustering as a function of stellar mass in \textsc{Uchuu} lightcones agrees with that estimated from the observed data. 
The results shown in the right panel of Fig.~\ref{fig:Uchuu_mass} demonstrate the accuracy of the method adopted in this work for modelling the incomplete eBOSS galaxy population. Despite the noise inherent to a low population statistic, there is a good agreement between the data and \textsc{Uchuu}, both samples presenting the same trend with stellar mass.

It is important to note that the eBOSS galaxy mass information was extracted from the \citet{comparat_maraston} eBOSS catalog, which does not share the exact same footprint or n(z) as the public eBOSS data and \textsc{Uchuu} lightcones. 
Additionally, \citet{comparat_maraston} does not provide its own random catalog, leading us to use the public random catalog for analysing the 2PCF. Again, this catalog does not present the same footprint and n(z). All these differences may introduce errors and discrepancies in the statistics.
To mitigate these effects when computing the 2PCF for different stellar mass thresholds, we have selected galaxies from the \citet{comparat_maraston} catalog that satisfy $\mathrm{DEC}>42~\mathrm{deg}$ and $\mathrm{RA}<225~\mathrm{deg}$. This subset aligns more closely with the footprint and n(z) of public data, as well as the random catalog. We do implement this filtering in the \textsc{Uchuu}-eBOSS lightcones as well.

\subsubsection{Large-scale bias predictions}

\begin{table}
\centering
    \begin{tabular}{cccc}
        \hline
        Sample & $z_\mathrm{med}$ & $b^{\mathrm{data}}$ & $b^{\mathrm{Uchuu}}$ \\
        \hline
        LOWZ-N  & $0.32$ & $1.87 \pm 0.03$ & $1.87 \pm 0.02$  \\
        CMASS-N & $0.57$ & $1.99 \pm 0.02$ & $2.00 \pm 0.02$  \\
        eBOSS-N & $0.70$ & $2.10 \pm 0.03$ & $2.08 \pm 0.02$  \\
        \hline
    \end{tabular}
    \caption{Bias factors of the BOSS and eBOSS samples. The first and second columns provide the sample name and its corresponding median redshift. The last two columns display the large-scale bias factors measured from the 2PCF monopole for both the observations and \textsc{Uchuu} lightcones. These bias factors are obtained as the mean value of $b(\mathrm{s})$ within the range of $15$ to $30\hMpc$.}
\label{tab:bias_output}
\end{table}

\begin{figure*}
    \centering
    \includegraphics[width=\linewidth]{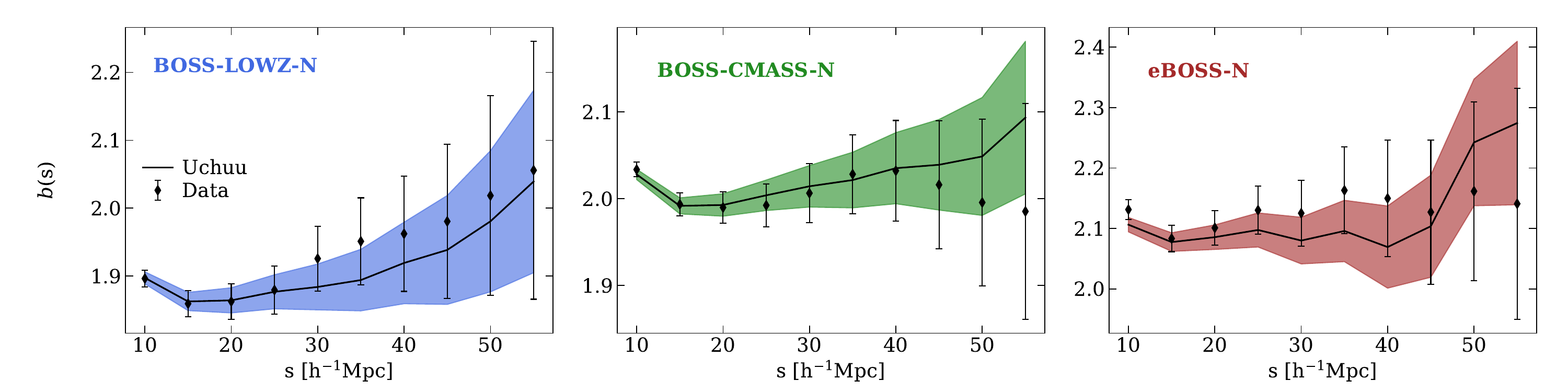}
    \caption{Measurements of the galaxy bias at the median redshift of the CMASS-N (left panel), LOWZ-N (middle panel), and eBOSS-N (right panel) samples are presented. For details on the bias estimation, refer to the text. In the case of the observations, the dark matter clustering is assumed to be the same as that in the \textsc{Uchuu} simulation.
    Observed data measurements are represented by points. The solid lines show the mean bias derived from the \textsc{Uchuu} lightcones for each of the samples. 
    Data error bars represent the standard deviation of our set of 2000 \textsc{GLAM-Uchuu} lightcones at the 1-$\sigma$ level, and the shaded area corresponding to  \textsc{Uchuu} indicates the error on the mean, $\sigma/\sqrt{2}$.
    }
    \label{fig:Uchuu_bias}
\end{figure*}
In Fig.~\ref{fig:Uchuu_bias}, we study the galaxy bias, $b(\mathrm{s})$, through our theoretical predictions of the 2PCF monopole. We then proceed to compare these predictions with observed data.
For each of the three studied samples, the large-scale bias at their respective median redshift, $z_\mathrm{med}$, is derived by solving the equation below, which gives a simple linear perturbation theory prediction of the RSD, applicable on large scales \citep[see][]{Kaiser1987,Hamilton1998}:
\begin{equation}
    \xi_0(\mathrm{s}) = b(\mathrm{s})^2 \left(1 + \frac{2}{3}\beta + \frac{1}{5}\beta^2 \right) \xi_\mathrm{lin}(\mathrm{s}).
\end{equation}
In the equation, we have $\beta=\Omega_{\mathrm{m}, z_\mathrm{med}}^{0.55}/b(\mathrm{s})$, where $\xi_0(\mathrm{s})$ represents the measured 2PCF monopole (derived from data and lightcones in each case), and $\xi_\mathrm{lin}(\mathrm{s})$ corresponds to the dark matter linear 2PCF from the \textsc{Uchuu} simulation at $z_\mathrm{med}$. 

Our \textsc{Uchuu} BOSS/eBOSS lightcones reproduce the $b(\mathrm{s})$ values from observed data within the uncertainties.
The BOSS/eBOSS bias factors, $b$, from both observed data and \textsc{Uchuu} lightcones, are presented in Table~\ref{tab:bias_output}. The bias factors have been obtained as the mean value of $b(\mathrm{s})$ measured between $15$ and $30\hMpc$. This table clearly shows how the bias increases with the median redshift of the sample. This is in agreement with the results reported in \citet{Zhou_2020}, where a similar behaviour was observed when examining the evolution of large-scale bias with redshift for a DESI-type LRG population selected from the Legacy Survey imaging dataset \citep{Dey_2019}.

\subsubsection{Performance of \textsc{Uchuu}-COMB LRG lightcones}

\begin{figure*}
    \centering
    \begin{subfigure}[c]{\linewidth}
        \includegraphics[width=\linewidth]{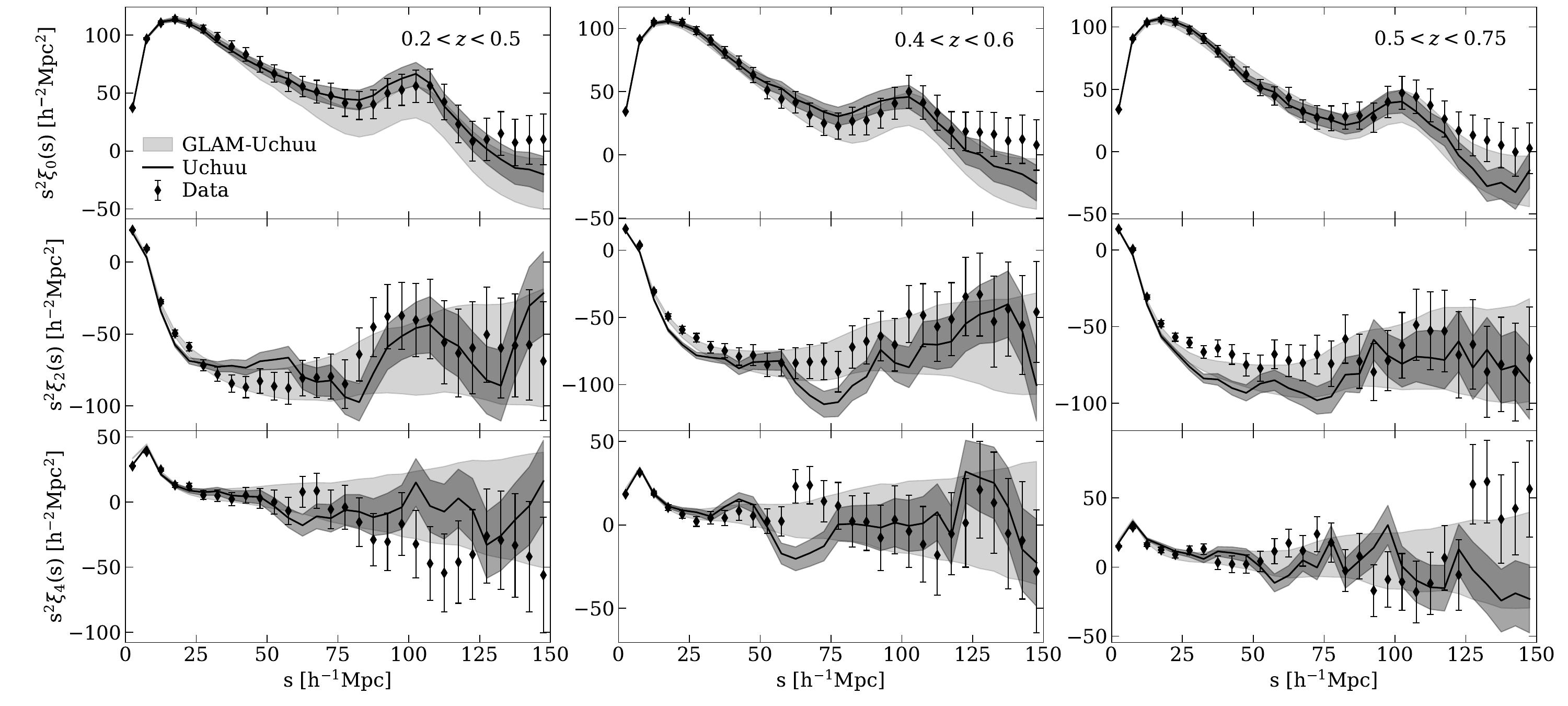}
        \caption{Large scale.}
    \end{subfigure}
    \begin{subfigure}[c]{\linewidth}
        \includegraphics[width=\linewidth]{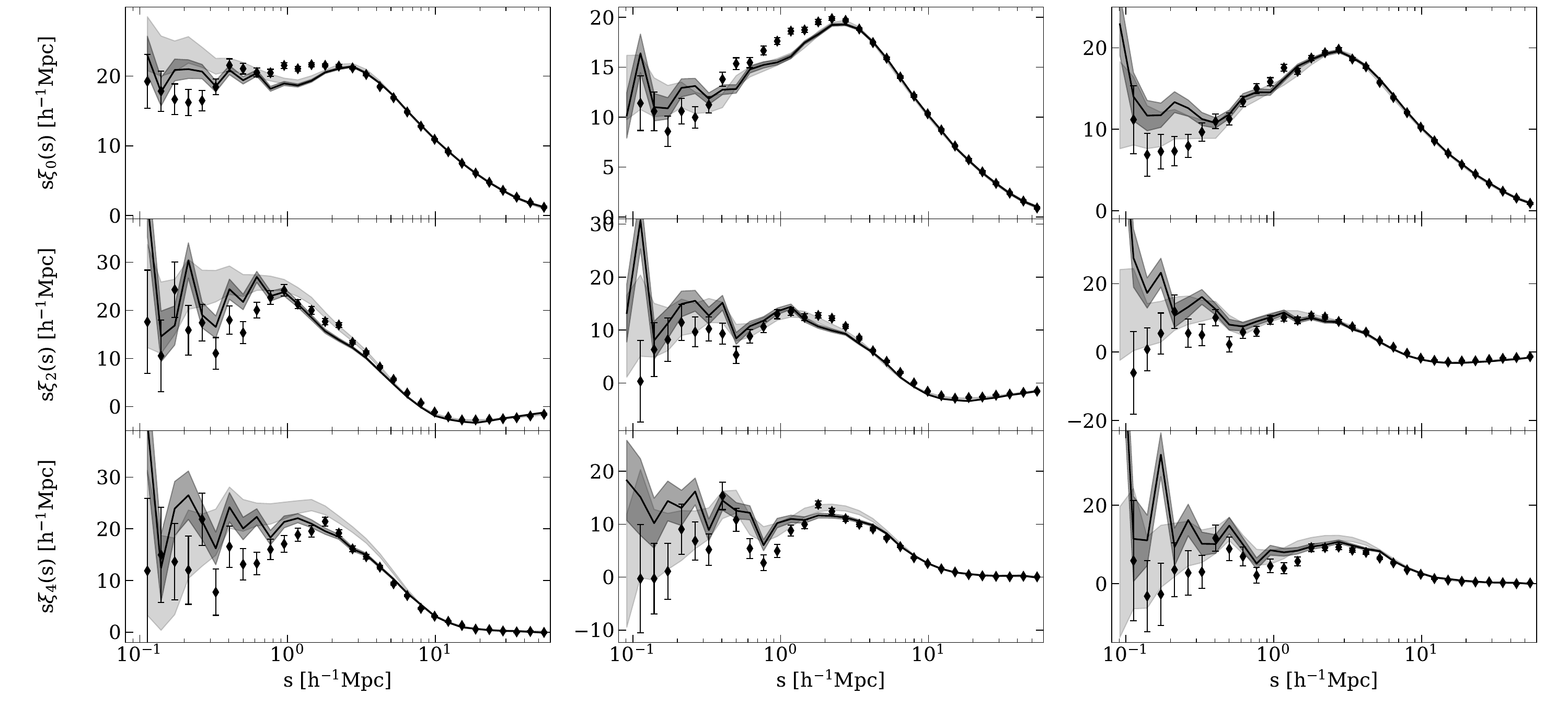}
        \caption{Small scale.}
    \end{subfigure}
    \caption{Measurements of the monopole, quadrupole and hexadecapole moments of the 2PCF for the COMB-N$+$S sample in three redshift ranges of interest: $0.2<z<0.5$ (left column), $0.4<z<0.6$ (middle column) and $0.5<z<0.75$ (right column). 
    Observed data measurements are represented as data points. The solid line shows the mean 2PCF derived from \textsc{Uchuu}. We additionally show the standard deviation of \textsc{GLAM-Uchuu} around its mean in a light shaded area.
    Data error bars represent the standard deviation of our set of 2000 \textsc{GLAM-Uchuu} lightcones, 1-$\sigma$, while the dark shaded area for \textsc{Uchuu} is the error on the mean, $\sigma/\sqrt{2}$.
    Panel (a) shows the 2PCF measurements from $0$ to $150\hMpc$ in $5\hMpc$ bins, while panel (b) shows results up to $60\hMpc$ using equally spaced logarithmic bins.}
    \label{fig:UchuuComb_TPCF}
\end{figure*}
\begin{figure*}
    \centering
    \includegraphics[width=\linewidth]{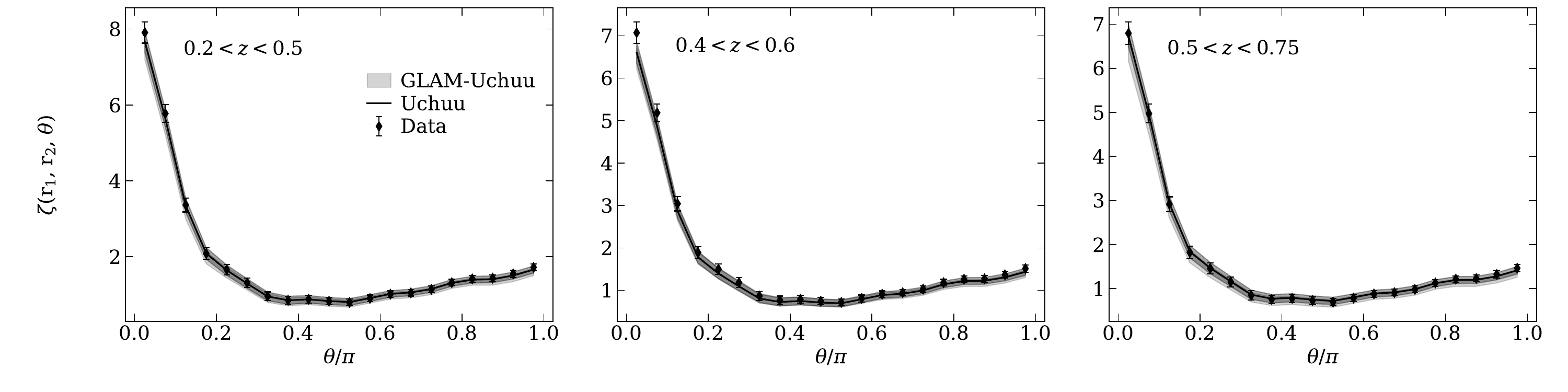}
    \caption{Measurements of the 3PCF for the COMB-N$+$S sample in three redshift ranges of interest: $0.2<z<0.5$ (left panel), $0.4<z<0.6$ (middle panel) and $0.5<z<0.75$ (right panel). 
    Observed data measurements are represented by data points. The solid line shows the mean power spectrum derived from \textsc{Uchuu}. We additionally show the standard deviation of \textsc{GLAM-Uchuu} around its mean in a light shaded area.
    Data error bars represent the standard deviation of our set of 2000 \textsc{GLAM-Uchuu} lightcones, 1-$\sigma$, while the dark shaded area for \textsc{Uchuu} is the error on the mean, $\sigma/\sqrt{2}$.
    The 3PCF is presented for $r_{1}=20\hMpc$ and $r_{2}=25\hMpc$, and $\theta$ is the angle between $r_{1}$ and $r_{2}$, in radians.}
    \label{fig:UchuuComb_3PCF}
\end{figure*}
\begin{figure*}
    \centering
    \includegraphics[width=\linewidth]{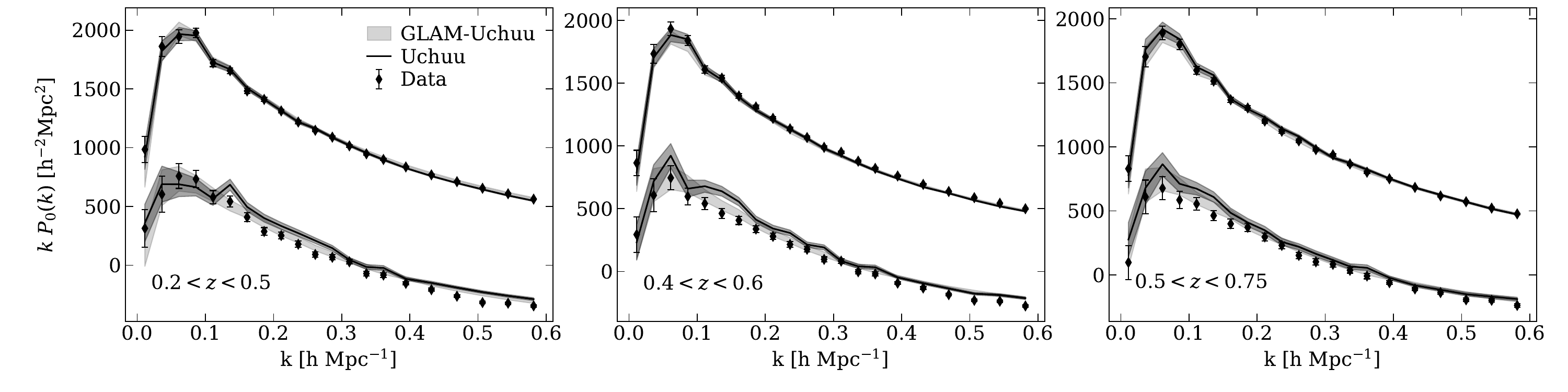}
    \caption{Measurements of the power spectrum monopole (upper curve) and quadrupole (lower curve) for the COMB-N$+$S sample in three redshift ranges of interest: $0.2<z<0.5$ (left panel), $0.4<z<0.6$ (middle panel) and $0.5<z<0.75$ (right panel). 
    Observed data measurements are represented by data points. The solid line show the mean power spectrum derived from \textsc{Uchuu}. We additionally show the standard deviation of \textsc{GLAM-Uchuu} around its mean in a light shaded area.
    Data error bars represent the standard deviation of our set of 2000 \textsc{GLAM-Uchuu} lightcones, 1-$\sigma$, while the dark shaded area for \textsc{Uchuu} is the error on the mean, $\sigma/\sqrt{2}$.
    The power spectrum is measured from $\mathrm{k}=0.005$ to $0.6\hMpcInv$ in bins of $0.03\hMpcInv$.}
    \label{fig:UchuuComb_PK}
\end{figure*}
Finally, we assess the ability of our \textsc{Uchuu}-COMB lightcones to reproduce the observed BOSS clustering statistics in both hemispheres, N$+$S, as well as the clustering evolution with redshift.
The monopole, quadrupole and hexadecapole moments of the 2PCF measurements are shown in Fig.~\ref{fig:UchuuComb_TPCF} for both the observed data and mean \textsc{Uchuu} measurements. 
We additionally show the \textsc{GLAM-Uchuu}-COMB measurements with a light shaded area to demonstrate their behavior and to elucidate the source of potential discrepancies between \textsc{Uchuu} and the data.
From left to right, the panels show the redshift intervals over which the 2PCF has been measured: $0.2<z<0.5$, $0.4<z<0.6$, and $0.5<z<0.75$. 
Correspondingly, Fig.~\ref{fig:UchuuComb_3PCF} presents results for the 3PCF. The power spectrum monopole and quadrupole measurements are shown in Fig.~\ref{fig:UchuuComb_PK} for $\mathrm{k}=0.005-0.6\hMpcInv$.

As expected, our \textsc{Uchuu} lightcones are able to fully reproduce with high accuracy the clustering evolution observed in the BOSS N$+$S data measurements. 
Residuals from the 2PCF monopole between the CMASS-N fiber collision scale, $4.2\hMpc$, and $25\hMpc$ ($100\hMpc$) exhibit values below $0.9\%$ ($8\%$) for $0.2<z<0.5$ sample, less than $1.5\%$ ($15\%$) for $0.4<z<0.6$ sample, and under $0.9\%$ ($10\%$) for $0.5<z<0.75$ sample.
For the 3PCF, all the data points are in agreement within 1-$\sigma$ errors with the mean of the \textsc{Uchuu} lightcones.
Regarding the power spectrum monopole between $k>0.03$ and $k<0.6\hMpcInv$, the residuals remain within $2.6, 4.5, 3.4\%$ for the three samples (left to right).
We interpret the discrepancies in the 2PCF and power spectrum multipoles as arising from the same reasons as discussed in Sections~\ref{sec:red}~and~\ref{sec:four}.
Furthermore, the findings from \textsc{GLAM-Uchuu} validate that certain discrepancies between \textsc{Uchuu} and the data can be ascribed to statistical fluctuations.
Our results are in agreement with previous works, both in configuration \citep[see][]{Chuang_2017} and Fourier space \citep[see][]{Beutler_2016}.

The results presented in this section allow us to reach the conclusion that our theoretical predictions, based on the standard Planck cosmology model and built by generating lightcones from the \textsc{Uchuu} simulation, exhibit a remarkable ability to reproduce, with a high accuracy, the observed data across all studied samples: LOWZ, CMASS, COMB, and eBOSS, spanning both for Northern and Southern hemispheres. This reaffirms the high quality of the \textsc{Uchuu} simulation and highlights the reliability of our methodology. Notably, our approach relies on just one free parameter, the scatter $\sigma_\mathrm{SHAM}$, and covers various aspects, including adopting a complete SMF in the \textsc{SHAM} process and addressing the treatment of galaxy population incompleteness and fiber collision systematics.

\subsection{Covariance errors from \textsc{GLAM-Uchuu} lightcones}
\label{sec:glam_vs_other}

\begin{figure*}
    \centering
    \includegraphics[width=\linewidth]{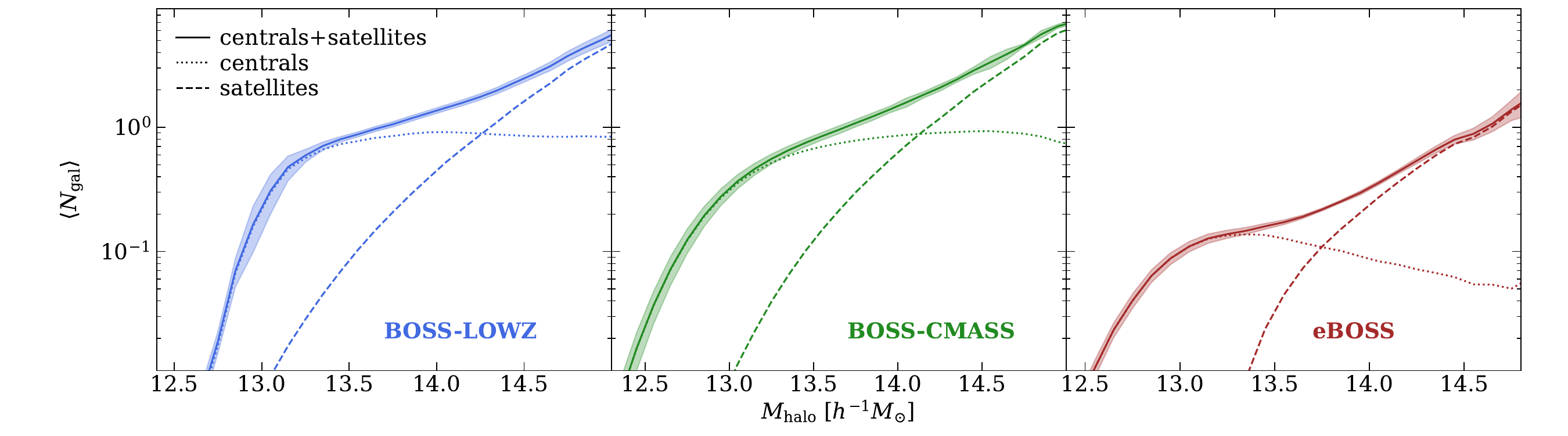}
    \caption{The HOD measured in the LOWZ (left panel), CMASS (middle panel) and eBOSS (right panel) samples, as determined from the mean of the \textsc{Uchuu} galaxy cubic boxes. 
    The solid lines represent the combined HOD of centrals and satellites, while the dotted and dashed lines show the mean halo occupancy for centrals and satellites, respectively.
    The shaded area shows the standard deviation computed using 2000 \textsc{GLAM-Uchuu} lightcones.
    }
    \label{fig:Uchuu_hod}
\end{figure*}
As described in Section~\ref{sec:glam_hod}, we populate the \textsc{GLAM} halos with galaxies using the HOD method. This statistic is obtained from the \textsc{Uchuu} galaxy cubic boxes and is shown in Fig.~\ref{fig:Uchuu_hod}.
By construction, the \textsc{GLAM-Uchuu} HOD agrees with that of the high-resolution \textsc{Uchuu} simulation. The forms of the HOD we recover are consistent with those found in previous BOSS and eBOSS LRG HOD studies \citep[e.g.][]{nuza_2013,Alam_2020}.

In this section, we analyze the covariance errors obtained from the \textsc{GLAM-Uchuu} covariance lightcones and present a comparison of these results with \textsc{MD-Patchy} and \textsc{EZmock}. For this analysis, we focus on the CMASS- and eBOSS-N samples, as the southern hemisphere samples have considerably smaller effective areas compared to the northern ones. Additionally, the LOWZ and COMB samples have significant discontinuities in their footprints.

For a fair comparison of \textsc{GLAM-Uchuu} results with the $2048$ \textsc{MD-Patchy} lightcones, we (randomly) select 2000 of those lightcones for the analysis carried out in this section. In contrast, given the availability of only 1000 \textsc{EZmock} lightcones, we randomly select an equivalent number from our set of 2000 \textsc{GLAM-Uchuu} lightcones.

\subsubsection{Covariances in configuration space}
\label{sec:glam_vs_other:tpcf}

\begin{figure}
    \centering
    \includegraphics[width=\linewidth]{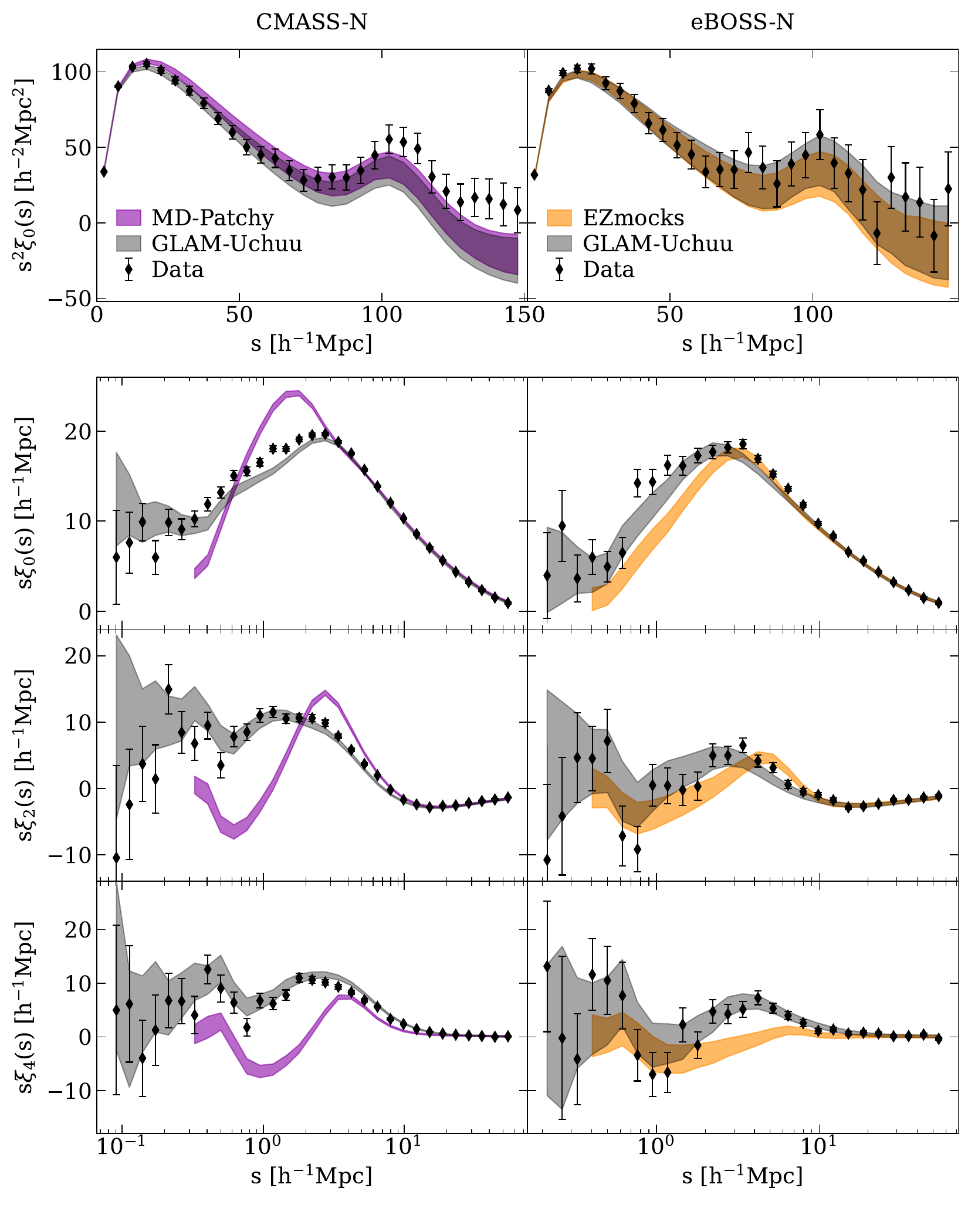}
    \caption{Measurements of the 2PCF monopole, quadrupole and hexadecapole moments for CMASS-N (left column) and eBOSS-N (right column). 
    The shaded regions represent the standard deviation around its mean for \textsc{GLAM-Uchuu} (in black), \textsc{MD-Patchy} (purple, left column), and \textsc{EZmock} (orange, right column).
    The points with error bars show the measurements from the observed data, where the errors correspond to the 1-$\sigma$ scatter derived from \textsc{GLAM-Uchuu} lightcones.
    The upper panels show the 2PCF measurements from $0$ to $150\hMpc$ in $5\hMpc$ bins. The lower panels show the smaller scales up to $60\hMpc$ in equally-spaced logarithmic bins.
    It is evident that the \textsc{MD-patchy} and \textsc{EZmock} results exhibit significant discrepancies with the data between $0.1$ to $30\hMpc$, further emphasizing the need for generating the \textsc{GLAM-Uchuu} lightcones.}
    \label{fig:glamother_tpcf}
\end{figure}

\begin{figure}
    \centering
    \begin{subfigure}[c]{\linewidth}
        \includegraphics[width=\linewidth]{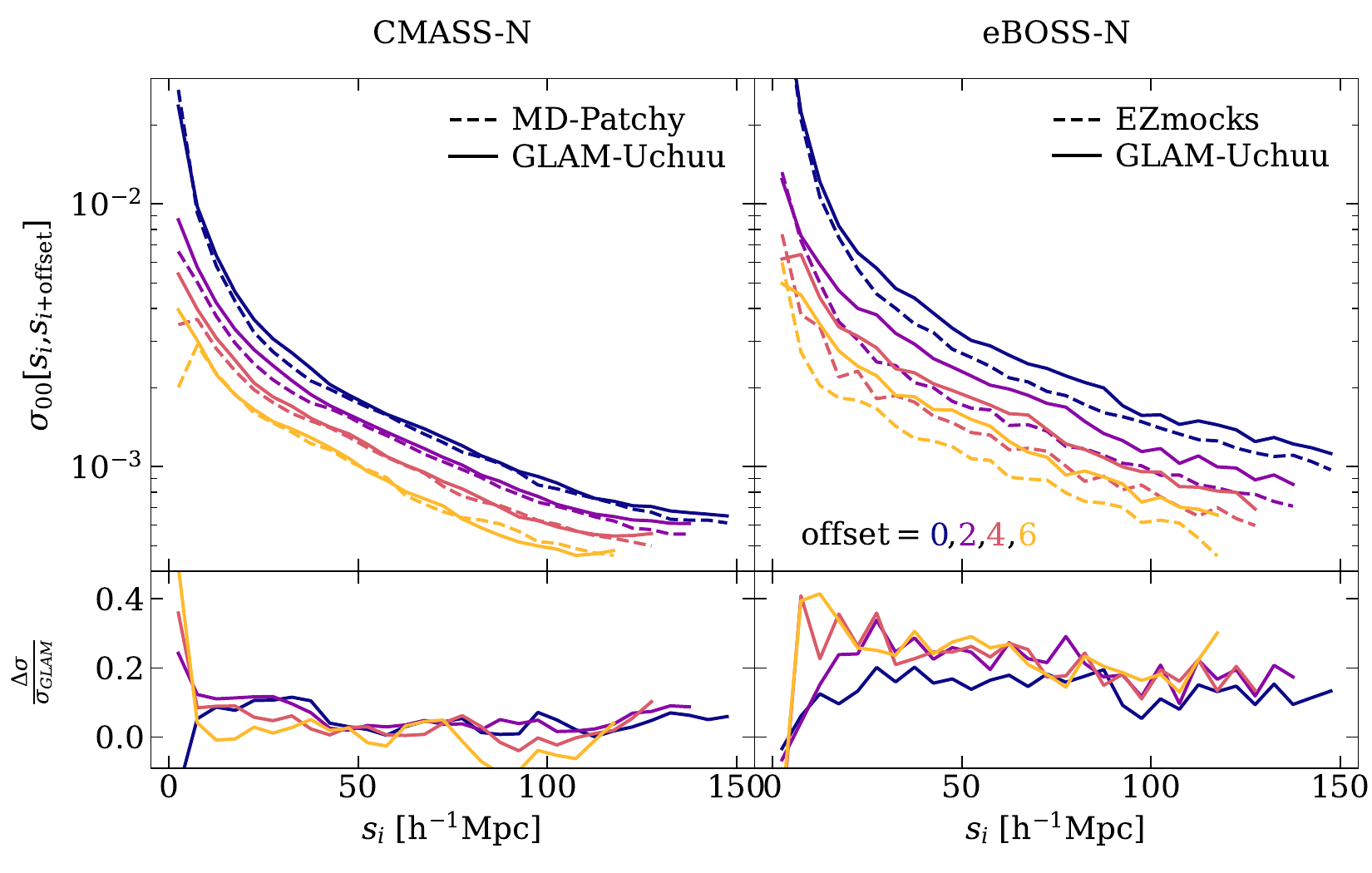}
        \caption{Covariance matrix diagonals.}
    \end{subfigure}
    \begin{subfigure}[c]{\linewidth}
        \includegraphics[width=\linewidth]{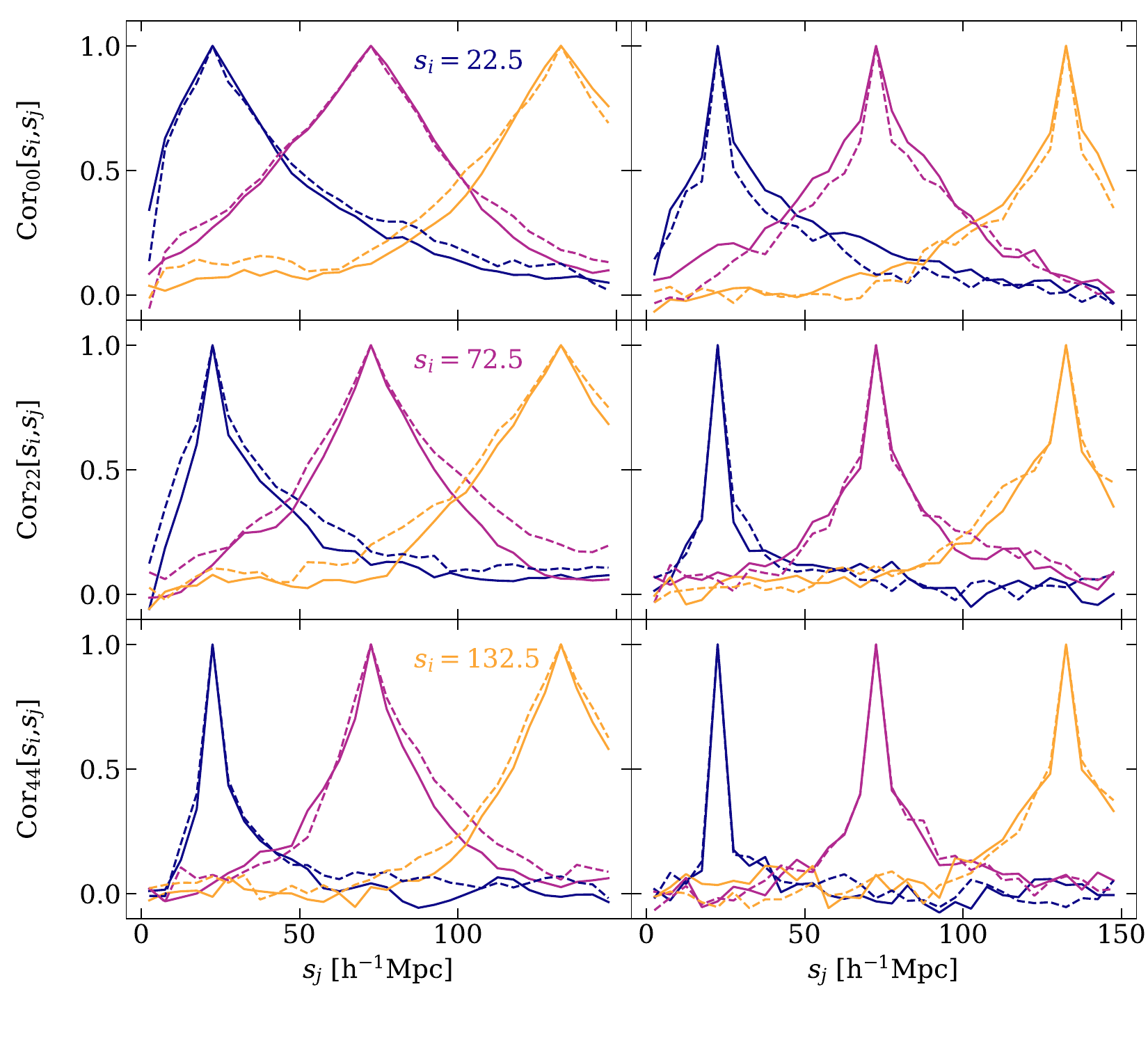}
        \caption{Correlation matrix rows.}
    \end{subfigure}
    \caption{Panel (a): Comparison of the covariance matrices of the 2PCF monopole, $\sigma_{00}=\sqrt{\mathrm{Cov}_{00}}$, for CMASS-N and eBOSS-N (left and right columns respectively). For both samples, \textsc{GLAM-Uchuu} is represented by solid lines, while \textsc{MD-Patchy} (left column) and \textsc{EZmock} (right column) are shown with dashed lines. 
    Different colours represent distinct terms of the covariance matrices: diagonal term in blue and second, fourth and sixth off-diagonal terms in purple, pink, and orange respectively. The bottom panel shows the residuals between covariances. 
    Panel (b): Slices through the correlation matrices of the monopole, quadrupole and hexadecapole ($\ell=0,2,4$) moments of the 2PCF for CMASS-N and eBOSS-N (left and right columns respectively). \textsc{GLAM-Uchuu} is depicted with solid lines in both columns. \textsc{MD-Patchy} (left column), and \textsc{EZmock} (right column) are indicated by dashed lines.
    Different colours represent distinct $\mathrm{s}_i$ values in units of $\hMpc$: $\mathrm{s}_i=22.5, 72.5$~and~$132.5$ in purple, pink, and yellow respectively.}
    \label{fig:glamother_details}
\end{figure}

\begin{figure*}
    \centering
    \includegraphics[width=\linewidth]{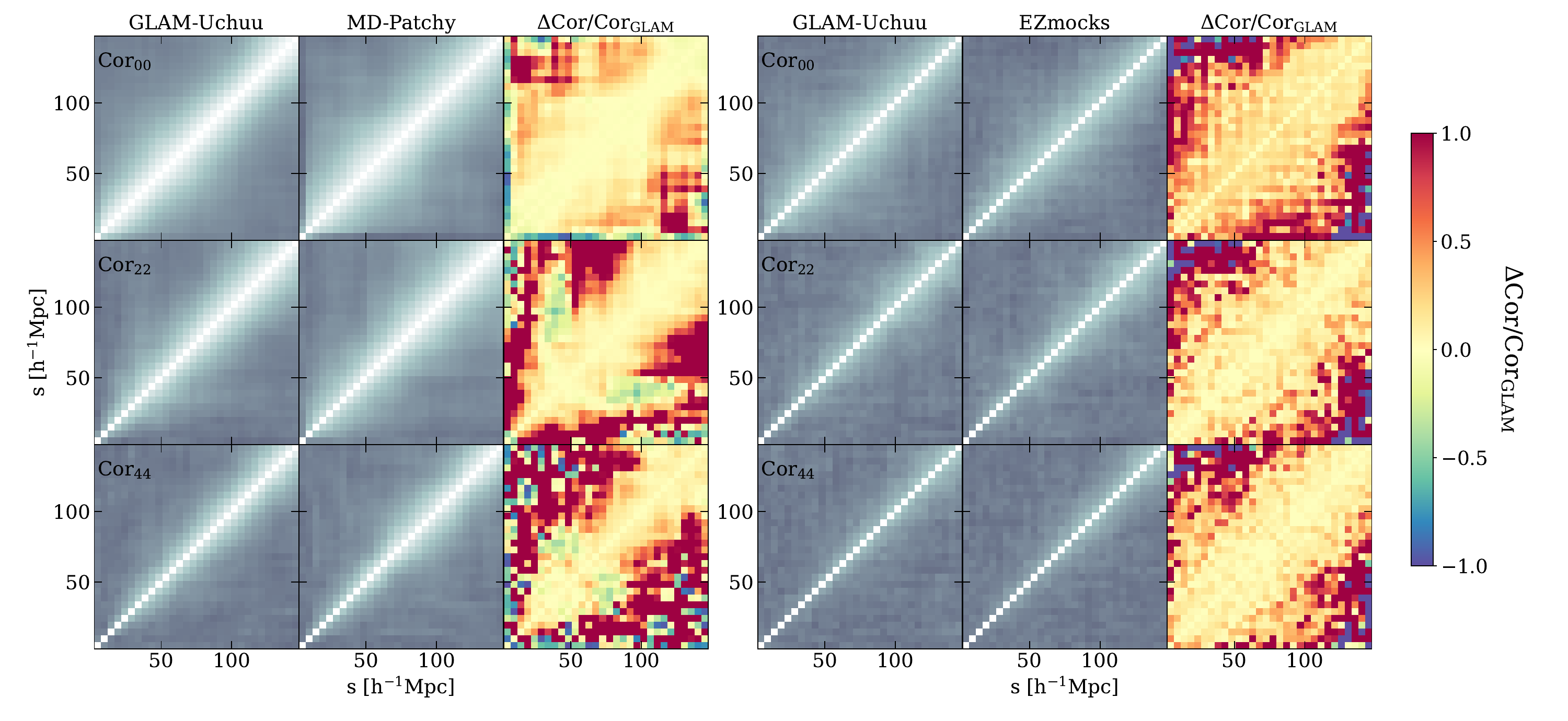}
    \caption{\textit{Left panel}: Correlation matrices (Equation~\ref{eq:corr_matrix}) of the 2PCF monopole, quadrupole and hexadecapole moments (first, second and third rows, respectively) obtained from \textsc{GLAM-Uchuu} (first column) and \textsc{MD-Patchy} (second column). The residuals, $r$, between \textsc{GLAM-Uchuu} and \textsc{MD-patchy}, are shown in the third column. Here, a positive value of $r$ indicates that the values of the \textsc{GLAM-Uchuu} correlation matrix are lower than those of \textsc{MD-patchy}, while $r<0$ means the opposite. 
    \textit{Right panel}: We examine the eBOSS-N sample covariances. The layout is identical to the left plot, but \textsc{EZmock} is used instead of \textsc{MD-Patchy}.}
    \label{fig:glamother_corr}
\end{figure*}

We compare the mean 2PCF multipole moments measured from the \textsc{GLAM-Uchuu} runs, along with those from \textsc{MD-patchy} and \textsc{EZmock}, in Fig.~\ref{fig:glamother_tpcf}.
The left column shows the analysis for the CMASS-N sample, while the right column corresponds to the eBOSS-N sample. 
Overall, \textsc{GLAM-Uchuu} accurately reproduces the observed CMASS-N and eBOSS-N data statistics across all studied multipoles and scales. However, when examining the results of \textsc{MD-patchy} and \textsc{EZmock}, it becomes evident that the their multipole moments exhibit strange features and discrepancies on scales below $20\hMpc$. 
The CMASS-N study also reveals that the resolution of \textsc{GLAM-Uchuu} lightcones is better than that of \textsc{MD-patchy} and \textsc{EZmock}, reaching down to scales of $0.1\hMpc$, in comparison to $0.3\hMpc$ and $0.4\hMpc$ for the other lightcones.

Furthermore, the behaviour observed in \textsc{GLAM-Uchuu} confirms that some of the variations previously noted in \textsc{Uchuu} multipole moments can be attributed to statistical fluctuations, as these discrepancies diminish as the number of lightcones increases. 
For scales below $60\hMpc$, the 2PCF multipoles of \textsc{GLAM} remain within 1-$\sigma$ of the data for both BOSS and eBOSS, whereas \textsc{Uchuu} shows variations between 1 and 2-$\sigma$ (see Fig.~\ref{fig:Uchuu_TPCF}~(b)).
The improved consistency of the quadrupole and hexadecapole measurements compared to the monopole, as the number of lightcones increases, could potentially be associated with volumetric effects and harmonic modes within the data. The $\ell=2$ and $\ell=4$ moments may exhibit greater sensitivity to fluctuations arising from the geometric characteristics and configuration of the observed volume. Increasing the number of lightcones and averaging the results could help mitigate some of these influences.

The covariance matrix of the 2PCF multipoles, $\xi_\ell(s)$, can be defined as follows:
\begin{equation}
    \begin{split}
        \mathrm{Cov}_{\ell\ell'}[s_i,s_j]&=\\
        & \frac{\mathcal{C}}{N_\mathrm{m}-1}\sum_{p}^{N_\mathrm{m}}\left[ \xi_{\ell}^{p}(s_i)-\overline{\xi}_{\ell}(s_i)\right]\left[ \xi_{\ell'}^{p}(s_j)-\overline{\xi}_{\ell'}(s_j)\right],
    \end{split}
    \label{eq:cov_matrix}
\end{equation}
where $N_\mathrm{m}$ is the number of lightcones, $\xi_{\ell}^{p}$ indicates the $l$-order 2PCF multipole of the $p$-th lightcone, and $\overline{\xi}_{\ell}$ denotes the mean 2PCF multipole across all lightcones. The term $\mathcal{C}$ accounts for the dependence of $\mathrm{Cov}[s_i,s_j]$ on the size of the simulation box. When the effective volume of the observed sample, $V_\mathrm{eff}$, is smaller than the simulation volume, $V_\mathrm{sim}$, then $\mathcal{C}=1$. Conversely, if $V_\mathrm{eff}>V_\mathrm{sim}$, the covariance matrix must be scaled, such that $\mathcal{C}\neq1$. Specifically, for the \textsc{GLAM-Uchuu}-CMASS-N covariance matrix, $\mathcal{C}\simeq 0.6$ (see Appendix~\ref{App:repTest} for details).
To further investigate the variations between lightcones, one can also examine the correlation matrices of the 2PCF multipoles:
\begin{equation}
    \mathrm{Cor}_{\ell\ell'}[\mathrm{s}_i,\mathrm{s}_j]= \frac{\mathrm{Cov}_{\ell\ell'}[\mathrm{s}_i,\mathrm{s}_j]}{\sqrt{\mathrm{Cov}_{\ell\ell'}[\mathrm{s}_i,\mathrm{s}_i]\mathrm{Cov}_{\ell\ell'}[\mathrm{s}_j,\mathrm{s}_j]}}.
    \label{eq:corr_matrix}
\end{equation}

Fig.~\ref{fig:glamother_details}~(a) provides a comparison of various components of the 2PCF monopole covariance matrices examined in this study. Specifically, it shows the diagonal and second, fourth, and sixth off-diagonal terms. 
In the left column, we compare results from \textsc{GLAM-Uchuu}-N with those from \textsc{MD-Patchy}-CMASS-N. 
The values of the \textsc{GLAM-Uchuu} covariance matrix are consistently higher than those of \textsc{MD-Patchy} across all slices, with a discrepancy between the two that fluctuates within $10-15\%$, as shown by the residuals in the bottom panel.
We compare \textsc{GLAM-Uchuu} and \textsc{EZmock} for the eBOSS-N sample in the right column. The trend is the same: the values of the \textsc{GLAM-Uchuu} elements are higher than those of \textsc{EZmock}. The disagreement in this case is larger for some values of $s$, reaching up to $30\%$.
Discrepancies within the $10\%$ range are anticipated in both cases, and only discrepancies above this value are considered relevant (see Appendix~\ref{App:repTest}). Differences above this percentage are due to the different predictions of each simulation.

In Fig.~\ref{fig:glamother_corr}, we report the derived correlation matrices for the monopole, quadrupole, and hexadecapole moments of the 2PCF.
In the left plot, we show the correlation matrices obtained for the CMASS-N sample using \textsc{GLAM-Uchuu}, \textsc{MD-Patchy}, and the residuals between the two. The right plot follows the same, but for the eBOSS-N sample, replacing \textsc{MD-Patchy} with \textsc{EZmock}. In both cases, the residuals show distinct patterns in $\mathrm{Cor}_{00}$, $\mathrm{Cor}_{22}$ and $\mathrm{Cor}_{44}$. 
To gain a deeper insight into the degree of correlation and the structure of these matrices, Fig.~\ref{fig:glamother_details}~(b) displays cross-sections through the covariance matrices that highlight the behaviour of their non-diagonal terms.
We make a comparison for three separation values: $\mathrm{s}_i = 22.5, 72.5, 132.5\hMpc$, as indicated by the colour scheme in the plot.
In general, we observe a stronger correlation among bins located near the diagonal for the CMASS-N sample compared to the eBOSS-N, resulting in broader peaks in the former case. This correlation tends to weaker as we consider higher order multipoles.
When examining individual cases, we notice that the off-diagonal elements of \textsc{MD-Patchy} display a higher level of correlation across all three multipoles when compared with \textsc{GLAM-Uchuu}. 
In comparison with \textsc{EZmock}, while \textsc{GLAM-Uchuu} presents more highly correlated off-diagonal elements for the monopole, it demonstrates similar results to \textsc{EZmock} for the quadrupole and hexadecapole.

\subsubsection{Covariances in Fourier space}
\label{sec:glam_vs_other:pk}

\begin{figure}
    \centering
    \includegraphics[width=\linewidth]{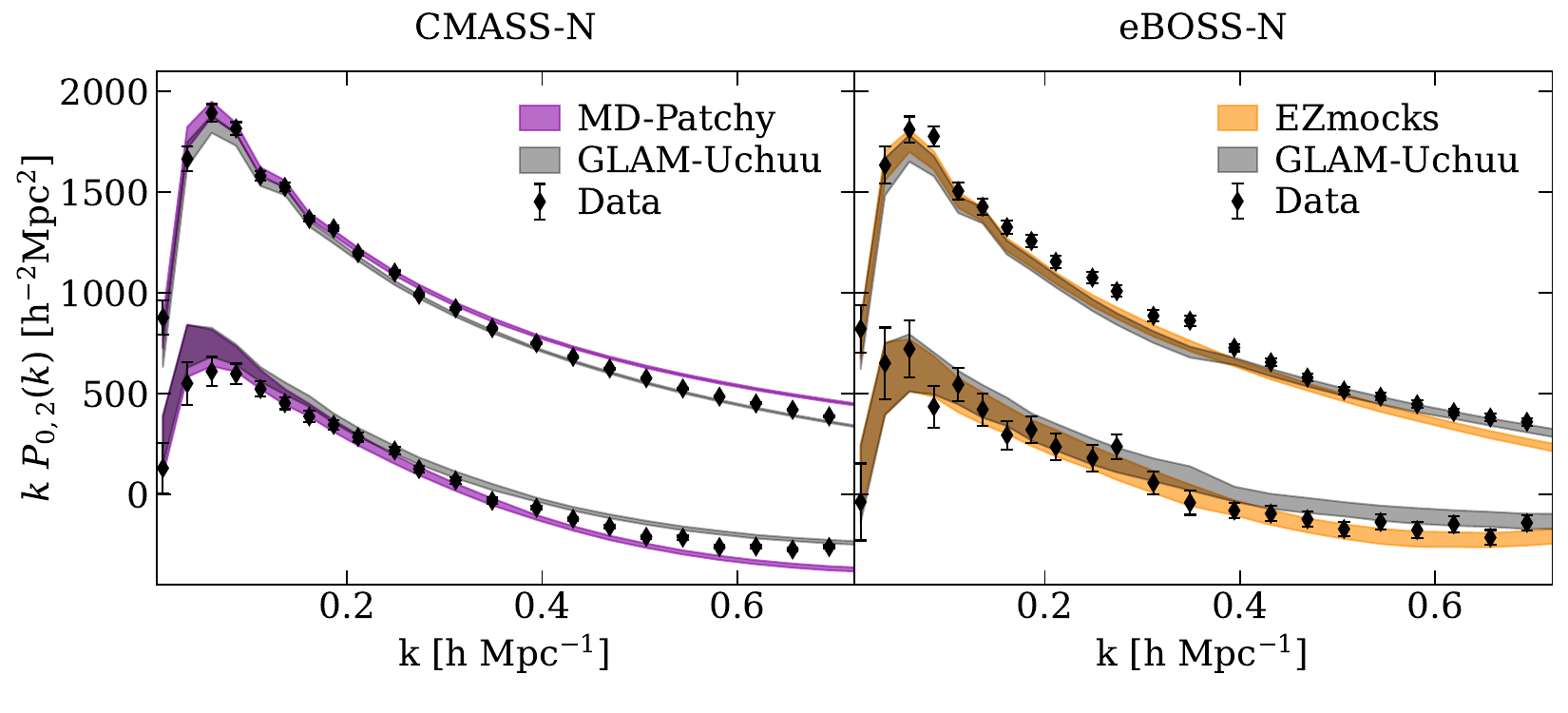}
    \caption{Measurements of the monopole (upper curve) and quadrupole (lower curve) moments of the power spectrum for the CMASS-N (left panel) and eBOSS-N (right panel) samples. 
    The shaded regions represent the standard deviation around the mean for \textsc{GLAM-Uchuu} (in black), \textsc{MD-Patchy} (purple), and \textsc{EZmock} (orange) lightcone. 
    The points with error bars show the measurements from the observed data, where the errors correspond to the 1-$\sigma$ scatter derived from \textsc{GLAM-Uchuu} lightcones.
    We measure the power spectrum over the range $\mathrm{k}=0.005$ to $0.72\hMpcInv$, using $0.03\hMpcInv$ bins.
    The results from \textsc{GLAM-Uchuu} show good agreement with the observed clustering statistics for both CMASS-N and eBOSS-N. For $k >0.4\hMpcInv$, the multipoles from \textsc{MD-Patchy} and \textsc{EZmock} begin to deviate from the observed data.}
    \label{fig:glamother_pk}
\end{figure}

\begin{figure}
    \centering
    \begin{subfigure}[c]{\linewidth}
        \includegraphics[width=\linewidth]{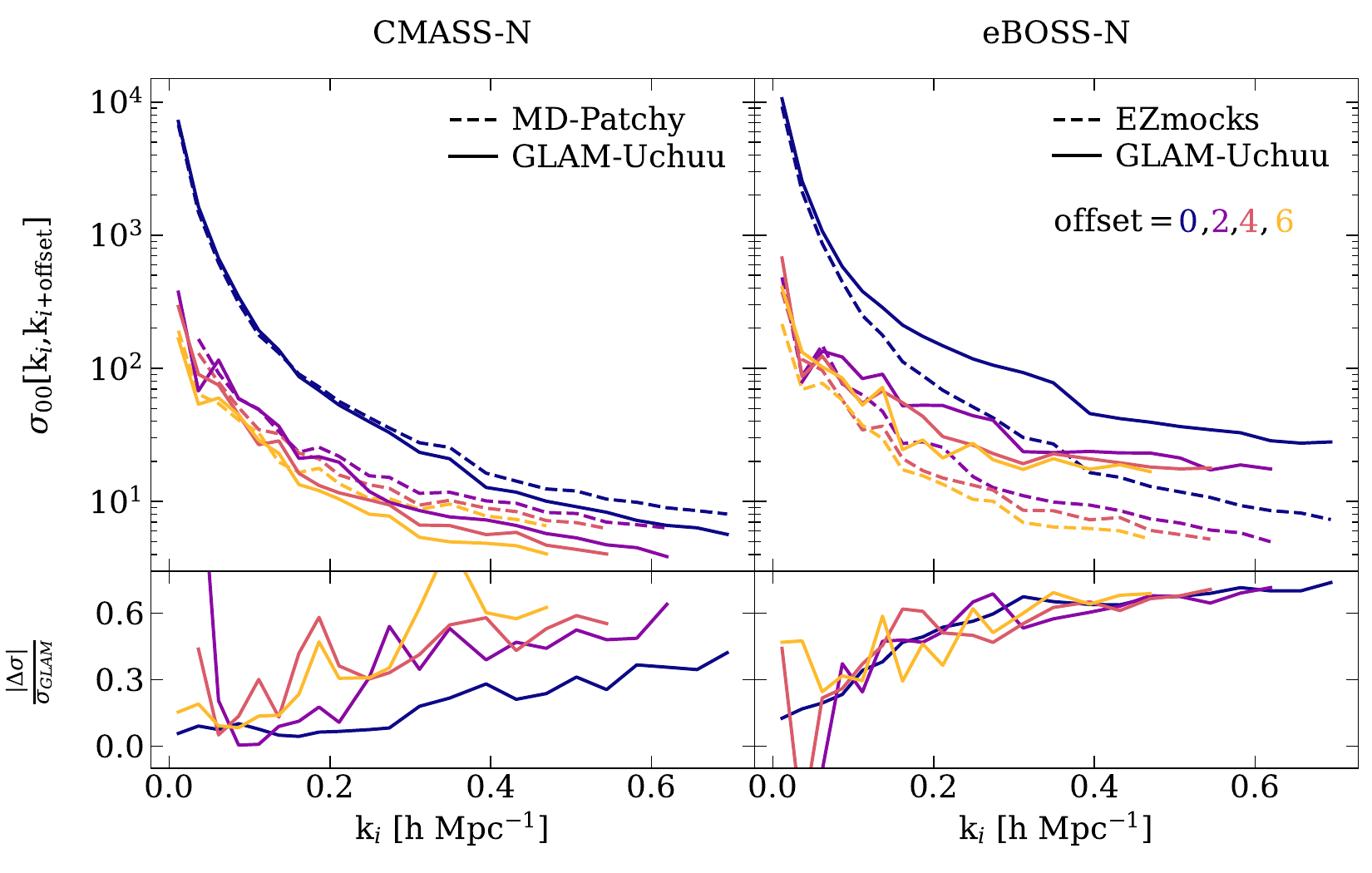}
        \caption{Covariance matrix diagonals.}
    \end{subfigure}
    \begin{subfigure}[c]{\linewidth}
        \includegraphics[width=\linewidth]{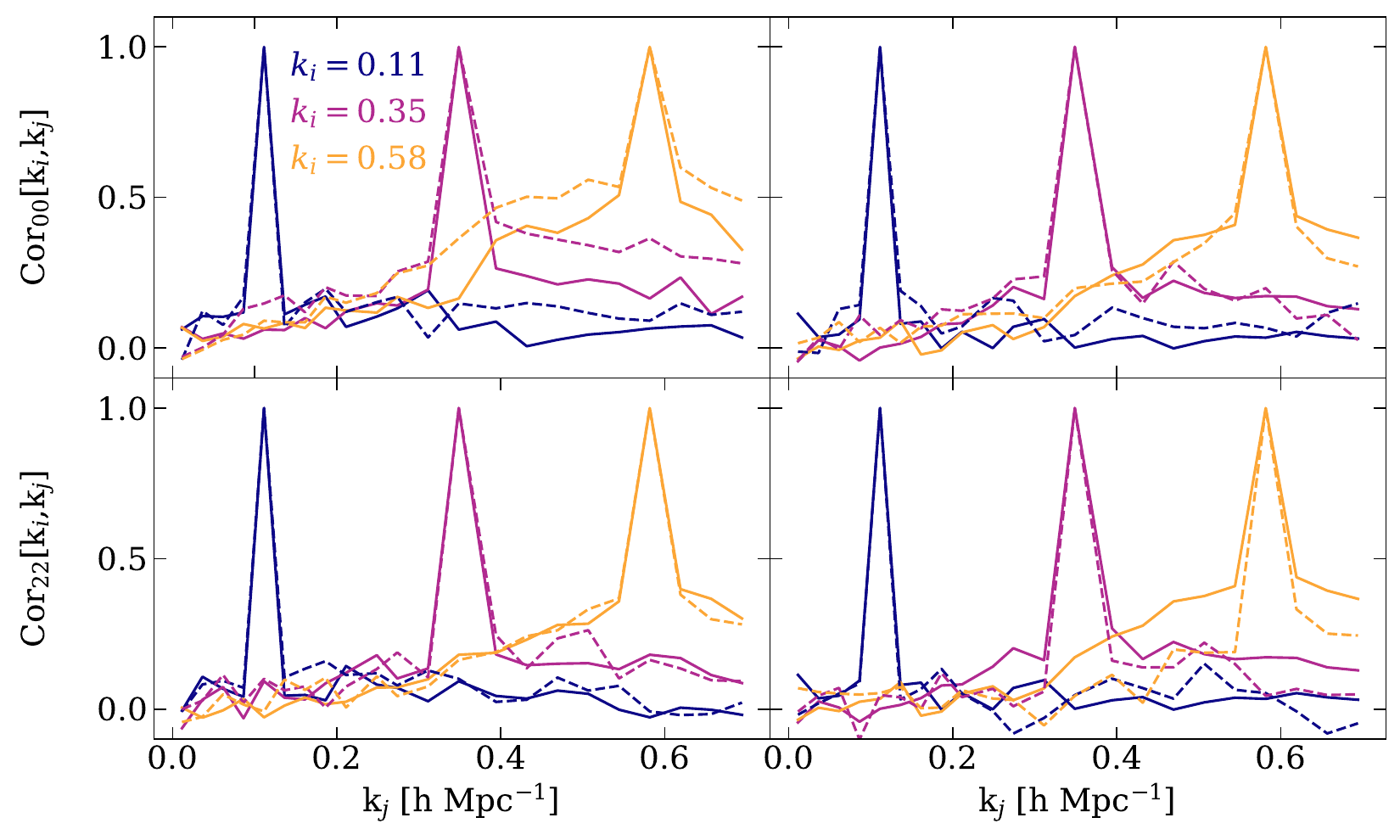}
        \caption{Correlation matrix rows.}
    \end{subfigure}
    \caption{Panel (a): Comparison of the covariance matrices of the power spectrum monopole,  $\sigma_{00}=\sqrt{\mathrm{Cov}_{00}}$, for CMASS-N and eBOSS-N (left and right columns respectively). For both samples, \textsc{GLAM-Uchuu} is represented by solid lines, while \textsc{MD-Patchy} (left column) and \textsc{EZmock} (right column) are shown with dashed lines. 
    Various colours represent different terms of the matrices: diagonal terms in blue, and second, fourth and sixth off-diagonal terms in purple, pink and orange, respectively. The bottom panel displays the absolute residuals between the covariances.
    Panel (b): This panel shows slices through the correlation matrices of the power spectrum monopole and quadrupole for CMASS-N and eBOSS-N (again, left and right columns respectively). \textsc{GLAM-Uchuu} is depicted with solid lines in both columns. \textsc{MD-Patchy} (left column), and \textsc{EZmock} (right column) are indicated by dashed lines. Different colours correspond to distinct $\mathrm{k}_i$ values in units of $\hMpcInv$: $\mathrm{k}_i=0.11, 0.35$~and~$0.58$ in purple, pink, and yellow respectively.}
    \label{fig:glamother_details_pk}
\end{figure}

\begin{figure*}
    \centering
    \includegraphics[width=\linewidth]{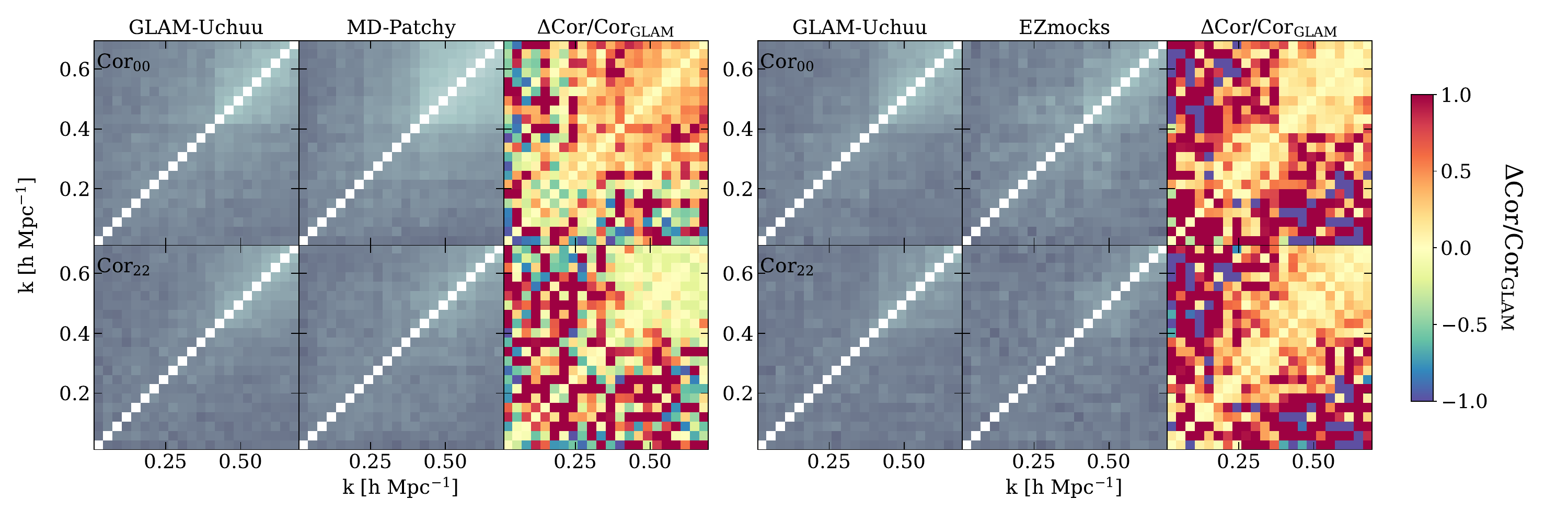}
    \caption{\textit{Left panel}: Correlation matrices (Equation~\ref{eq:corr_matrix}) for the power spectrum monopole and quadrupole (first and second rows, respectively) obtained from \textsc{GLAM-Uchuu} (first column) and \textsc{MD-Patchy} (second column). The third column displays the residuals, $r$, between \textsc{GLAM-Uchuu} and \textsc{MD-Patchy}. Here, $r>0$ indicates that the values of the \textsc{GLAM-Uchuu} correlation matrix are lower than those of \textsc{MD-Patchy}, and $r<0$ indicates the opposite.
    \textit{Right panel}: A similar study is conducted for the eBOSS-N sample, using the same layout as the left plot. However, \textsc{EZmock} replaces \textsc{MD-Patchy} in the comparison.}
    \label{fig:glamother_corr_pk}
\end{figure*}

We analyze our \textsc{GLAM} lightcones in Fourier space using the same methods and equations ($\xi_{\ell}(\mathrm{s})^{p}\rightarrow \mathrm{P}_{\ell}(\mathrm{k})^{p}$) as presented in the configuration space section.

Fig.~\ref{fig:glamother_pk} shows the mean monopole and quadrupole moments of the power spectrum measured from  \textsc{GLAM-Uchuu}, along with those from \textsc{MD-Patchy} and \textsc{EZmock}. Error bars represent the uncertainty in the observed data, with errors estimated by the 1-$\sigma$ scatter derived from \textsc{GLAM-Uchuu} lightcones. 
The \textsc{GLAM-Uchuu} results are in good agreement with the observed CMASS-N and eBOSS-N clustering statistics for all the studied multipoles and scales. 
For k values above $0.4\hMpcInv$, the \textsc{MD-Patchy} and \textsc{EZmock} multipoles deviates from the data. This trend is consistent with that demonstrated in Fig.~\ref{fig:glamother_tpcf}.

In Fig.~\ref{fig:glamother_details_pk}~(a), we present the diagonal and second, fourth and sixth off-diagonal terms of the power spectrum monopole covariance matrices. 
In the left column, we compare \textsc{GLAM-Uchuu} and \textsc{MD-Patchy} for the CMASS-N sample, while the right column compares \textsc{GLAM-Uchuu} and \textsc{EZmock} for the eBOSS-N sample. 
The diagonal values of \textsc{GLAM-Uchuu} are slightly higher than those of \textsc{MD-Patchy} for low k values. However, this trend inverts for $\mathrm{k}\gtrsim0.25\hMpcInv$.
When comparing with \textsc{EZmock}, \textsc{GLAM-Uchuu} slices are higher at all scales, with a discrepancy between the two that increases with $\mathrm{k}_i$, reaching values above $60\%$ for $\mathrm{k}_i\geq0.3\hMpcInv$. 
As in configuration space analysis, discrepancies within $10\%$ are expected, with discrepancies above this value attributed to the distinct predictions of each simulation.

We present the derived correlation matrices for the monopole and quadrupole moments of the power spectrum in Fig.~\ref{fig:glamother_corr_pk}.
Following the same format as shown in Fig.~\ref{fig:glamother_corr}, the left plot displays the correlation matrices obtained for the CMASS-N sample using \textsc{GLAM-Uchuu} and \textsc{MD-Patchy}. In the right plot, the eBOSS-N sample is analysed with \textsc{EZmock} instead of \textsc{MD-Patchy}. 
The residuals, $r$, are plotted in the third columns, showing again some patterns evident in both monopole and quadrupole moments. 
We show three cross-sections through these correlation matrices at $k_{i}=0.11, 0.35, 0.58\hMpcInv$ in Fig.~\ref{fig:glamother_details_pk}~(b).
In general, the correlation of the analysed cuts are very similar among \textsc{GLAM-Uchuu}, \textsc{MD-Patchy} and \textsc{EZmock}, particularly for the terms near the diagonal. For the bins located farther from the diagonal, noise becomes prominent and the signal amplitude is considerably low, posing challenges in determining clear trend.
In the CMASS monopole case however, we clearly observe that \textsc{MD-Patchy} display a higher level of correlation than \textsc{GLAM-Uchuu}. 

The analysis presented in this section confirms the reliability of the methodology used in generating the \textsc{GLAM-Uchuu} covariance lightcones, as the accurately reproduce the clustering measurements from observations.
Comparisons of the \textsc{GLAM-Uchuu} lightcones with results from \textsc{MD-Patchy} and \textsc{EZmock} prove insightful. These approximate approaches exhibit peculiar behaviors in the multipole moments of the 2PCF, clearly demonstrating the advantage of using $N$-body simulations over less accurate methods.
We have verified that the diagonal elements of the \textsc{GLAM-Uchuu} covariance matrices generally exceed those obtained from \textsc{MD-Patchy} and \textsc{EZmock}. 
Consequently, the errors we estimate for the observed 2PCFs and power spectrum are larger than previously assumed. 
However, it is important to note that discrepancies within$10\%$ in the error estimation are anticipated, and only discrepancies above this threshold are considered relevant as they are, presumably, due to the different predictions of each simulation.
The discovery of discrepancies greater than $10\%$ leads to increased uncertainties in cosmological constraints derived from BOSS and eBOSS data compared to those using \textsc{MD-patchy} and \textsc{EZmock} covariances. This includes impacts on BAO distance estimates and the estimation of cosmological parameters \citep[see][and references therein]{Alam2015,Alam_2021}, as well as implications for uncovering primordial non-Gaussianity \citep[see][]{Beutler_2016, Merz_2021, Mueller_2022}. 
Certain studies employing \textsc{MD-patchy} and \textsc{EZmock} covariance matrices have reported tensions in the inferred value of $\sigma_{8}$ when compared with Planck measurements \citep[see][]{Wadekar_2020, Chen_2022, Kobayashi_2022}. 
This tension could potentially be mitigated or even resolved with larger uncertainties factored into the estimated parameters.

\subsection{Performance with other cosmologies}
\label{sec:glam_cosmos}

Up to this point, we have demonstrated that our \textsc{Uchuu} clustering  model, based on standard Planck cosmology, accurately reproduces the observed clustering data while accounting for relevant uncertainties. However, it is also worthwhile to assess the performance of other cosmological models.
For this purpose, we use our \textsc{GLAM} simulation runs for different cosmologies as presented in Section~\ref{sec:glam_simu}. These models, summarized in Table~\ref{tab:simus_info}, include \textsc{GLAM-PMILL}, \textsc{GLAM-PMILLnoBAO} and \textsc{GLAM-Abacus}. It is important to note that their cosmological parameters do not considerably deviate from the Planck15 cosmology adopted for \textsc{Uchuu}.
These newly generated lightcones are produced following the same methodology as the \textsc{GLAM-Uchuu} lightcones, described in Section~\ref{sec:glam_light}. 
Regardless of the specific cosmology, galaxies are populated in all \textsc{GLAM} halo cubic boxes using the HODs obtained from the \textsc{Uchuu} galaxy boxes, which are based on the PL15 cosmology. 
Given the minor difference among the various explored cosmological parameters, creating distinct HODs for each cosmology would yield marginal improvements at the cost of considerable additional complexity.

\begin{figure}
    \centering
    \includegraphics[width=\linewidth]{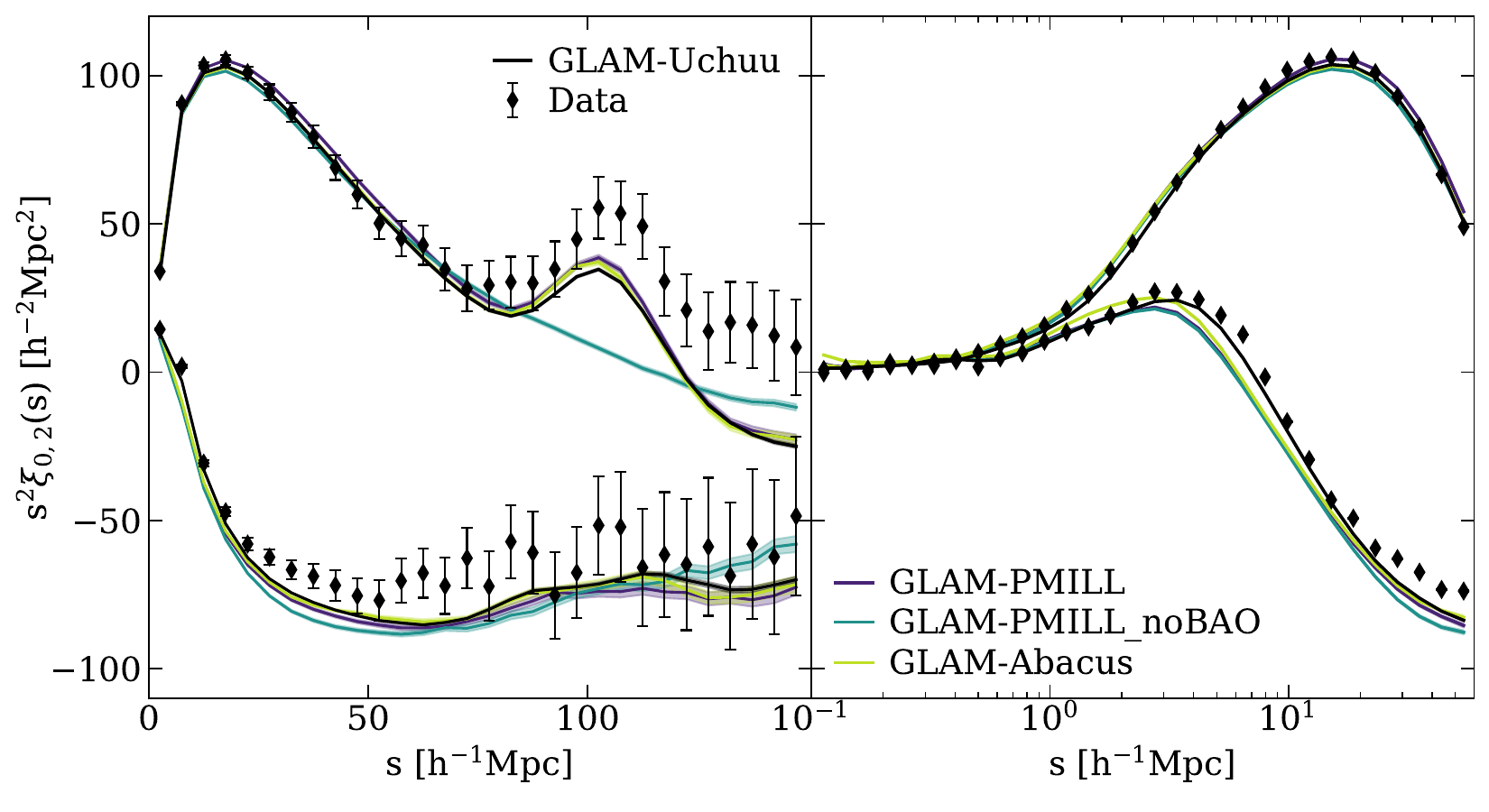}
    \caption{Measurements of the monopole and quadrupole moments of the 2PCF for the CMASS-N are presented using different sets of \textsc{GLAM} simulations (see Table~\ref{tab:simus_info}).
    The mean 2PCF derived from the \textsc{GLAM-PMILL} lightcones is represented by solid purple lines, \textsc{GLAM-PMILLnoBAO} in blue, \textsc{GLAM-Abacus} in green, and \textsc{GLAM-Uchuu}in black. The shaded regions indicate the errors of the mean, estimated from the diagonal components of their respective covariance matrices.
    Observational data are shown as points with error bars, where these bars represent the 1-$\sigma$ errors estimated using the \textsc{GLAM-Uchuu} lightcones.
    The 2PCF measurements span from $0$ to $150\hMpc$ in $5\hMpc$ bins (left plot), and up to $60\hMpc$ in equally spaced logarithmic bins (right plot).}
    \label{fig:CosmoGLAM_TPCF}
\end{figure}
The impact of different cosmologies on the large-scale distribution of galaxies is depicted in Fig.~\ref{fig:CosmoGLAM_TPCF}, where the mean monopole and quadrupole moments of the 2PCF obtained from the \textsc{GLAM-PMILL}, \textsc{GLAM-PMILLnoBAO}, and \textsc{GLAM-Abacus} lightcones for the CMASS-N sample are shown.
For reference, results from observed data and theoretical predictions calculated from the mean of the \textsc{GLAM-Uchuu} lightcones are also presented.
Generally, except for \textsc{GLAM-PMILLnoBAO}, all \textsc{GLAM} lightcones effectively replicate the observational measurements.
The absence of the BAO feature in the \textsc{GLAM-PMILLnoBAO} monopole, alongside its BAO-inclusive counterpart \textsc{GLAM-PMILL}, allows us to assess the statistical significance of BAO detection. 
Despite the measured residuals and associated uncertainties, this study does not definitively distinguish between the presence or absence of massive neutrinos. Although \textsc{GLAM-Uchuu} achieves better results (residuals below $3\%$ up to $50\hMpc$) compared to \textsc{GLAM-Abacus}, which includes the effect of massive neutrinos (resulting in residuals of $4\%$), both models are capable of reasonably reproducing the observed data within the error margins.

\section{Conclusions}
\label{sec:conclusions}
\input{conclusions}

\section*{Acknowledgements}
\input{thanks}


\section*{Data Availability}

The datasets included in this work consist of the following: $2~\mathrm{(lightcones)}\times8~\mathrm{(samples)}$ \textsc{Uchuu} lightcones, the  $2000\times8$ \textsc{GLAM-Uchuu} lightcones, the $100\times8$ \textsc{GLAM-PMILL}, \textsc{GLAM-PMILLnoBAO} and \textsc{GLAM-Abacus} lightcones, along with the BOSS and eBOSS catalogues. These datasets are publicly available at \url{https://www.skiesanduniverses.org/Products/}. A comprehensive list and brief description of the  catalogue columns can be found in Appendix~\ref{App:mock_columns}.


\bibliographystyle{mnras}
\bibliography{references}


\appendix

\section{Content of the public catalogues}
\label{App:mock_columns}

Below is a list of the columns of each data set, along with a short description.

\subsection{\textsc{Uchuu} lightcones}

Each \textsc{Uchuu} lightcone has the following columns:
\begin{itemize}
    \item \texttt{ra}: right ascension (degrees).
    \item \texttt{dec}: declination (degrees).
    \item \texttt{z\_cos}: cosmological redshift.
    \item \texttt{z\_obs}: observed redshift (accounting for peculiar velocities).
    \item \texttt{nz}: number density of galaxies, n(z), at the redshift of the galaxy ($h^3\mathrm{Mpc}^{-3}$).
    \item \texttt{logMstll}: logarithm ($\log_{10}$) of the galaxy stellar mass (M$_\odot$).
    \item \texttt{poly\_id}: healpix pixel ID containing the galaxy.
    \item \texttt{poly\_w}: fibre-collision completeness in the healpix pixel containing the galaxy.
    \item \texttt{w\_cp}: fibre collision correction weight (NNW).
    \item \texttt{w\_fkp}: FKP weight.
    \item \texttt{w\_npcf}: only in COMB sample lightcones. Extra weight to adjust the contribution of each sample (LOWZ and CMASS). 
\end{itemize}
In addition, the following columns describe the properties of the DM host (sub)halos of galaxies. These are taken from the original values in the \textsc{Uchuu} simulation halo catalogues:
\begin{itemize}
    \item \texttt{Mvir\_all}: mass enclosed within the virial overdensity, including unbound particles ($\hMsun$).
    \item \texttt{Rvir}: virial radius ($\hkpc$).
    \item \texttt{rs}: scale radius of a fitted NFW profile (comoving $\hkpc$).
    \item \texttt{Cvir}: dark matter concentration within Rvir.
    \item \texttt{Vpeak}: peak value of $V_\mathrm{max}$ over the halo history (physical $\si{\kilo\meter\per\second}$).
    \item \texttt{id}: ID that identify each (sub)halo (unique in the whole simulation).
    \item \texttt{pid}: ID of the parent central halo for satellite halos, -1 for central halos.
    \item \texttt{x, y, z}: 3D position coordinates. The coordinate system of the \textsc{Uchuu} cubic box has been shifted so that the observer is at the origin, and there are periodic replications of the box (comoving $\hMpc$).
    \item \texttt{vx, vy, vz}: 3D velocity (physical $\kms$).
    \item \texttt{vrms}: velocity dispersion (physical $\kms$).
    \item \texttt{vlos}: velocity vector projected along line-of-sight, with the observer positioned at the origin ($\kms$).
\end{itemize}

Random catalogues, containing $40$ times the number of galaxies of an individual catalogue, are also provided.

\subsection{\textsc{GLAM} lightcones}

Each \textsc{GLAM} lightcone has the following columns:
\begin{itemize}
    \item \texttt{gal\_id}: indicates whether the galaxy is central or a satellite (1 for centrals, 2 for satellites).
    \item \texttt{ra}
    \item \texttt{dec}
    \item \texttt{z\_obs}
    \item \texttt{nz}
    \item \texttt{logMstll}
    \item \texttt{poly\_id}
    \item \texttt{poly\_w}
    \item \texttt{w\_cp}
    \item \texttt{w\_fkp}
    \item \texttt{w\_npcf}
\end{itemize}
In addition, the following columns describe the DM host halo (not subhalos) properties of galaxies, taken from the original values found in the \textsc{GLAM} simulation halo catalogues,
\begin{itemize}
    \item \texttt{Mtotal}: halo virial mass ($\hMsun$).
    \item \texttt{Rvir}
    \item \texttt{rs}
\end{itemize}

Random catalogues, containing $30$ times the number of galaxies of an individual catalogue, are also provided.

\subsection{BOSS and eBOSS data samples}
In order to facilitate any analysis, we also make available the observed galaxy samples used is this work. The columns are as follows: 
\begin{itemize}
    \item \texttt{RA}: right ascension (degrees).
    \item \texttt{DEC}: declination (degrees).
    \item \texttt{Z}: measured redshift, including fibre collisions.
    \item \texttt{NZ}: number density of galaxies, n(z), at the redshift of the galaxy ($h^3\mathrm{Mpc}^{-3}$).
    \item \texttt{WEIGHT\_CP}: fibre collision correction weight.
    \item \texttt{WEIGHT\_NOZ}: redshift failures weight.
    \item \texttt{WEIGHT\_SYSTOT}: total angular systematic weight.
    \item \texttt{WEIGHT\_FKP}: FKP weight.
\end{itemize}

Random catalogues, containing $40$ times the number of galaxies of an individual catalogue, are also provided.

\section{Impact of simulation volume on covariance matrices}
\label{App:repTest}
\begin{figure}
    \centering
    \includegraphics[width=8.1cm]{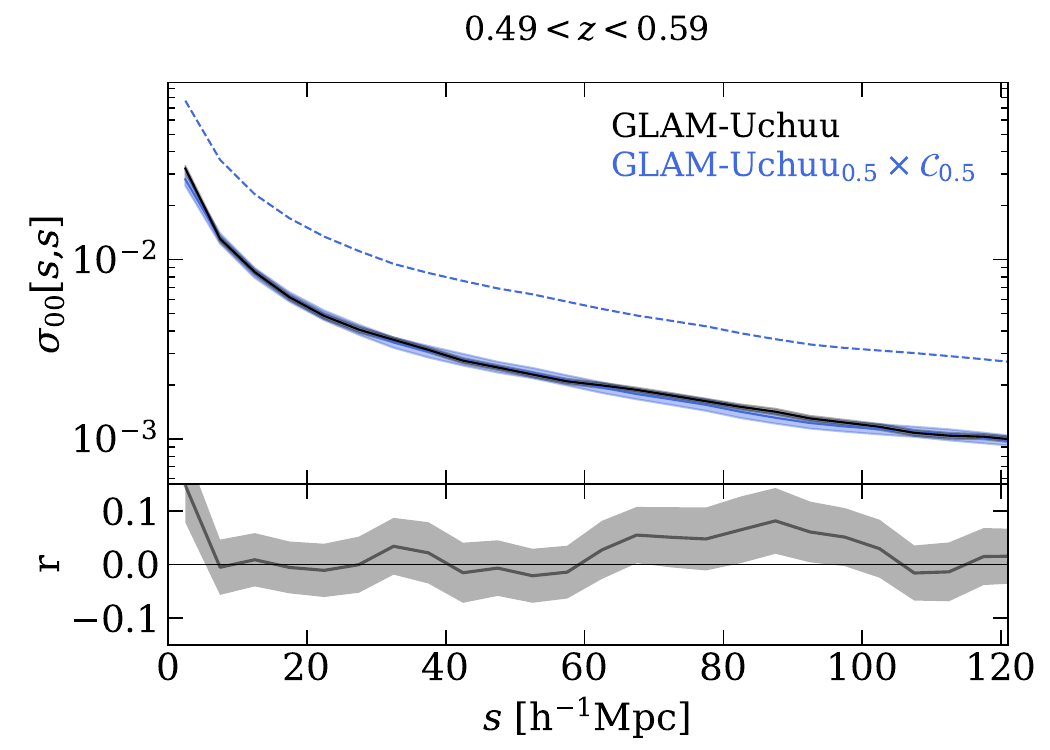}
    \caption{Diagonal components of the covariance matrix for the 2PCF monopole, $\sigma_{00}=\sqrt{\mathrm{Cov}_{00}}$, of the CMASS-N subsample within $0.49<z<0.59$ are displayed. The black solid curve represents estimates for \textsc{GLAM-Uchuu}. Due to the large volume, these results are not affected by the simulation box size. For reference, we include the \textsc{GLAM-Uchuu$_{0.5}$} covariance without box size correction (blue dashed line). The shaded areas indicate the 1-$\sigma$ standard deviation for our set of lightcones. The bottom panel shows the residuals between \textsc{GLAM-Uchuu} and \textsc{GLAM-Uchuu$_{0.5}~\mathrm{x}~\mathcal{C}_{0.5}$} covariances.} 
    \label{fig:repTest}
\end{figure}

Covariances of correlation functions and power spectrum measure the statistical uncertainties in clustering statistics. They are widely used to estimate statistical errors in measured clustering signals and, ultimately, in cosmological parameters. The main sources of these errors are well understood \citep[e.g.][]{Li2014,Mohammed2017,GLAMsimu,KlypinPrada2019}.
Three terms define the covariances of the power spectrum: (1) Gaussian fluctuations due to finite number of Fourier harmonics, (2) non-linear fluctuations that are not related to the number of harmonics, and (3) fluctuations due to waves longer than the simulation box. Similar terms exist for the correlation functions, with the first term resulting from Gaussian fluctuations in the number of pairs at a given separation.
The first two terms of covariances scale inversely with the volume of the observational sample \citep{Mohammed2017,GLAMsimu}. The third term, associated with the superscale modes, is very small for simulations with box sizes $\gtrsim 1\hGpc$ \citep{GLAMsimu,KlypinPrada2019} and is neglected in this analysis.

In order to re-scale covariances, one must know the effective volume covered by either observational samples or simulated galaxy distributions.
While it's straightforward to determine the volume for configurations with volume-limited samples or simulations with cubic boxes, the process becomes more complex with typical observational samples from large-scale galaxy surveys. 
In these cases, the number density of selected galaxies varies with distance, $n(z)$, and the effective volume can be expressed as:
\begin{equation}
V_{\rm eff} = \int{\left( \frac{n(z)~P_\mathrm{FKP}}{1+n(z)P_\mathrm{FKP}} \right)^2 \mathrm{d}V} 
\label{eq:Veff}
\end{equation}
where $P_\mathrm{FKP} = 10^4~h^{-3}\mathrm{Mpc}^{3}$. 
In this equation, the denominator aims to partially compensate for the incompleteness of the selection function by assigning larger contribution (a higher weight) to pairs of objects with small $n(z)$ \citep{Feldman94}. 
However, this weighting does not change the number of pairs, which ultimately defines the covariance matrix. 
For instance, even if a pair of galaxies at a large distance is weighted 10 times more than a pair at a small distance, it still counts as only one pair, and its contribution to the overall noise is equivalent to that of the pair at the smaller distance.

The effective volume for covariances, $V_{\rm eff, cov}$, is defined as a volume that i) contains the same number of objects as the observed sample, $N$; ii) has a constant number density; and iii) provides the same number of pairs as in a random catalog with number density $n(z)$. 
If $n_0$ is the average number density in $V_{\rm eff, cov}$, then conditions i) and iii) can be expressed as follows:
\[ N = n_0 V_{\text{eff, cov}} = \int n(z) ~ dV, \hspace{0.25cm} \mathrm{and} \hspace{0.25cm} n_0^2 V_{\text{eff, cov}} = \int n^2(z) ~ dV. \]
By combining these relations, we can determine the effective volume for covariances:
\begin{equation}
V_{\rm eff, cov} = \frac{\left[\int n(z)dV\right]^2}{\int n^2(z)dV}.
\label{eq:Veffcov}
\end{equation}

When the effective volume of a lightcone, $V_\mathrm{eff}$, is smaller than the volume of the simulation, $V_\mathrm{sim}$, the resulting covariances remain unaffected by the simulation volume. In this scenario, no re-scaling of covariances is needed. However, if $V_\mathrm{eff}>V_\mathrm{sim}$, box replications are required to cover the desired volume.
In such cases, the covariance matrix should be scaled according to Equation~\ref{eq:cov_matrix}, with $\mathcal{C}$ defined as $\mathcal{C}=V_\mathrm{sim}/V_\mathrm{eff, cov}$.
This section aims to test this re-scaling procedure. To achieve this, we construct 500 lightcones using \textsc{GLAM-Uchuu} simulations with box sizes $0.5~h^{-1}\mathrm{Gpc}$ and $2000^3$ dark matter particles, denoted as \textsc{GLAM-Uchuu$_{0.5}$}. Both \textsc{GLAM-Uchuu} (with $1\hGpc$ boxes) and \textsc{GLAM-Uchuu}$_{0.5}$ simulations are conducted using the same cosmological parameters. 

For this test, we estimate the CMASS-N correlation functions for galaxies within the redshift range $0.49<z<0.59$, equivalent to an effective volume (for covariances) of $V_\mathrm{eff}=0.55\,h^{-3}$Gpc$^3$ ($V_\mathrm{eff, cov}=0.89~h^{-3}\mathrm{Gpc}^3$), as defined by Equation~\ref{eq:Veff}~(\ref{eq:Veffcov}).
This $V_\mathrm{eff}$ is smaller than $V_\mathrm{sim}=1~h^{-3}\mathrm{Gpc}^3$, so the covariances of \textsc{GLAM-Uchuu} do not need re-scaling. However, it is much larger than $V_\mathrm{sim}=0.5~h^{-3}\mathrm{Gpc}^3$ of \textsc{GLAM-Uchuu}$_{0.5}$ simulations, which require significant re-scaling. The required re-scaling factor for this scenario is calculated to be $\mathcal{C}_{0.5}=0.5^3/0.89\approx 0.14$.

In Fig.~\ref{fig:repTest} we present the diagonal components of the 2PCF monopole covariance matrix, $\sigma_{00}=\sqrt{\mathrm{Cov}_{00}}$, for the \textsc{GLAM-Uchuu} and \textsc{GLAM-Uchuu$_{0.5}$} lightcones. 
For the smaller box simulation, \textsc{GLAM-Uchuu}$_{0.5}$, the covariance without box size correction surpasses that estimated by \textsc{GLAM-Uchuu}. This increase is expected due to the reduced number of independent pairs available in smaller simulation volumes compared to the larger observational sample.
After applying the re-scaling factor, we find very good agreement between both simulations, with residuals within $\sim 5-10\%$. Importantly, there's no indication that the re-scaling depends on the distance $s$. Despite the covariances of the correlation function changing by almost a factor of 100 from the smallest to the largest scale, the results remain consistent within the $1\sigma$ statistical error.

For samples discussed in the paper, only CMASS-N requires re-scaling.
For this sample, we estimate that the effective volume for covariances is $V_{\rm eff, cov} = 1.68~h^{-3}\mathrm{Gpc}^3$. In this case, we correct the \textsc{GLAM-Uchuu} covariance matrix using factor $\mathcal{C}\approx 0.6$.


\bsp
\label{lastpage}
\end{document}

%% file: abstract.tex
This study investigates the clustering and bias of Luminous Red Galaxies (LRG) in the BOSS-LOWZ, -CMASS, -COMB, and eBOSS samples, using two types of simulated lightcones: (i) high-fidelity lightcones from \textsc{Uchuu} $N$-body simulation, employing SHAM technique to assign LRG to (sub)halos, and (ii) 16000 covariance lightcones from \textsc{GLAM-Uchuu} $N$-body simulations, including LRG using HOD data from \textsc{Uchuu}.
Our results indicate that \textsc{Uchuu} and \textsc{GLAM} lightcones closely replicate BOSS/eBOSS data, reproducing correlation function and power spectrum across scales from redshifts $0.2$ to $1.0$, from $2$ to $150\hMpc$ in configuration space, from $0.005$ to $0.7\hMpcInv$ in Fourier space, and across different LRG stellar masses. 
Furthermore, comparing with existing \textsc{MD-Patchy} and \textsc{EZmock} BOSS/eBOSS lightcones based on approximate methods, our \textsc{GLAM-Uchuu} lightcones provide more precise clustering estimates. We identify significant deviations from observations within $20\hMpc$ scales in \textsc{MD-Patchy} and \textsc{EZmock}, with our covariance matrices indicating that these methods underestimate errors by between $10\%$ and $60\%$.
Lastly, we explore the impact of cosmology on galaxy clustering. Our findings suggest that, given the current level of uncertainties in BOSS/eBOSS data, distinguishing models with and without massive neutrino effects on LSS is challenging.
This paper highlights the \textsc{Uchuu} and \textsc{GLAM-Uchuu} simulations' robustness in verifying the accuracy of Planck cosmological parameters, providing a strong foundation for enhancing lightcone construction in future LSS surveys. We also demonstrate that generating thousands of galaxy lightcones is feasible using $N$-body simulations with adequate mass and force resolution.

%% file: intro.tex
One of the main goals in cosmology is to ascertain the underlying physical principles governing the formation and evolution of the large-scale structure (LSS) in the Universe. 
In the current paradigm this structure originates from primordial density fluctuations seeded on classical scales during inflation, which subsequently grow through gravitational instability. The Baryon Acoustic Oscillations (BAO) are a unique component of these fluctuations, imprinted on the matter distribution later in cosmic history. They represent the distance travelled by sounds waves prior to matter-radiation decoupling \citep[see][for the first BAO detections using the SDSS and 2dFGRS galaxy redshift surveys]{Eisenstein2005,Cole2005}. 

The latest SDSS-III BOSS \citep[][]{BOSS_1} and SDSS-IV eBOSS \citep[][]{Dawson2016} surveys mapped the distribution of luminous red galaxies (LRG), quasars, and emission line galaxies to measure the characteristic scale imprinted by the BAO in their clustering signal, and to determine the redshift-distance relation over the redshift interval $z=0.2$ to $1$. These BAO measurements, together with redshift-space distortions (RSD), have allowed the constraint of some cosmological parameters with high precision \citep[see][]{Alam_2021}. In particular, for the LRG galaxy samples which are the subject of this investigation, the BOSS Data Release 12 \citep[DR12;][]{Alam2015} provides the redshifts of 1 million LRG over $10000$~deg$^2$ of the sky, covering the redshift interval from $0.2$ to $0.8$. Additionally, eBOSS Data Release 16 \citep[DR16;][]{Ahumada2020} provides a sample with around $180000$ LRG in the redshift interval between $z=0.6$ and $1$, over an area of $4000$~deg$^2$. These figures will be largely superseded by the ongoing DESI survey, which is expected to measure spectra for an order of magnitude more galaxies, thereby tightening the constraints on cosmological parameters \citep[see][]{DESI}.

In order to compare a given cosmological model, and in particular the predictions of the standard $\Lambda$CDM model with observational data, it is essential to generate high-fidelity galaxy lightcones from cosmological simulations \citep[e.g.][]{delatorre13,Smith2017, chian,Yu_2022}. 
These lightcones must be able to fully capture the properties of the observed galaxy sample, rangingfrom the angular footprint to clustering. 
Different methods are available to generate such lightcones.
Hydrodynamic simulations are arguably the most accurate way to model the formation and evolution of galaxies. However, these calculations are computationally infeasible for the volumes required in LSS studies (e.g. \citealt{Frenk_1999,Springel_hidro}, though see the recent large volume runs in the FLAMINGO suite of \citealt{Flamingo2023} and the MillenniumTNG run presented by \citealt{Hernandez-Aguayo:2022xcl,Pakmor:2022yyn}). A cheaper option is to use a semi-analytical model to predict the galactic content of dark matter (DM) halos in lightcones  \citep{Kitzbichler2007,Merson:2019,Doris:2019,Barrera:2022jgo}; however, it is still challenging to tune such models to reproduce clustering measurements as closely as is possible using empirical methods for populating halos with galaxies. 

Dark matter only $N$-body simulations are not only a computationally cheaper option, but also enable us to simulate large volumes. 
These simulations model the gravitational interactions of a system of $N$ collisionless particles over time and are able to follow the growth of LSS deep into the nonlinear regime.
To produce galaxy lightcones, the DM halos extracted from these simulations must be populated with galaxies. Empirical methods are the most popular ones, as they are simple to apply and their parameters can be rapidly chosen to reproduce the clustering and abundance of a target galaxy sample.
A common method is the halo occupation distribution, \citep[HOD;][]{PeacockSmith,wechsler:2000bf,BerlindWeinberg,Kravtsov}, which models the probability that a halo of mass $M_\mathrm{halo}$ hosts $N_\mathrm{gal}$ galaxies. These models have a number of free parameters that allow the observed clustering to be replicated with simulated lightcones.
Subhalo abundance matching \citep[SHAM;][]{Vale2006,Behroozi_2010,Trujillo,Reddick_2013,Guo-HAM_vs_HOD,Masaki2023}  
assumes a one-to-one relation between galaxy luminosity (or stellar mass) and a given halo property, such as halo mass or circular velocity, to which a scatter is applied (e.g. \citealt{Sergio}). Other extensions of the basic SHAM model have also been proposed and applied to build mock catalogues (e.g. \citealt{Contreras:2021, Contreras:2023, Contreras:2023b}). The SHAM model relies on the assumption that more luminous (massive) galaxies inhabit more massive halos.  
The galaxy lightcones generated using this method reproduce the observed luminosity (or stellar mass) function and, by construction, recover the observed galaxy clustering. 

Prior to this study, simulated galaxy lightcones have been generated for both the BOSS and eBOSS-LRG samples. In the context of BOSS, the most commonly used LRG lightcones are the \textsc{MultiDark-Patchy} ones \citep[\textsc{MD-Patchy};][]{Kitaura_2016}, based on the high-fidelity \textsc{BigMultiDark} lightcone \citep[\textsc{BigMD};][]{Sergio}. On the other hand, for the eBOSS-LRG, \textsc{EZmock} lightcones are employed \citep[][]{Zhao_2021}.
In the case of \textsc{BigMD} lightcones, the SHAM method was used to populate LRG within halos drawn from the \textsc{BigMD} Planck $N$-body simulation \citep{Klypin_MD}. Subsequent steps required for generating the lightcones, including those for \textsc{MD-patchy},  were carried out using the SUrvey GenerAtoR code \citep[\textsc{\textsc{SUGAR}}; as outlined in][]{Sergio}. 
Both the \textsc{Patchy} and \textsc{EZmock} methodologies, primarily designed for covariance matrices, make use of approximate models.
The \textsc{Patchy} code generates fields for dark matter density and peculiar velocity on a mesh, employing Gaussian fluctuations and implementing
the Augmented Lagrangian Perturbation Theory scheme \citep[ALPT, as described in][]{Patchy_code}. 
On the other hand, the \textsc{EZmock} approach adopts the Zel'dovich approximation \citep{Zeldovich} to create density fields at specific redshifts. Subsequently, galaxies are added to the density field in both the \textsc{Patchy} and \textsc{EZmock} methods using a model based on the bias of LRG. The parameters of this bias model are determined by fitting the LRG two-point and three-point correlation functions (2PCF, 3PCF). For the \textsc{Patchy} method, the 2PCF of the high-fidelity \textsc{BigMD}-BOSS lightcone is fitted, while for the \textsc{EZmock} method, the observed eBOSS 2PCF is employed for this fitting process. 

There are two primary reasons for generating new simulated lightcones for the BOSS and eBOSS surveys. Firstly, in this work we make use of the \textsc{Uchuu} $N$-body cosmological simulation to construct high-fidelity LRG lightcones for both surveys, adopting the SHAM method to populate (sub)halos with LRG.
\textsc{Uchuu} is a large simulation containing a 2.1 trillion particles, set within a $(2~h^{-1}\mathrm{Gpc})^3$ volume. This unprecedented numerical resolution within such a large volume enables the resolution of DM halos and subhalos down to the domain of dwarf galaxies \citep[see][for details]{Ishiyama21}. This surpasses the numerical resolution of \textsc{BigMD}, which was employed for constructing BOSS-LRG lightcones \citep{Sergio}.
Consequently, we anticipate achieving clustering signals at smaller scales, further enhancing the statistics of low-mass galaxies.
In addition, \textsc{Uchuu} adopts the Planck-2015 cosmology \citep[hereafter PL15;][]{planck15}, making it perfectly well-suited for assessing the clustering predictions of the standard $\Lambda$CDM model against data from BOSS and eBOSS. 

Secondly, alternative methods to $N$-body cosmological simulations, such as those adopted to generate the \textsc{MD-Patchy} and \textsc{EZmock} covariance matrices for BOSS and eBOSS, are more efficient but less accurate than $N$-body simulations, exhibiting considerably lower precision on scales below $20~h^{-1}\mathrm{Mpc}$, as highlighted by our investigation.
\textsc{MD-patchy} and EZmock are not based on a real $N$-body simulation, and they fail to generate a precise matter density field compared to a full $N$-body simulation. Consequently, it is unclear whether the resulting galaxy catalogues are feasible in a real universe, raising uncertainty about whether the resulting galaxy lightcones can produce accurate covariance error estimates for the BOSS and eBOSS clustering statistics. 
Moreover, a recent study by \citet{Yu_2023} reported some evidence of model specification errors in \textsc{MD-patchy}. Here, we generate BOSS and eBOSS lightcones for estimating covariance error by utilizing GLAM $N$-body cosmological simulations, that fully encode the nonlinear gravitational evolution \citep[see][for details]{GLAMsimu}. In contrast to \textsc{Uchuu}, the GLAM simulations can only be used to resolve distinct halos (not subhalos), prompting us to employ the HOD approach for populating GLAM halos with LRG. We obtain this HOD from the galaxy catalogues constructed using \textsc{Uchuu} for BOSS and eBOSS surveys. 
In addition, previous generation of covariance lightcones required the specification of a substantial number of parameters (five for \textsc{MD-patchy} and six for EZmock), resulting in a considerable degree of parameter degeneracy. In the \textsc{GLAM} lightcones we have only one free parameter: the scatter parameter introduced in the \textsc{SHAM} method applied in \textsc{Uchuu} that quantifies the scatter between the galaxy stellar mass and the proxy of halo mass.

This paper is organized as follows. In Sections~\ref{sec:galaxy_data}~and~\ref{sec:cosmo_simu}, we introduce the BOSS and eBOSS galaxy samples, along with the \textsc{Uchuu} and GLAM simulations used in this study. 
Section~\ref{sec:uchuu_light} outlines the methodology employed to generate the \textsc{Uchuu} lightcones for BOSS and eBOSS. 
The construction of the GLAM covariance lightcones is described in Section~\ref{sec:glam_light}.
Our results and their discussion, presented in Section~\ref{sec:results}, are divided into three parts: comparison of observations to our theoretical predictions based on the Planck cosmology as determined from the \textsc{Uchuu} lightcones (Section~\ref{sec:provingPL15}), a study of the performance of our GLAM lightcones compared to previous ones (Section~\ref{sec:glam_vs_other}), and an exploration of the effect that cosmology has on the distribution of galaxies using different GLAM runs (Section~\ref{sec:glam_cosmos}).

%% file: conclusions.tex
We have generated and analyzed simulated lightcones for luminous red galaxies in BOSS-LOWZ, -CMASS, -COMB and eBOSS samples, all based on the Planck cosmology. Our study encompasses two types of lightcones:
\begin{enumerate}
    \item[(a)] High-fidelity LRG lightcones derived from the high-resolution and large-volume \textsc{Uchuu} $N$-body simulation, which tracks the evolution of halos and subhalos. Galaxies are assigned to (sub)halos using the SHAM technique, which uses one free parameter, $\sigma_{\mathrm{SHAM}}$ (see Equation~\ref{eq:sigma_sham}), and the observed stellar mass function. We obtain the corresponding halo occupancy, determining the number of galaxies in each dark matter halo, for these lightcones. This HOD information is then used to populate distinct halos in lower-resolution \textsc{GLAM-Uchuu} simulations.
    \item[(b)] Covariance LRG lightcones based on two thousand \textsc{GLAM-Uchuu} simulations. Galaxies are assigned to distinct halos using the HOD method, with HOD parameters obtained from \textsc{Uchuu} lightcones.
\end{enumerate}

It is crucial to highlight that we do not modify our methods to make our results fit the observed clustering of BOSS and eBOSS galaxies. This is in contrast to approximate methods such as \textsc{MD-Patchy} and \textsc{EZmock}, which employ fitting techniques with numerous free parameters. 
Our \textsc{Uchuu} and \textsc{GLAM-Uchuu} LRG lightcones are generated without fitting any observed clustering data and without any free parameter: they rely solely on theoretical predictions.
The success of this approach depends on high-fidelity cosmological simulations, such as \textsc{Uchuu}. 
We also demonstrate that generating thousands of galaxy lightcones is feasible using $N$-body simulations with adequate mass and force resolution, such as those that we  have generated using \textsc{GLAM} and used here. This capability enables us to accurately estimate covariance errors, impact of systematics and identify potential tensions between model and data in future research endeavors.

Throughout this paper, we have studied various aspects of LRG clustering in the BOSS and eBOSS surveys and assess how well our theoretical predictions for Planck cosmology match the data. The main results, based on both our high-fidelity \textsc{Uchuu} and covariance \textsc{GLAM-Uchuu} LRG lightcones, are summarised as follows:
\begin{enumerate}
    \item Overall, our \textsc{Uchuu} lightcones accurately reproduces the correlation function and power spectrum measurements of the observed data for all the studied samples.
    We find certain discrepancies on large scales, primary caused by cosmic variance and the substantial impact of observational systematics. On smaller scales, below $\sim1\hMpc$, these deviations can be attributed to the non-application of PIP weights in our analysis. 
    In configuration space, our findings indicate that the residuals of the BOSS 2PCF monopole remain within $10\%$ ($15\%$ for eBOSS) up to $100\hMpc$ and fall below $2.5\%$ up to $25\hMpc$. 
    We also find a good agreement for the 3PCF, with our theoretical predictions determined from the mean of the \textsc{Uchuu} lightcones and the observational measurements within 1-$\sigma$ errors.
    Moving to Fourier space, we observe that the power spectrum monopole residuals between $k=0.03$ and $k=0.6\hMpcInv$ are below $12.5\%$ for eBOSS, and drop below $5\%$ and $4\%$ for BOSS-LOWZ and -CMASS, respectively.
    \item From the dependence of clustering on stellar mass, we conclude that the observed LRG population in both BOSS samples is complete at the massive end. 
    We also validate the accuracy of the method adopted in this work for modeling the overall incompleteness of the eBOSS galaxy population.  
    \item The scale-dependent galaxy bias of our \textsc{Uchuu} lightcones is in a good agreement with observations. Additionally, we observe an increase in the bias factor with redshift.
    \item We check whether our theoretical predictions reproduce the observed evolution of clustering with redshift by analyzing the 2PCF, the 3PCF, and the power spectrum of the COMB sample in three redshift bins. In general, we find good agreement between the \textsc{Uchuu} (and \textsc{GLAM-Uchuu}) lightcones and the observed data. We report 2PCF monopole residuals that remain below $1.5\%$ within $25\hMpc$, and under $12\%$ within $100\hMpc$, and within $4.5\%$ for the monopole power spectrum between $k=0.03$ and $k=0.6$. 
    For the 3PCF, we again find good agreement (discrepancies within 1-$sigma$ errors). Any potential discrepancies can be attributed to the same factors mentioned earlier. 
    \item Strikingly, the \textsc{MD-Patchy} and \textsc{EZmock} LRG lightcones created for BOSS and eBOSS, respectively, display peculiar features in their clustering multipoles that deviate from observations within $20\hMpc$ and for $\mathrm{k}>0.4\hMpcInv$. In contrast, our $N$-body simulation-based lightcones agree with the data even on scales as small $0.1\hMpc$. 
    \item When comparing the BOSS-CMASS covariance matrices, we find that \textsc{GLAM-Uchuu} and \textsc{MD-Patchy} estimates agree, in general, within $15\%$. On the other hand, \textsc{EZmock} estimates for eBOSS deviate up to $60\%$ from \textsc{GLAM-Uchuu}. However, we note that discrepancies of up to $10\%$ are to be expected, with those exceeding this value being attributed to the distinct predictions of each simulation (approximate methods versus $N$-body simulations). Regarding the correlation matrices, we notice that \textsc{GLAM-Uchuu} presents lower correlation compared to \textsc{MD-Patchy}, but similar to \textsc{EZmock}.
    \item Finally, we explore the impact of cosmology on galaxy clustering using various \textsc{GLAM} simulations. Our analysis reveals that it is not feasible to discern between a universe with or without massive neutrinos within the current BOSS/eBOSS uncertainties. 
\end{enumerate} 

In conclusion, our theoretical predictions, based on the Planck cosmology and derived from the high-fidelity \textsc{Uchuu} lightcones, confirm the accuracy of Planck cosmological parameters in explaining observations from the LSS BOSS/eBOSS surveys, and demonstrate the robustness of the \textsc{Uchuu} simulation. These findings have significant implications for refining lightcone construction methodologies and advancing our understanding of clustering measurements. Moreover, we demonstrate that both \textsc{MD-Patchy} and \textsc{EZmock} LRG lightcones systematically underestimate covariance errors by $\approx10-60\%$ compared to \textsc{GLAM-Uchuu}. This may have important implications for cosmological parameter inferences derived from BOSS and eBOSS, underscoring the critical importance of utilising high-fidelity simulations to enhance the reliability and accuracy of cosmological analyses.

A similar study, as presented here, can be expanded by using the Modify Gravity version of GLAM \citep[MG-GLAM; see][]{HA2022,Ruan2022} for analyzing BOSS/eBOSS clustering, potentially revealing novel insights into the nature of gravity. In the near future, we plan to generate \textsc{Uchuu} LRG lightcones for the first year of DESI data. We will then construct their covariance matrices within the \textsc{GLAM-Uchuu} framework, as outlined in this paper. This effort will enhance the precision of the cosmological analyses conducted using the DESI data.

%% file: thanks.tex
JE, FP and AK acknowledge financial support from the Spanish MICINN funding grant PGC2018-101931-B-I00. TI has been supported by IAAR Research Support Program in Chiba University Japan, MEXT/JSPS KAKENHI (Grant Number JP19KK0344 and JP21H01122), MEXT as ``Program for Promoting Researches on the Supercomputer Fugaku'' (JPMXP1020200109 and JPMXP1020230406), and JICFuS. AS, BL and CMB acknowledge support from the Science Technology Facilities Council through ST/X001075/1. CHA acknowledges support from the Excellence Cluster ORIGINS which is funded by the Deutsche Forschungsgemeinschaft (DFG, German Research Foundation) under Germany's Excellence Strategy -- EXC-2094 -- 390783311. JE thanks Oliver Philcox for his assistance in the calculation of the 3PCF.

The Uchuu simulation was carried out on the Aterui II supercomputer at CfCA-NAOJ. We thank IAA-CSIC, CESGA, and RedIRIS in Spain for hosting the Uchuu data releases in the \textsc{Skies \& Universes} site for cosmological simulations. The analysis done in this paper have made use of NERSC at LBNL and \textit{skun6}@IAA-CSIC computer facility managed by IAA-CSIC in Spain (MICINN EU-Feder grant EQC2018-004366-P).

This work used the DiRAC@Durham facility managed by the Institute for Computational Cosmology on behalf of the STFC DiRAC HPC Facility (www.dirac.ac.uk). The equipment was funded by BEIS capital funding via STFC capital grants ST/K00042X/1, ST/P002293/1, ST/R002371/1 and ST/S002502/1, Durham University and STFC operations grant ST/R000832/1. DiRAC is part of the National e-Infrastructure.

%% file: main.bbl
\begin{thebibliography}{}
\makeatletter
\relax
\def\mn@urlcharsother{\let\do\@makeother \do\$\do\&\do\#\do\^\do\_\do\%\do\~}
\def\mn@doi{\begingroup\mn@urlcharsother \@ifnextchar [ {\mn@doi@} {\mn@doi@[]}}
\def\mn@doi@[#1]#2{\def\@tempa{#1}\ifx\@tempa\@empty \href {http://dx.doi.org/#2} {doi:#2}\else \href {http://dx.doi.org/#2} {#1}\fi \endgroup}
\def\mn@eprint#1#2{\mn@eprint@#1:#2::\@nil}
\def\mn@eprint@arXiv#1{\href {http://arxiv.org/abs/#1} {{\tt arXiv:#1}}}
\def\mn@eprint@dblp#1{\href {http://dblp.uni-trier.de/rec/bibtex/#1.xml} {dblp:#1}}
\def\mn@eprint@#1:#2:#3:#4\@nil{\def\@tempa {#1}\def\@tempb {#2}\def\@tempc {#3}\ifx \@tempc \@empty \let \@tempc \@tempb \let \@tempb \@tempa \fi \ifx \@tempb \@empty \def\@tempb {arXiv}\fi \@ifundefined {mn@eprint@\@tempb}{\@tempb:\@tempc}{\expandafter \expandafter \csname mn@eprint@\@tempb\endcsname \expandafter{\@tempc}}}

\bibitem[\protect\citeauthoryear{{Ahumada} et~al.,}{{Ahumada} et~al.}{2020}]{Ahumada2020}
{Ahumada} R.,  et~al., 2020, \mn@doi [\apjs] {10.3847/1538-4365/ab929e}, \href {https://ui.adsabs.harvard.edu/abs/2020ApJS..249....3A} {249, 3}

\bibitem[\protect\citeauthoryear{{Alam} et~al.,}{{Alam} et~al.}{2015}]{Alam2015}
{Alam} S.,  et~al., 2015, \mn@doi [\apjs] {10.1088/0067-0049/219/1/12}, \href {https://ui.adsabs.harvard.edu/abs/2015ApJS..219...12A} {219, 12}

\bibitem[\protect\citeauthoryear{{Alam} et~al.,}{{Alam} et~al.}{2017}]{Alam_clust}
{Alam} S.,  et~al., 2017, \mn@doi [\mnras] {10.1093/mnras/stx721}, \href {https://ui.adsabs.harvard.edu/abs/2017MNRAS.470.2617A} {470, 2617}

\bibitem[\protect\citeauthoryear{{Alam}, {Peacock}, {Kraljic}, {Ross}  \& {Comparat}}{{Alam} et~al.}{2020}]{Alam_2020}
{Alam} S.,  {Peacock} J.~A.,  {Kraljic} K.,  {Ross} A.~J.,   {Comparat} J.,  2020, \mn@doi [\mnras] {10.1093/mnras/staa1956}, \href {https://ui.adsabs.harvard.edu/abs/2020MNRAS.497..581A} {497, 581}

\bibitem[\protect\citeauthoryear{{Alam} et~al.,}{{Alam} et~al.}{2021}]{Alam_2021}
{Alam} S.,  et~al., 2021, \mn@doi [\prd] {10.1103/PhysRevD.103.083533}, \href {https://ui.adsabs.harvard.edu/abs/2021PhRvD.103h3533A} {103, 083533}

\bibitem[\protect\citeauthoryear{{Aung} et~al.,}{{Aung} et~al.}{2023}]{Aung2023}
{Aung} H.,  et~al., 2023, \mn@doi [\mnras] {10.1093/mnras/stac3514}, \href {https://ui.adsabs.harvard.edu/abs/2023MNRAS.519.1648A} {519, 1648}

\bibitem[\protect\citeauthoryear{Barrera et~al.,}{Barrera et~al.}{2022}]{Barrera:2022jgo}
Barrera M.,  et~al., 2022, The MillenniumTNG Project: Semi-analytic galaxy formation models on the past lightcone (\mn@eprint {arXiv} {2210.10419})

\bibitem[\protect\citeauthoryear{{Baugh} et~al.,}{{Baugh} et~al.}{2019}]{Pmill_cosmo}
{Baugh} C.~M.,  et~al., 2019, \mn@doi [\mnras] {10.1093/mnras/sty3427}, \href {https://ui.adsabs.harvard.edu/abs/2019MNRAS.483.4922B} {483, 4922}

\bibitem[\protect\citeauthoryear{{Behroozi}, {Conroy}  \& {Wechsler}}{{Behroozi} et~al.}{2010}]{Behroozi_2010}
{Behroozi} P.~S.,  {Conroy} C.,   {Wechsler} R.~H.,  2010, \mn@doi [\apj] {10.1088/0004-637X/717/1/379}, \href {https://ui.adsabs.harvard.edu/abs/2010ApJ...717..379B} {717, 379}

\bibitem[\protect\citeauthoryear{{Behroozi}, {Wechsler}  \& {Wu}}{{Behroozi} et~al.}{2013a}]{Behroozi13}
{Behroozi} P.~S.,  {Wechsler} R.~H.,   {Wu} H.-Y.,  2013a, \mn@doi [\apj] {10.1088/0004-637X/762/2/109}, \href {https://ui.adsabs.harvard.edu/abs/2013ApJ...762..109B} {762, 109}

\bibitem[\protect\citeauthoryear{{Behroozi}, {Wechsler}, {Wu}, {Busha}, {Klypin}  \& {Primack}}{{Behroozi} et~al.}{2013b}]{Behroozi2013b}
{Behroozi} P.~S.,  {Wechsler} R.~H.,  {Wu} H.-Y.,  {Busha} M.~T.,  {Klypin} A.~A.,   {Primack} J.~R.,  2013b, \mn@doi [\apj] {10.1088/0004-637X/763/1/18}, \href {http://adsabs.harvard.edu/abs/2013ApJ...763...18B} {763, 18}

\bibitem[\protect\citeauthoryear{{Berlind} \& {Weinberg}}{{Berlind} \& {Weinberg}}{2002}]{BerlindWeinberg}
{Berlind} A.~A.,  {Weinberg} D.~H.,  2002, \mn@doi [\apj] {10.1086/341469}, \href {https://ui.adsabs.harvard.edu/abs/2002ApJ...575..587B} {575, 587}

\bibitem[\protect\citeauthoryear{{Beutler} et~al.,}{{Beutler} et~al.}{2017}]{Beutler_2016}
{Beutler} F.,  et~al., 2017, \mn@doi [\mnras] {10.1093/mnras/stw2373}, \href {https://ui.adsabs.harvard.edu/abs/2017MNRAS.464.3409B} {464, 3409}

\bibitem[\protect\citeauthoryear{{Bianchi} \& {Percival}}{{Bianchi} \& {Percival}}{2017}]{Bianchi_2017}
{Bianchi} D.,  {Percival} W.~J.,  2017, \mn@doi [\mnras] {10.1093/mnras/stx2053}, \href {https://ui.adsabs.harvard.edu/abs/2017MNRAS.472.1106B} {472, 1106}

\bibitem[\protect\citeauthoryear{{Chaves-Montero}, {Angulo}, {Schaye}, {Schaller}, {Crain}, {Furlong}  \& {Theuns}}{{Chaves-Montero} et~al.}{2016}]{Jonas}
{Chaves-Montero} J.,  {Angulo} R.~E.,  {Schaye} J.,  {Schaller} M.,  {Crain} R.~A.,  {Furlong} M.,   {Theuns} T.,  2016, \mn@doi [\mnras] {10.1093/mnras/stw1225}, \href {https://ui.adsabs.harvard.edu/abs/2016MNRAS.460.3100C} {460, 3100}

\bibitem[\protect\citeauthoryear{{Chen}, {Vlah}  \& {White}}{{Chen} et~al.}{2022}]{Chen_2022}
{Chen} S.-F.,  {Vlah} Z.,   {White} M.,  2022, \mn@doi [\jcap] {10.1088/1475-7516/2022/02/008}, \href {https://ui.adsabs.harvard.edu/abs/2022JCAP...02..008C} {2022, 008}

\bibitem[\protect\citeauthoryear{{Chuang} et~al.,}{{Chuang} et~al.}{2017}]{Chuang_2017}
{Chuang} C.-H.,  et~al., 2017, \mn@doi [\mnras] {10.1093/mnras/stx1641}, \href {https://ui.adsabs.harvard.edu/abs/2017MNRAS.471.2370C} {471, 2370}

\bibitem[\protect\citeauthoryear{{Cole} et~al.,}{{Cole} et~al.}{2005}]{Cole2005}
{Cole} S.,  et~al., 2005, \mn@doi [\mnras] {10.1111/j.1365-2966.2005.09318.x}, \href {https://ui.adsabs.harvard.edu/abs/2005MNRAS.362..505C} {362, 505}

\bibitem[\protect\citeauthoryear{Comparat et~al.,}{Comparat et~al.}{2017}]{comparat_maraston}
Comparat J.,  et~al., 2017, Stellar population properties for 2 million galaxies from SDSS DR14 and DEEP2 DR4 from full spectral fitting, \mn@doi{10.48550/ARXIV.1711.06575}, \url {https://arxiv.org/abs/1711.06575}

\bibitem[\protect\citeauthoryear{{Contreras}, {Angulo}  \& {Zennaro}}{{Contreras} et~al.}{2021}]{Contreras:2021}
{Contreras} S.,  {Angulo} R.~E.,   {Zennaro} M.,  2021, \mn@doi [\mnras] {10.1093/mnras/stab2560}, \href {https://ui.adsabs.harvard.edu/abs/2021MNRAS.508..175C} {508, 175}

\bibitem[\protect\citeauthoryear{{Contreras}, {Angulo}, {Chaves-Montero}, {White}  \& {Aric{\`o}}}{{Contreras} et~al.}{2023a}]{Contreras:2023}
{Contreras} S.,  {Angulo} R.~E.,  {Chaves-Montero} J.,  {White} S. D.~M.,   {Aric{\`o}} G.,  2023a, \mn@doi [\mnras] {10.1093/mnras/stad122}, \href {https://ui.adsabs.harvard.edu/abs/2023MNRAS.520..489C} {520, 489}

\bibitem[\protect\citeauthoryear{{Contreras} et~al.,}{{Contreras} et~al.}{2023b}]{Contreras:2023b}
{Contreras} S.,  et~al., 2023b, \mn@doi [\mnras] {10.1093/mnras/stac3699}, \href {https://ui.adsabs.harvard.edu/abs/2023MNRAS.524.2489C} {524, 2489}

\bibitem[\protect\citeauthoryear{{DESI Collaboration} et~al.,}{{DESI Collaboration} et~al.}{2022}]{DESI}
{DESI Collaboration} et~al., 2022, \mn@doi [\aj] {10.3847/1538-3881/ac882b}, \href {https://ui.adsabs.harvard.edu/abs/2022AJ....164..207D} {164, 207}

\bibitem[\protect\citeauthoryear{{Davis} \& {Peebles}}{{Davis} \& {Peebles}}{1983}]{Peebles}
{Davis} M.,  {Peebles} P.~J.~E.,  1983, \mn@doi [\apj] {10.1086/160884}, \href {https://ui.adsabs.harvard.edu/abs/1983ApJ...267..465D} {267, 465}

\bibitem[\protect\citeauthoryear{{Dawson} et~al.,}{{Dawson} et~al.}{2013}]{BOSS_1}
{Dawson} K.~S.,  et~al., 2013, \mn@doi [\aj] {10.1088/0004-6256/145/1/10}, \href {https://ui.adsabs.harvard.edu/abs/2013AJ....145...10D} {145, 10}

\bibitem[\protect\citeauthoryear{{Dawson} et~al.,}{{Dawson} et~al.}{2016}]{Dawson2016}
{Dawson} K.~S.,  et~al., 2016, \mn@doi [\aj] {10.3847/0004-6256/151/2/44}, \href {https://ui.adsabs.harvard.edu/abs/2016AJ....151...44D} {151, 44}

\bibitem[\protect\citeauthoryear{{Dey} et~al.,}{{Dey} et~al.}{2019}]{Dey_2019}
{Dey} A.,  et~al., 2019, \mn@doi [\aj] {10.3847/1538-3881/ab089d}, \href {https://ui.adsabs.harvard.edu/abs/2019AJ....157..168D} {157, 168}

\bibitem[\protect\citeauthoryear{Dong-Páez et~al.,}{Dong-Páez et~al.}{2022}]{chian}
Dong-Páez C.~A.,  et~al., 2022, The Uchuu-SDSS galaxy lightcones: a clustering, RSD and BAO study (\mn@eprint {arXiv} {2208.00540})

\bibitem[\protect\citeauthoryear{{Eisenstein} et~al.,}{{Eisenstein} et~al.}{2005}]{Eisenstein2005}
{Eisenstein} D.~J.,  et~al., 2005, \mn@doi [\apj] {10.1086/466512}, \href {https://ui.adsabs.harvard.edu/abs/2005ApJ...633..560E} {633, 560}

\bibitem[\protect\citeauthoryear{{Feldman}, {Kaiser}  \& {Peacock}}{{Feldman} et~al.}{1994}]{Feldman94}
{Feldman} H.~A.,  {Kaiser} N.,   {Peacock} J.~A.,  1994, \mn@doi [\apj] {10.1086/174036}, \href {https://ui.adsabs.harvard.edu/abs/1994ApJ...426...23F} {426, 23}

\bibitem[\protect\citeauthoryear{{Frenk} et~al.,}{{Frenk} et~al.}{1999}]{Frenk_1999}
{Frenk} C.~S.,  et~al., 1999, \mn@doi [\apj] {10.1086/307908}, \href {https://ui.adsabs.harvard.edu/abs/1999ApJ...525..554F} {525, 554}

\bibitem[\protect\citeauthoryear{{Guo}, {Zehavi}  \& {Zheng}}{{Guo} et~al.}{2012}]{fibre_coll}
{Guo} H.,  {Zehavi} I.,   {Zheng} Z.,  2012, \mn@doi [\apj] {10.1088/0004-637X/756/2/127}, \href {https://ui.adsabs.harvard.edu/abs/2012ApJ...756..127G} {756, 127}

\bibitem[\protect\citeauthoryear{{Guo} et~al.,}{{Guo} et~al.}{2016}]{Guo-HAM_vs_HOD}
{Guo} H.,  et~al., 2016, \mn@doi [\mnras] {10.1093/mnras/stw845}, \href {https://ui.adsabs.harvard.edu/abs/2016MNRAS.459.3040G} {459, 3040}

\bibitem[\protect\citeauthoryear{{Hamilton}}{{Hamilton}}{1998}]{Hamilton1998}
{Hamilton} A.~J.~S.,  1998, in {Hamilton} D.,  ed.,  Astrophysics and Space Science Library Vol. 231, The Evolving Universe. p.~185 (\mn@eprint {arXiv} {astro-ph/9708102}), \mn@doi{10.1007/978-94-011-4960-0_17}

\bibitem[\protect\citeauthoryear{Hand, Li, Slepian  \& Seljak}{Hand et~al.}{2017}]{Hand2017}
Hand N.,  Li Y.,  Slepian Z.,   Seljak U.,  2017, \mn@doi [J. Cosmol. Astropart. Phys.] {10.1088/1475-7516/2017/07/002}, 2017, 002

\bibitem[\protect\citeauthoryear{{Hayashi}, {Navarro}, {Taylor}, {Stadel}  \& {Quinn}}{{Hayashi} et~al.}{2003}]{Hayashi_2003}
{Hayashi} E.,  {Navarro} J.~F.,  {Taylor} J.~E.,  {Stadel} J.,   {Quinn} T.,  2003, \mn@doi [\apj] {10.1086/345788}, \href {https://ui.adsabs.harvard.edu/abs/2003ApJ...584..541H} {584, 541}

\bibitem[\protect\citeauthoryear{Hern{\'{a}}ndez-Aguayo, Prada, Baugh  \& Klypin}{Hern{\'{a}}ndez-Aguayo et~al.}{2021}]{CesarGLAM}
Hern{\'{a}}ndez-Aguayo C.,  Prada F.,  Baugh C.~M.,   Klypin A.,  2021, \mn@doi [Monthly Notices of the Royal Astronomical Society] {10.1093/mnras/stab434}, 503, 2318

\bibitem[\protect\citeauthoryear{{Hern{\'a}ndez-Aguayo}, {Ruan}, {Li}, {Arnold}, {Baugh}, {Klypin}  \& {Prada}}{{Hern{\'a}ndez-Aguayo} et~al.}{2022}]{HA2022}
{Hern{\'a}ndez-Aguayo} C.,  {Ruan} C.-Z.,  {Li} B.,  {Arnold} C.,  {Baugh} C.~M.,  {Klypin} A.,   {Prada} F.,  2022, \mn@doi [\jcap] {10.1088/1475-7516/2022/01/048}, \href {https://ui.adsabs.harvard.edu/abs/2022JCAP...01..048H} {2022, 048}

\bibitem[\protect\citeauthoryear{Hern\'andez-Aguayo et~al.}{Hern\'andez-Aguayo et~al.}{2023}]{Hernandez-Aguayo:2022xcl}
Hern\'andez-Aguayo C.,  et~al., 2023, \mn@doi [Mon. Not. Roy. Astron. Soc.] {10.1093/mnras/stad1657}, 524, 2556

\bibitem[\protect\citeauthoryear{Huterer, Cunha  \& Fang}{Huterer et~al.}{2013}]{Huterer13}
Huterer D.,  Cunha C.~E.,   Fang W.,  2013, \mn@doi [Monthly Notices of the Royal Astronomical Society] {10.1093/mnras/stt653}, 432, 2945

\bibitem[\protect\citeauthoryear{{Ishiyama}, {Fukushige}  \& {Makino}}{{Ishiyama} et~al.}{2009}]{Ishiyama09}
{Ishiyama} T.,  {Fukushige} T.,   {Makino} J.,  2009, \mn@doi [\pasj] {10.1093/pasj/61.6.1319}, \href {https://ui.adsabs.harvard.edu/abs/2009PASJ...61.1319I} {61, 1319}

\bibitem[\protect\citeauthoryear{{Ishiyama}, {Nitadori}  \& {Makino}}{{Ishiyama} et~al.}{2012}]{Ishiyama12}
{Ishiyama} T.,  {Nitadori} K.,   {Makino} J.,  2012, arXiv e-prints, \href {https://ui.adsabs.harvard.edu/abs/2012arXiv1211.4406I} {p. arXiv:1211.4406}

\bibitem[\protect\citeauthoryear{{Ishiyama} et~al.,}{{Ishiyama} et~al.}{2021}]{Ishiyama21}
{Ishiyama} T.,  et~al., 2021, \mn@doi [\mnras] {10.1093/mnras/stab1755}, \href {https://ui.adsabs.harvard.edu/abs/2021MNRAS.506.4210I} {506, 4210}

\bibitem[\protect\citeauthoryear{{Kaiser}}{{Kaiser}}{1987}]{Kaiser1987}
{Kaiser} N.,  1987, \mn@doi [\mnras] {10.1093/mnras/227.1.1}, \href {https://ui.adsabs.harvard.edu/abs/1987MNRAS.227....1K} {227, 1}

\bibitem[\protect\citeauthoryear{{Kitaura}, {Yepes}  \& {Prada}}{{Kitaura} et~al.}{2014}]{Patchy_code}
{Kitaura} F.~S.,  {Yepes} G.,   {Prada} F.,  2014, \mn@doi [\mnras] {10.1093/mnrasl/slt172}, \href {https://ui.adsabs.harvard.edu/abs/2014MNRAS.439L..21K} {439, L21}

\bibitem[\protect\citeauthoryear{{Kitaura} et~al.,}{{Kitaura} et~al.}{2016}]{Kitaura_2016}
{Kitaura} F.-S.,  et~al., 2016, \mn@doi [\mnras] {10.1093/mnras/stv2826}, \href {https://ui.adsabs.harvard.edu/abs/2016MNRAS.456.4156K} {456, 4156}

\bibitem[\protect\citeauthoryear{{Kitzbichler} \& {White}}{{Kitzbichler} \& {White}}{2007}]{Kitzbichler2007}
{Kitzbichler} M.~G.,  {White} S.~D.~M.,  2007, \mn@doi [\mnras] {10.1111/j.1365-2966.2007.11458.x}, \href {https://ui.adsabs.harvard.edu/abs/2007MNRAS.376....2K} {376, 2}

\bibitem[\protect\citeauthoryear{{Klypin} \& {Prada}}{{Klypin} \& {Prada}}{2018}]{GLAMsimu}
{Klypin} A.,  {Prada} F.,  2018, \mn@doi [\mnras] {10.1093/mnras/sty1340}, \href {https://ui.adsabs.harvard.edu/abs/2018MNRAS.478.4602K} {478, 4602}

\bibitem[\protect\citeauthoryear{{Klypin} \& {Prada}}{{Klypin} \& {Prada}}{2019}]{KlypinPrada2019}
{Klypin} A.,  {Prada} F.,  2019, \mn@doi [\mnras] {10.1093/mnras/stz2194}, \href {https://ui.adsabs.harvard.edu/abs/2019MNRAS.489.1684K} {489, 1684}

\bibitem[\protect\citeauthoryear{Klypin, Yepes, Gottlöber, Prada  \& Heß}{Klypin et~al.}{2016}]{Klypin_MD}
Klypin A.,  Yepes G.,  Gottlöber S.,  Prada F.,   Heß S.,  2016, \mn@doi [Monthly Notices of the Royal Astronomical Society] {10.1093/mnras/stw248}, 457, 4340

\bibitem[\protect\citeauthoryear{{Kobayashi}, {Nishimichi}, {Takada}  \& {Miyatake}}{{Kobayashi} et~al.}{2022}]{Kobayashi_2022}
{Kobayashi} Y.,  {Nishimichi} T.,  {Takada} M.,   {Miyatake} H.,  2022, \mn@doi [\prd] {10.1103/PhysRevD.105.083517}, \href {https://ui.adsabs.harvard.edu/abs/2022PhRvD.105h3517K} {105, 083517}

\bibitem[\protect\citeauthoryear{{Kravtsov}, {Berlind}, {Wechsler}, {Klypin}, {Gottl{\"o}ber}, {Allgood}  \& {Primack}}{{Kravtsov} et~al.}{2004}]{Kravtsov}
{Kravtsov} A.~V.,  {Berlind} A.~A.,  {Wechsler} R.~H.,  {Klypin} A.~A.,  {Gottl{\"o}ber} S.,  {Allgood} B.,   {Primack} J.~R.,  2004, \mn@doi [\apj] {10.1086/420959}, \href {https://ui.adsabs.harvard.edu/abs/2004ApJ...609...35K} {609, 35}

\bibitem[\protect\citeauthoryear{{Landy} \& {Szalay}}{{Landy} \& {Szalay}}{1993}]{Landy93}
{Landy} S.~D.,  {Szalay} A.~S.,  1993, \mn@doi [\apj] {10.1086/172900}, \href {https://ui.adsabs.harvard.edu/abs/1993ApJ...412...64L} {412, 64}

\bibitem[\protect\citeauthoryear{{Li}, {Hu}  \& {Takada}}{{Li} et~al.}{2014}]{Li2014}
{Li} Y.,  {Hu} W.,   {Takada} M.,  2014, \mn@doi [\prd] {10.1103/PhysRevD.89.083519}, \href {https://ui.adsabs.harvard.edu/abs/2014PhRvD..89h3519L} {89, 083519}

\bibitem[\protect\citeauthoryear{Maksimova, Garrison, Eisenstein, Hadzhiyska, Bose  \& Satterthwaite}{Maksimova et~al.}{2021}]{Abacus}
Maksimova N.~A.,  Garrison L.~H.,  Eisenstein D.~J.,  Hadzhiyska B.,  Bose S.,   Satterthwaite T.~P.,  2021, \mn@doi [Monthly Notices of the Royal Astronomical Society] {10.1093/mnras/stab2484}, 508, 4017

\bibitem[\protect\citeauthoryear{{Maraston} et~al.,}{{Maraston} et~al.}{2013}]{Maraston_2013}
{Maraston} C.,  et~al., 2013, \mn@doi [\mnras] {10.1093/mnras/stt1424}, \href {https://ui.adsabs.harvard.edu/abs/2013MNRAS.435.2764M} {435, 2764}

\bibitem[\protect\citeauthoryear{{Masaki}, {Kashino}, {Ishikawa}  \& {Lin}}{{Masaki} et~al.}{2023}]{Masaki2023}
{Masaki} S.,  {Kashino} D.,  {Ishikawa} S.,   {Lin} Y.-T.,  2023, \mn@doi [\mnras] {10.1093/mnras/stad1808}, \href {https://ui.adsabs.harvard.edu/abs/2023MNRAS.523.5280M} {523, 5280}

\bibitem[\protect\citeauthoryear{{Merson}, {Smith}, {Benson}, {Wang}  \& {Baugh}}{{Merson} et~al.}{2019}]{Merson:2019}
{Merson} A.,  {Smith} A.,  {Benson} A.,  {Wang} Y.,   {Baugh} C.,  2019, \mn@doi [\mnras] {10.1093/mnras/stz1204}, \href {https://ui.adsabs.harvard.edu/abs/2019MNRAS.486.5737M} {486, 5737}

\bibitem[\protect\citeauthoryear{{Merz} et~al.,}{{Merz} et~al.}{2021}]{Merz_2021}
{Merz} G.,  et~al., 2021, \mn@doi [\mnras] {10.1093/mnras/stab1887}, \href {https://ui.adsabs.harvard.edu/abs/2021MNRAS.506.2503M} {506, 2503}

\bibitem[\protect\citeauthoryear{{Mohammad} et~al.,}{{Mohammad} et~al.}{2020}]{Mohammad_2020}
{Mohammad} F.~G.,  et~al., 2020, \mn@doi [\mnras] {10.1093/mnras/staa2344}, \href {https://ui.adsabs.harvard.edu/abs/2020MNRAS.498..128M} {498, 128}

\bibitem[\protect\citeauthoryear{{Mohammed}, {Seljak}  \& {Vlah}}{{Mohammed} et~al.}{2017}]{Mohammed2017}
{Mohammed} I.,  {Seljak} U.,   {Vlah} Z.,  2017, \mn@doi [\mnras] {10.1093/mnras/stw3196}, \href {https://ui.adsabs.harvard.edu/abs/2017MNRAS.466..780M} {466, 780}

\bibitem[\protect\citeauthoryear{{Moustakas} et~al.,}{{Moustakas} et~al.}{2013}]{Moustakas_Primus}
{Moustakas} J.,  et~al., 2013, \mn@doi [\apj] {10.1088/0004-637X/767/1/50}, \href {https://ui.adsabs.harvard.edu/abs/2013ApJ...767...50M} {767, 50}

\bibitem[\protect\citeauthoryear{{Mueller} et~al.,}{{Mueller} et~al.}{2022}]{Mueller_2022}
{Mueller} E.-M.,  et~al., 2022, \mn@doi [\mnras] {10.1093/mnras/stac812}, \href {https://ui.adsabs.harvard.edu/abs/2022MNRAS.514.3396M} {514, 3396}

\bibitem[\protect\citeauthoryear{{Norberg}, {Baugh}, {Gazta{\~n}aga}  \& {Croton}}{{Norberg} et~al.}{2009}]{norberg2009}
{Norberg} P.,  {Baugh} C.~M.,  {Gazta{\~n}aga} E.,   {Croton} D.~J.,  2009, \mn@doi [\mnras] {10.1111/j.1365-2966.2009.14389.x}, \href {https://ui.adsabs.harvard.edu/abs/2009MNRAS.396...19N} {396, 19}

\bibitem[\protect\citeauthoryear{Nuza et~al.,}{Nuza et~al.}{2013}]{nuza_2013}
Nuza S.~E.,  et~al., 2013, \mn@doi [Monthly Notices of the Royal Astronomical Society] {10.1093/mnras/stt513}, 432, 743

\bibitem[\protect\citeauthoryear{{Oogi} et~al.,}{{Oogi} et~al.}{2022}]{Oogi2023}
{Oogi} T.,  et~al., 2022, \mn@doi [arXiv e-prints] {10.48550/arXiv.2207.14689}, \href {https://ui.adsabs.harvard.edu/abs/2022arXiv220714689O} {p. arXiv:2207.14689}

\bibitem[\protect\citeauthoryear{Pakmor et~al.}{Pakmor et~al.}{2023}]{Pakmor:2022yyn}
Pakmor R.,  et~al., 2023, \mn@doi [Mon. Not. Roy. Astron. Soc.] {10.1093/mnras/stac3620}, 524, 2539

\bibitem[\protect\citeauthoryear{{Peacock} \& {Smith}}{{Peacock} \& {Smith}}{2000}]{PeacockSmith}
{Peacock} J.~A.,  {Smith} R.~E.,  2000, \mn@doi [\mnras] {10.1046/j.1365-8711.2000.03779.x}, \href {https://ui.adsabs.harvard.edu/abs/2000MNRAS.318.1144P} {318, 1144}

\bibitem[\protect\citeauthoryear{{Peacock} et~al.,}{{Peacock} et~al.}{2001}]{Peacock_2001}
{Peacock} J.~A.,  et~al., 2001, \mn@doi [\nat] {10.1038/35065528}, \href {https://ui.adsabs.harvard.edu/abs/2001Natur.410..169P} {410, 169}

\bibitem[\protect\citeauthoryear{{Percival}, {Verde}  \& {Peacock}}{{Percival} et~al.}{2004}]{Percival_weights}
{Percival} W.~J.,  {Verde} L.,   {Peacock} J.~A.,  2004, \mn@doi [\mnras] {10.1111/j.1365-2966.2004.07245.x}, \href {https://ui.adsabs.harvard.edu/abs/2004MNRAS.347..645P} {347, 645}

\bibitem[\protect\citeauthoryear{{Planck Collaboration} et~al.,}{{Planck Collaboration} et~al.}{2014}]{planck13}
{Planck Collaboration} et~al., 2014, \mn@doi [\aap] {10.1051/0004-6361/201321591}, \href {https://ui.adsabs.harvard.edu/abs/2014A&A...571A..16P} {571, A16}

\bibitem[\protect\citeauthoryear{{Planck Collaboration} et~al.,}{{Planck Collaboration} et~al.}{2016}]{planck15}
{Planck Collaboration} et~al., 2016, \mn@doi [\aap] {10.1051/0004-6361/201525830}, \href {https://ui.adsabs.harvard.edu/abs/2016A&A...594A..13P} {594, A13}

\bibitem[\protect\citeauthoryear{{Prada}, {Behroozi}, {Ishiyama}, {Klypin}  \& {P{\'e}rez}}{{Prada} et~al.}{2023}]{Prada2023}
{Prada} F.,  {Behroozi} P.,  {Ishiyama} T.,  {Klypin} A.,   {P{\'e}rez} E.,  2023, \mn@doi [arXiv e-prints] {10.48550/arXiv.2304.11911}, \href {https://ui.adsabs.harvard.edu/abs/2023arXiv230411911P} {p. arXiv:2304.11911}

\bibitem[\protect\citeauthoryear{{Prakash} et~al.,}{{Prakash} et~al.}{2016}]{eboss_lrgs}
{Prakash} A.,  et~al., 2016, \mn@doi [\apjs] {10.3847/0067-0049/224/2/34}, \href {https://ui.adsabs.harvard.edu/abs/2016ApJS..224...34P} {224, 34}

\bibitem[\protect\citeauthoryear{{Reddick}, {Wechsler}, {Tinker}  \& {Behroozi}}{{Reddick} et~al.}{2013}]{Reddick_2013}
{Reddick} R.~M.,  {Wechsler} R.~H.,  {Tinker} J.~L.,   {Behroozi} P.~S.,  2013, \mn@doi [\apj] {10.1088/0004-637X/771/1/30}, \href {https://ui.adsabs.harvard.edu/abs/2013ApJ...771...30R} {771, 30}

\bibitem[\protect\citeauthoryear{Reid et~al.,}{Reid et~al.}{2015}]{Reid_lss}
Reid B.,  et~al., 2015, SDSS-III Baryon Oscillation Spectroscopic Survey Data Release 12: galaxy target selection and large scale structure catalogues, \mn@doi{10.48550/ARXIV.1509.06529}, \url {https://arxiv.org/abs/1509.06529}

\bibitem[\protect\citeauthoryear{{Rodr{\'\i}guez-Torres} et~al.,}{{Rodr{\'\i}guez-Torres} et~al.}{2016}]{Sergio}
{Rodr{\'\i}guez-Torres} S.~A.,  et~al., 2016, \mn@doi [\mnras] {10.1093/mnras/stw1014}, \href {https://ui.adsabs.harvard.edu/abs/2016MNRAS.460.1173R} {460, 1173}

\bibitem[\protect\citeauthoryear{{Ross} et~al.,}{{Ross} et~al.}{2017}]{Ashely_boss_lss}
{Ross} A.~J.,  et~al., 2017, \mn@doi [\mnras] {10.1093/mnras/stw2372}, \href {https://ui.adsabs.harvard.edu/abs/2017MNRAS.464.1168R} {464, 1168}

\bibitem[\protect\citeauthoryear{{Ross} et~al.,}{{Ross} et~al.}{2020}]{Ashely_eboss_lss}
{Ross} A.~J.,  et~al., 2020, \mn@doi [\mnras] {10.1093/mnras/staa2416}, \href {https://ui.adsabs.harvard.edu/abs/2020MNRAS.498.2354R} {498, 2354}

\bibitem[\protect\citeauthoryear{{Ruan}, {Hern{\'a}ndez-Aguayo}, {Li}, {Arnold}, {Baugh}, {Klypin}  \& {Prada}}{{Ruan} et~al.}{2022}]{Ruan2022}
{Ruan} C.-Z.,  {Hern{\'a}ndez-Aguayo} C.,  {Li} B.,  {Arnold} C.,  {Baugh} C.~M.,  {Klypin} A.,   {Prada} F.,  2022, \mn@doi [\jcap] {10.1088/1475-7516/2022/05/018}, \href {https://ui.adsabs.harvard.edu/abs/2022JCAP...05..018R} {2022, 018}

\bibitem[\protect\citeauthoryear{{Schaye} et~al.,}{{Schaye} et~al.}{2023}]{Flamingo2023}
{Schaye} J.,  et~al., 2023, \mn@doi [arXiv e-prints] {10.48550/arXiv.2306.04024}, \href {https://ui.adsabs.harvard.edu/abs/2023arXiv230604024S} {p. arXiv:2306.04024}

\bibitem[\protect\citeauthoryear{Sefusatti, Crocce, Scoccimarro  \& Couchman}{Sefusatti et~al.}{2016}]{Sefusatti2016}
Sefusatti E.,  Crocce M.,  Scoccimarro R.,   Couchman H. M.~P.,  2016, \mn@doi [Mon. Not. R. Astro. Soc.] {10.1093/mnras/stw1229}, 460, 3624

\bibitem[\protect\citeauthoryear{{Skibba}, {van den Bosch}, {Yang}, {More}, {Mo}  \& {Fontanot}}{{Skibba} et~al.}{2011}]{Skibba_2010}
{Skibba} R.~A.,  {van den Bosch} F.~C.,  {Yang} X.,  {More} S.,  {Mo} H.,   {Fontanot} F.,  2011, \mn@doi [\mnras] {10.1111/j.1365-2966.2010.17452.x}, \href {https://ui.adsabs.harvard.edu/abs/2011MNRAS.410..417S} {410, 417}

\bibitem[\protect\citeauthoryear{{Smith}, {Cole}, {Baugh}, {Zheng}, {Angulo}, {Norberg}  \& {Zehavi}}{{Smith} et~al.}{2017}]{Smith2017}
{Smith} A.,  {Cole} S.,  {Baugh} C.,  {Zheng} Z.,  {Angulo} R.,  {Norberg} P.,   {Zehavi} I.,  2017, \mn@doi [\mnras] {10.1093/mnras/stx1432}, \href {https://ui.adsabs.harvard.edu/abs/2017MNRAS.470.4646S} {470, 4646}

\bibitem[\protect\citeauthoryear{{Smith}, {Cole}, {Grove}, {Norberg}  \& {Zarrouk}}{{Smith} et~al.}{2022}]{Smith22b}
{Smith} A.,  {Cole} S.,  {Grove} C.,  {Norberg} P.,   {Zarrouk} P.,  2022, \mn@doi [\mnras] {10.1093/mnras/stac2519}, \href {https://ui.adsabs.harvard.edu/abs/2022MNRAS.516.4529S} {516, 4529}

\bibitem[\protect\citeauthoryear{{Springel}}{{Springel}}{2005}]{Springel_hidro}
{Springel} V.,  2005, \mn@doi [\mnras] {10.1111/j.1365-2966.2005.09655.x}, \href {https://ui.adsabs.harvard.edu/abs/2005MNRAS.364.1105S} {364, 1105}

\bibitem[\protect\citeauthoryear{{Stoppacher} et~al.,}{{Stoppacher} et~al.}{2019}]{Doris:2019}
{Stoppacher} D.,  et~al., 2019, \mn@doi [\mnras] {10.1093/mnras/stz797}, \href {https://ui.adsabs.harvard.edu/abs/2019MNRAS.486.1316S} {486, 1316}

\bibitem[\protect\citeauthoryear{{Szapudi} \& {Szalay}}{{Szapudi} \& {Szalay}}{1998}]{3pcf_estim}
{Szapudi} I.,  {Szalay} A.~S.,  1998, \mn@doi [\apjl] {10.1086/311146}, \href {https://ui.adsabs.harvard.edu/abs/1998ApJ...494L..41S} {494, L41}

\bibitem[\protect\citeauthoryear{{Trujillo-Gomez}, {Klypin}, {Primack}  \& {Romanowsky}}{{Trujillo-Gomez} et~al.}{2011}]{Trujillo}
{Trujillo-Gomez} S.,  {Klypin} A.,  {Primack} J.,   {Romanowsky} A.~J.,  2011, \mn@doi [\apj] {10.1088/0004-637X/742/1/16}, \href {https://ui.adsabs.harvard.edu/abs/2011ApJ...742...16T} {742, 16}

\bibitem[\protect\citeauthoryear{{Vale} \& {Ostriker}}{{Vale} \& {Ostriker}}{2006}]{Vale2006}
{Vale} A.,  {Ostriker} J.~P.,  2006, \mn@doi [\mnras] {10.1111/j.1365-2966.2006.10605.x}, \href {https://ui.adsabs.harvard.edu/abs/2006MNRAS.371.1173V} {371, 1173}

\bibitem[\protect\citeauthoryear{{Wadekar}, {Ivanov}  \& {Scoccimarro}}{{Wadekar} et~al.}{2020}]{Wadekar_2020}
{Wadekar} D.,  {Ivanov} M.~M.,   {Scoccimarro} R.,  2020, \mn@doi [\prd] {10.1103/PhysRevD.102.123521}, \href {https://ui.adsabs.harvard.edu/abs/2020PhRvD.102l3521W} {102, 123521}

\bibitem[\protect\citeauthoryear{Wechsler, Somerville, Bullock, Kolatt, Primack, Blumenthal  \& Dekel}{Wechsler et~al.}{2001}]{wechsler:2000bf}
Wechsler R.~H.,  Somerville R.~S.,  Bullock J.~S.,  Kolatt T.~S.,  Primack J.~R.,  Blumenthal G.~R.,   Dekel A.,  2001, \mn@doi [Astrophys. J.] {10.1086/321373}, 554, 85

\bibitem[\protect\citeauthoryear{{Yu} et~al.,}{{Yu} et~al.}{2022}]{Yu_2022}
{Yu} J.,  et~al., 2022, \mn@doi [\mnras] {10.1093/mnras/stac2176}, \href {https://ui.adsabs.harvard.edu/abs/2022MNRAS.516...57Y} {516, 57}

\bibitem[\protect\citeauthoryear{{Yu}, {Seljak}, {Li}  \& {Singh}}{{Yu} et~al.}{2023}]{Yu_2023}
{Yu} B.,  {Seljak} U.,  {Li} Y.,   {Singh} S.,  2023, \mn@doi [\jcap] {10.1088/1475-7516/2023/04/057}, \href {https://ui.adsabs.harvard.edu/abs/2023JCAP...04..057Y} {2023, 057}

\bibitem[\protect\citeauthoryear{{Zel'dovich}}{{Zel'dovich}}{1970}]{Zeldovich}
{Zel'dovich} Y.~B.,  1970, \aap, \href {https://ui.adsabs.harvard.edu/abs/1970A&A.....5...84Z} {5, 84}

\bibitem[\protect\citeauthoryear{{Zhao} et~al.,}{{Zhao} et~al.}{2021}]{Zhao_2021}
{Zhao} C.,  et~al., 2021, \mn@doi [\mnras] {10.1093/mnras/stab510}, \href {https://ui.adsabs.harvard.edu/abs/2021MNRAS.503.1149Z} {503, 1149}

\bibitem[\protect\citeauthoryear{{Zhou} et~al.,}{{Zhou} et~al.}{2021}]{Zhou_2020}
{Zhou} R.,  et~al., 2021, \mn@doi [\mnras] {10.1093/mnras/staa3764}, \href {https://ui.adsabs.harvard.edu/abs/2021MNRAS.501.3309Z} {501, 3309}

\bibitem[\protect\citeauthoryear{{de la Torre} et~al.,}{{de la Torre} et~al.}{2013}]{delatorre13}
{de la Torre} S.,  et~al., 2013, \mn@doi [\aap] {10.1051/0004-6361/201321463}, \href {https://ui.adsabs.harvard.edu/abs/2013A&A...557A..54D} {557, A54}

\makeatother
\end{thebibliography}
